\documentclass[journal,onecolumn]{IEEEtran}

\usepackage{cite}
\usepackage{amsmath,amssymb,amsfonts}
\usepackage{amsthm}
\usepackage[ruled,linesnumbered]{algorithm2e}
\usepackage{algorithmic}
\usepackage{graphicx}
\usepackage{textcomp}
\usepackage{dsfont}
\usepackage{xcolor}
\usepackage{bm}
\usepackage{soul}
\def\BibTeX{{\rm B\kern-.05em{\sc i\kern-.025em b}\kern-.08em
    T\kern-.1667em\lower.7ex\hbox{E}\kern-.125emX}}

\usepackage{tcolorbox}
\usepackage{siunitx}
\usepackage{etoolbox}
\usepackage{subfigure}
\usepackage{ifthen}
\newboolean{showcomments}
\setboolean{showcomments}{true}

\newcommand{\uts}[1]{\text{unpredictable task size}}
\newcommand{\pts}[1]{\text{predictable task size}}

\newcommand{\g}[3][\gamma]{D(#2,#3,#1)}
\newcommand{\newG}[3][\gamma,\lambda]{G(#2,#3,#1)}

\newcommand{\PTS}{\text{(PTS)}}
\newcommand{\UTS}{\text{(UTS)}}

\newcommand{\bap}{\ba^p}
\newcommand{\bau}{\ba^u}
\newcommand{\ba}{\boldsymbol{a}}

\usepackage{mathtools}

\newcommand{\bs}{\boldsymbol}

\DeclarePairedDelimiter{\ceil}{\lceil}{\rceil}
\DeclarePairedDelimiter{\set}{\{}{\}}

\DeclareMathOperator{\spansp}{span}

\newcommand{\taubar}{\bar{\tau}}
\newcommand{\w}{\omega}
\newcommand{\ymax}{y_{\max}}
\newcommand{\qmax}{q_{\max}}
\newcommand{\Yint}{\mathcal{Y}}

\newcommand{\Ebatch}{\varepsilon} 
\newcommand{\Etask}{W}
\newcommand{\E}{\mathbb{E}}
\newcommand{\Fbar}{\overline{F}}
\newcommand{\nn}{\nonumber\\}
\newcommand{\Tau}{\mathrm{T}}
\newcommand{\Rhat}{\hat{R}}
\newcommand{\btau}{\bs{\tau}}
\newcommand{\Wbar}{\bar{W}}
\newcommand{\Lbar}{\bar{L}}

\renewcommand{\ij}[2][i,j]{[#2]_{#1}}
\newcommand{\Tn}{\boldsymbol{\Tau}_{n}}
\newcommand{\T}{\boldsymbol{\Tau}}

\newtheorem{theorem}{Theorem}

\newtheorem{proposition}{Proposition}
\newtheorem{lemma}{Lemma}

\newtheorem{corollary}{Corollary}

\begin{document}

\title{Timely CPU Scheduling for Computation-intensive Status Updates
\thanks{
An earlier version of this paper was presented in part at the 2024 IEEE INFOCOM~\cite{10621420}.
}
}

\author{Mengqiu~Zhou,~\IEEEmembership{Student Member,~IEEE,}
Meng~Zhang,~\IEEEmembership{Member,~IEEE,}
Howard~H.~Yang,~\IEEEmembership{Member,~IEEE} and~Roy~D.~Yates~\IEEEmembership{Fellow,~IEEE}
\thanks{Mengqiu Zhou is with the College of Information Science and Electronic Engineering and ZJU-UIUC Institute, Zhejiang University, China (email: mengqiuzhou@zju.edu.cn).}
\thanks{Meng Zhang and Howard H. Yang are with the ZJU-UIUC Institute, Zhejiang University, China (e-mail: mengzhang@intl.zju.edu.cn; haoyang@intl.zju.edu.cn).}
\thanks{Roy D. Yates is with the Department of Electrical and Computer Engineering, Rutgers University, USA (email: ryates@winlab.rutgers.edu).}
}


\maketitle

\begin{abstract}
   The proliferation of mobile devices and real-time status updating applications has motivated the optimization of data freshness in the context of age of information (AoI). Meanwhile, increasing computational demands have inspired research on CPU scheduling. 
   Since prior CPU scheduling strategies have ignored data freshness and prior age-minimization strategies have considered only constant CPU speed,
   we formulate the first CPU scheduling problem as a constrained semi-Markov decision process (SMDP) problem with uncountable space, which aims to minimize the long-term average age of information, subject to an average CPU power constraint. 
   We optimize strategies that specify when the CPU sleeps and adapt the CPU speed (clock frequency) during the execution of update-processing tasks.
 We consider the age-minimal CPU scheduling problem 
for both predictable task size (PTS) and unpredictable task size (UTS) cases, where the task size is realized at the start (PTS) or at the completion (UTS) of the task, respectively.  To address the non-convex objective, we employ Dinkelbach's fractional programming method to transform our problem into an average cost SMDP. We develop a value-iteration-based algorithm and prove its convergence to obtain optimal policies and structural results 
for both the PTS and UTS systems. 
Compared to constant CPU speed, numerical results show that our proposed scheme can reduce the AoI by 50\% or more,  with increasing benefits under tighter power constraints. Further, for a given AoI target, the age-minimal CPU scheduling policy can reduce the energy consumption by 50\% or more, 
with greater AoI reductions when the task size distribution exhibits higher variance.
\end{abstract}

\section{Introduction}
\subsection{Background}
Data freshness is becoming increasingly significant due to the rapid growth of real-time applications. 
For instance, real-time updates from healthcare wearables are critical for monitoring the health of clients.
In the case of pneumonia patients, real-time oxygen metrics provided by oximeters play a vital role in their lung health management~\cite{teo2020early}.
Similarly, real-time traffic alerts are needed for transportation safety and real-time vehicular location and speed information is the key to reliable autonomous driving. An emerging metric to assess the freshness of data is the age of information (AoI)~\cite{yates2021age},
which measures the time that has elapsed since the most recent data update. 

Since high computer performance is critical for timely operation of  computation-intensive real-time applications,
the CPU can benefit from scheduling techniques to manage and process all activities~\cite{harki2020cpu}.
In this context,  various CPU scheduling techniques have been developed to guarantee system performance; e.g.,~\cite{goyal1996hierarchical} maximizes the throughput and~\cite{duda1999borrowed} supports low latency.
Since the CPU consumes considerable energy and often runs on ubiquitous battery-powered mobile computing devices (e.g., Internet-of-Things), energy management is critically important~\cite{8130358,9381665}, especially for edge and mobile devices~\cite{kadota2016minimizing, talak2020optimizing}. 
In addition, the growing demand for cloud infrastructure has drastically increased the energy consumption of data centers, making energy management a critical issue~\cite{berl2010energy}.
Extensive methods have been proposed for enhancing the energy management of cloud computing clusters, including energy-aware resource allocation algorithms that improve energy efficiency in data centers~\cite{beloglazov2012energy} and 
the Voltage Island Largest Capacity First (VILCF) algorithm for energy-efficient scheduling of periodic real-time tasks on multi-core processors~\cite{liu2016energy}.

To trade off power consumption and computer performance, researchers have proposed numerous solutions, including circuit and architectural techniques~\cite{venkatachalam2005power}. 
In particular, the \textit{Dynamic Voltage Scaling} (DVS) mechanism has been widely applied in computer systems to achieve power reductions subject to task-based performance guarantees~\cite{gruian2001hard,flautner2002vertigo}. 
DVS reduces power dissipation by reducing CPU voltage, as it exploits an important feature of CMOS-based processors that dynamic power $P$ is quadratic in voltage $V$ and linear in clock frequency $\omega$, i.e.,  $P\propto V^2\omega$. 
Moreover, as captured in the $\alpha$ approximation~\cite{gonzalez1997supply}, the frequency $\omega$ is proportional to $V^{\alpha-1}$, where the technology-dependent parameter $\alpha\in(1,2]$ accounts for velocity saturation. 
Consequently, the goal of DVS is to reduce energy consumption by dynamically adapting the voltage (therefore the execution speed) of the processor as a task progresses toward a completion deadline~\cite{yao1995scheduling,weiser1996scheduling}. 

Regrettably, the existing CPU scheduling schemes mainly focus on meeting deadlines of computation tasks without consideration of data freshness \cite{lee2009minimizing,mishra2014energy,haripriya2022energy,yuan2006energy}.
This motivates our study of {\em age-minimal CPU scheduling policies in an update-processing system}.
To this end, we 
optimize CPU scheduling strategies that specify when the CPU waits (i.e. sleeps) and adapt the CPU speed during the execution of update-processing tasks, subject to a power budget.
We will see that the interaction between the CPU scheduling and waiting strategy introduces significant challenges in the problem formulation.

In real-time computing systems, tasks are often divided into batches and executed by the processor batch-by-batch to enhance performance and efficiency.
However, the range of tasks includes data transfer where the number of batches in a task, which we define as the {\em task size}, is observable upon task initiation
as well as tasks such as blockchain mining
where the task size is observed only upon task completion. We refer to these cases as the predictable task size (PTS) and unpredictable task size (UTS) systems. 
While the latter UTS case was introduced in our prior work~\cite{10621420},
here we consider both scenarios.
Due to the structural differences in the system information of these two cases, we will see it is challenging to unify the state, action, and dynamic processes of these two scenarios.

\subsection{Contributions}
We formulate the age-minimal CPU scheduling problem as a constrained semi-Markov Decision Process (SMDP)~\cite{puterman2014markov} problem with uncountable state space and a fractional objective function for both 
PTS and UTS cases.

First, we observe that our problem combines two types of actions: selecting a waiting time and then, following the wait, selecting the CPU scaling strategy. Furthermore, the PTS system 
presents an additional challenge as the
extra observation of the task size is revealed at the end of the waiting period when the new updating task is generated. To address this issue, we design for PTS an action matrix that enumerates CPU strategies for all task sizes so that the CPU schedule is specified by the observation of the task size. With this approach, the task size state in the PTS system is mapped into the action dimension. This unifies the PTS and UTS problem formulations in that both systems permit the combined action (wait and CPU schedule) to be specified at the completion of the prior updating task based on the execution time of the previous task as the system state.
Moreover, this approach transforms the unified problem formulation into a standard SMDP.

In this unified problem setting, the difficulty of the constrained SMDP problem with uncountable state space is well known and optimality equations (e.g., Bellman's equation) cannot be applied directly~\cite{feinberg1994constrained,bertsekas2012dynamic}. 
We adopt a quantization method to overcome the issue of the uncountable state space. 
Different from Sun \textit{et al.}~\cite{sun2017update} which also studied this class of problem, the state transition probabilities of our SMDP depend on decisions and thus requires a value-iteration-based method to obtain optimal solutions. Our contributions include: 
\begin{itemize}
\item \textit{Problem Formulation:} To the best of our knowledge, this paper formulates the first age-minimal CPU scheduling problem via CPU speed adaptation and waiting strategies.

\item \textit{Unification of PTS and UTS systems:} To unify the PTS and UTS problem formulations, we introduce for PTS  an action matrix that incorporates
a CPU strategy for every task size combined with a selection process when the task is generated and the task size is learned.

\item \textit{Age-minimal CPU Scheduling:} We first prove the existence of a stationary randomized policy for a quantized state space that achieves asymptotic optimality in the uncountable state space as the quantization step size goes to zero. To overcome the nonconvexity of our fractional cost SMDP problem, we transform it into an average cost problem and apply the Generalized Benders Decomposition method.
To obtain the optimal solution that achieves the tradeoff between AoI and energy saving, we develop a value-iteration-based algorithm consisting of four loops and we construct a contraction mapping operator to prove its convergence in the uncountable space. 

\item \textit{Structure of the Optimal Policy:} We characterize the structure of the optimal policies for both PTS and UTS cases.  
Our structural results indicate that the optimal wait is a generalization of the waiting strategy found in \cite{sun2017update}.
Specifically, the optimal CPU strategy for UTS is to accelerate as the task progresses, with CPU speed adjustment for each batch occurring only when there is no waiting in the system.
For PTS, the optimal CPU strategy is to operate at a higher constant speed when handling tasks with larger task sizes and speed up for each task size scenario when there is no waiting.

\item \textit{Numerical Results:} 
By adopting a piece-wise constant approximation in numerical evaluation, we validate our analysis of the structure of the optimal age-minimal CPU scheduling policy for PTS and UTS cases. 
In addition, we also compare the age-minimal CPU scheduling scheme to four existing benchmarks in terms of long-term average AoI. 
Numerical results show that the age-minimal CPU scheduling scheme can reduce the AoI by $53\%$ in UTS and $55\%$ in PTS relative to the existing benchmarks
and that 
greater AoI reductions are achieved as the average power constraint becomes stricter.
\end{itemize}

\section{Related Work}
\label{sec:related work}
In this section, we briefly review the work related to our proposed age-minimal CPU scheduling. We focus on two areas: AoI which measures the freshness of data, and the CPU scheduling.

\subsection{Age of Information}
The age of information (AoI) was proposed as a data freshness metric as early as the 1990s in the studies of real-time databases~(e.g.,~\cite{segev1991optimal, adelberg1995applying}). 
In recent years, \cite{kaul2012real} has motivated many studies of data freshness in various settings,
such as in queueing systems (e.g.,~\cite{talak2020age,yates2018age,bedewy2019minimizing,kosta2021age}), wireless networks (e.g.,~\cite{sun2017update,arafa2019age,buyukates2020scaling,sun2019sampling}), and edge computing (e.g.,~\cite{kuang2020analysis,chiariotti2021peak,chen2021information,zou2021optimizing,chen2023joint}). 
 Talak \textit{et al.}~\cite{talak2020age} achieves the age-delay tradeoffs in a multi-server system where the updates are generated according to a Poisson process.
Yates \textit{et al.}~\cite{yates2018age} characterized the region of feasible status ages in a wide variety of multiple source service systems.
In~\cite{bedewy2019minimizing}, Bedewy \textit{et al.} investigates scheduling policies that minimize the age of information in single-hop queueing systems.
Kosta \textit{et al.} considers the information freshness in a discrete-time queueing system in~\cite{kosta2021age}.
In addition, Arafa \textit{et al.} designs the status update transmission policy sucthat the long-term average age of information is minimized for energy-harvesting sensors with finite batteries in~\cite{arafa2019age}. 
Buyukates \textit{et al.} study the AoI in a multiple source multiple destination setting with a focus on its scaling in large wireless networks in~\cite{buyukates2020scaling}.
In~\cite{sun2019sampling}, Sun \textit{et al.} investigate how to take samples at a data source for improving the received data freshness at a remote receiver.
Specifically, references in \cite{kuang2020analysis,chiariotti2021peak,chen2021information,zou2021optimizing,chen2023joint} mainly considered to optimize the task offloading, arrival probabilities, and/or channel allocation
to minimize the age of information for status update in mobile edge computing systems.
The most related study is \cite{sun2017update}, in which Sun \textit{et al.} characterized the optimal generation of fresh information updates.
\textit{However, all of these efforts assumed that servers have fixed service rates and did not consider state adaptive variable speed service policies.}

\subsection{CPU Scheduling}
There are many CPU scheduling methods that aim to reduce energy consumption while guaranteeing performance~\cite{6127864,kocot2023energy}.
Dynamic voltage scaling (DVS) is one of the classical power management techniques in computer architecture. 
It achieves energy savings by dynamically adapting the processor's voltage/frequency, while ensuring a task completion deadline is met. 
There are many existing works focus on the analysis of energy-saving CPU scheduling schemes determined by DVS, see, e.g.,~\cite{lee2009minimizing,mishra2014energy,haripriya2022energy, yuan2006energy}.
Studies in~\cite{lee2009minimizing,mishra2014energy,haripriya2022energy} addressed the task scheduling problem on different computing systems and minimized energy consumption while guaranteeing scheduling quality. 
In particular, 
\cite{yuan2006energy}
proposed a simple yet effective histogram technique to estimate the probability distribution of CPU cycle demands, and maximized the energy saving of DVS by accelerating CPU speed as the task progressed.

In addition, thermal-aware scheduling methods are widely used to save energy. In particular, scheduling tasks 
to manage the CPU's temperature can effectively avoid overheating, which reduces the need for energy-intensive cooling mechanisms while ensuring tasks are completed within their deadlines. For example, 
Choi \textit{et al.} \cite{choi2007thermal} enabled lowering of on-chip unit temperatures by changing the workload in a timely manner while 
Fisher \textit{et al.} \cite{4840574} proposed a global scheduling algorithm that exploits the flexibility of multicore platforms at low temperature.
\textit{However, these works focused only on adapting the processing speed to meet individual task deadlines, but did not consider AoI or data freshness.}

\section{System Model}
\label{sec:system model}
\begin{figure}[t]
\centering
\includegraphics[width=0.7\linewidth]{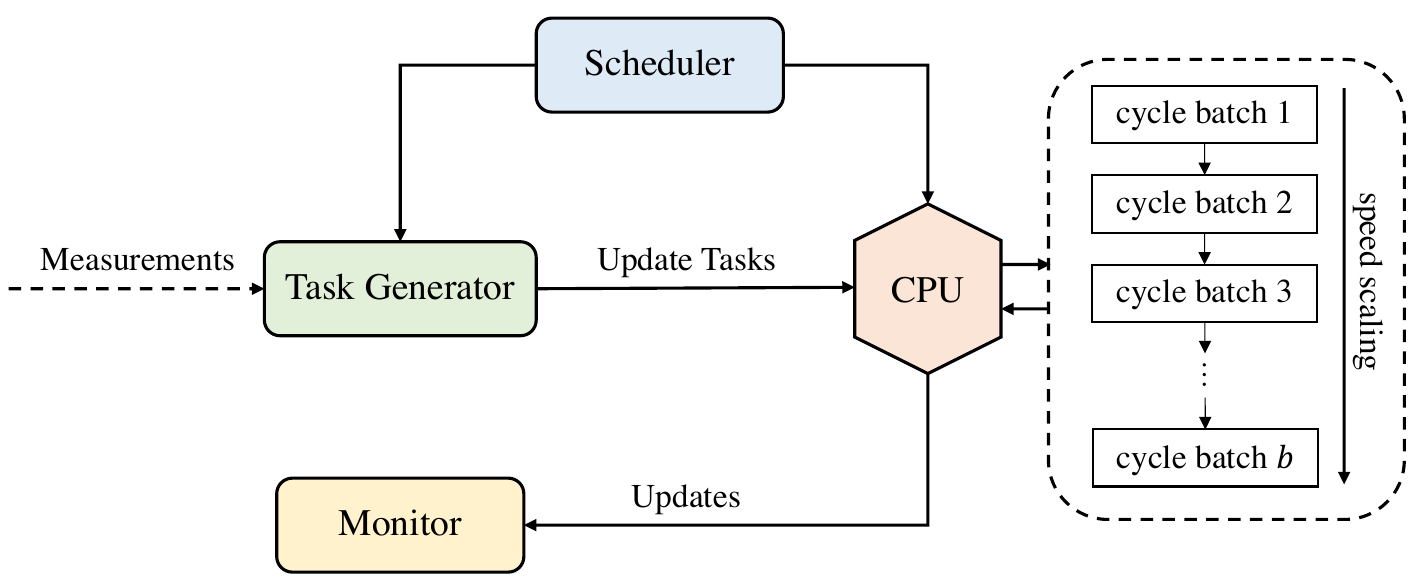}
\caption{The system model of age-minimal CPU scheduling.}\label{fig:model}
\end{figure}

In this section, we introduce an information update system with a CPU that operates at an adaptive clock frequency. As the first study addressing timely updating with CPU scheduling, we start with the system depicted in Fig.~\ref{fig:model} consisting of a monitor, a CPU, a scheduler, and a task generator, similar to~\cite{sun2017update,arafa2019timely}. The CPU has a single core with adaptive speed and is single-threaded.
We now describe models for the update task demand, the CPU speed and power, and the CPU scheduling policy in order to then formulate the timely  CPU scheduling problem.

\subsubsection*{Notation} We summarize key notation as follows.
For any positive integer $A$, $[A]\triangleq\{1,2,\cdots,A\}$  denotes the set of positive integers up to $A$. 
In addition, $\mathbb{R}_+$ denotes the set of all positive real numbers.
The notations $[x]^+\triangleq\max(x,0)$ and $[x]_a^b\triangleq\min(\max(x,a),b)$ are also used in this paper.

\subsection{Update Task Demand Model}\label{model}
We consider a \textit{generate-at-will} update processing model, as in \cite{sun2017update,arafa2019timely}.
The task generator observes an external measurement process and decides when to generate a (computation-intensive) fresh update. Associated with each update is an information processing task 
requiring various computational resources~\cite{yuan2006energy}.
Specifically, 
the random variable $C_n$ denotes the CPU demand (total cycle requirement) of the computation task corresponding to the $n$th update. Because it is impractical to adjust the CPU clock frequency on a per-cycle basis, we partition the CPU demand of a task into batches of $c$ cycles over which the clock frequency is held constant. 
For example, a video encoder task can be divided into several batches of $10^6$ cycles in practice~\cite{yuan2006energy}.
When  task $n$ has a CPU demand of $C_n$ cycles, the processor will execute $X_n=\ceil{C_n/c}$ batches.
Furthermore, 
we assume that $1\le X_n\le b$, where $b$ is the maximum possible number of batches in a task. 

Henceforth, it will be convenient to measure the CPU demand in batches and so we will refer to $X_n$ as the CPU demand of task $n$.  
In addition, the sequence of CPU demands $X_1, X_2,\ldots$ is assumed to be i.i.d.\footnote{The i.i.d. assumption has been widely adopted (e.g.,\cite{flautner2002vertigo,yuan2006energy}). 
We note that for Markovian CPU demand, the corresponding extension is conceptually straightforward by including demand $X_n$ in the state of the SMDP.
}
with each $X_n$ having cumulative distribution function  $F(x)\triangleq{\rm Pr}\left[X_n \le x \right]$.\footnote{When the demand distribution is unknown, it is possible to apply online estimation techniques of such distributions as in \cite{yuan2006energy}. This will be left for future work.}  
We will see that it also will be useful to define 
\begin{align}\label{eq:Fbar}
f(x)&\triangleq{\rm Pr}[X_n= x]\quad{\rm and}\nn
\Fbar(x)&\triangleq {\rm Pr}[X_n\ge x]=1-F(x-1),\quad x\in[b].
\end{align}

Given that real-world computational tasks can be divided into several batches and then be executed by processors accordingly, we consider the following two general cases regarding when knowledge of the task size $X_n$ is realized by the scheduler:
\begin{itemize}
    \item \textit{Predictable Task Size (PTS)}: The task size $X_n$ is realized at the start of task $n$. This scenario often appears in networked systems.
    For example, in the transmission of text, video, and other forms of information, 
    the task size of the task (i.e., the number of data packets
    to be transmitted)
    is known in advance~\cite{lee2001applied}. As a result, the system has access to the specific task size $X_n$ of the task before it starts and then it decides the appropriate processing speed.
     Another example is neural network training, where a predefined batch size hyperparameter is adopted to determine the number of data samples processed before updating the model's weights~\cite{lecun2015deep}.
     Similarly, in big data processing, frameworks like MapReduce (e.g., Hadoop~\cite{white2012hadoop}, Spark~\cite{armbrust2015spark}), operate on a fixed batch of input data at a time.
     This ensures efficient resource utilization while adhering to system configuration and memory constraints.
    \item \textit{Unpredictable Task Size (UTS)}: The task size $X_n$ is realized upon the completion of task $n$. 
This scenario arises in blockchain technology~\cite{zheng2018blockchain}.
    The proof-of-work computation to add a block is based on the evaluation of random guesses such that the required number of guesses 
    is a discrete uniform random variable. In this context, the process to add a block can be viewed as a task with an unpredictable task size such that each batch represents the evaluation of a guess. 
    Similarly, in wireless event-driven sensor networks, the amount of data generated depends on the nature of the detected event (e.g., seismic activity or fire)~\cite{akyildiz2002survey}. High-intensity events may produce significantly more data, making the task size unpredictable until the event concludes. 
\end{itemize}
By exploring both the PTS and UTS cases, we develop a CPU scheduling methodology that has general applicability across different scenarios.

\subsection{CPU Speed and Power}
\label{subsec:CPU power}
We consider a mobile device (e.g., a smartphone or an Internet-of-Things device) with a single processor, which has a single core and is single-threaded. 
We employ an \textit{ideal processor} model~\cite{yuan2006energy} 
that supports a continuous range of operating speeds (clock frequencies) $\w\in[\w_{\min},\w_{\max}]$.
\footnote{In practice, mobile devices often have a non-ideal processor, where the processor supports a discrete set of speeds, rather than a continuous range. Unlike the speed schedule for an ideal CPU, the speed schedule for a non-ideal CPU is NP-hard, which will be discussed in future work.}

The CPU power (energy consumption rate) typically consists of two parts: dynamic power consumption and static power consumption. A lower CPU speed, however, may increase the power of other resources such as memory. Since our goal is to reduce the total energy consumed by the whole device, we are more interested in the total power consumed by the device at different CPU speeds. 
We assume that the energy consumption and the delay incurred by CPU speed changes are negligible\footnote{For example, the energy consumption resulting from maximum speed variation of the lpARM processor~\cite{pering2000voltage} is negligibly small.}
and we ignore the static power and just focus on the dynamic power. Under these assumptions, the dynamic power consumption of a CMOS integrated circuit (e.g., a modern computer processor) with loading capacitance $C_L$, clock frequency $\omega$, and supply voltage $V$ is given by \cite{wolf2016physics},
\begin{align}
\label{eq:dvs}
    P=C_LV^2\omega.
\end{align}
By the $\alpha$ approximation~\cite{gonzalez1997supply}, the proportional relationship between clock frequency $\w$ and voltage $V$ is given by
\begin{align}
\label{eq:chip}
    \omega \propto \frac{(V-V_{\rm th})^\alpha}{V},
\end{align}
where $\alpha$ is a technology-dependent constant that accounts for velocity saturation and $V_{\rm th}$ is the threshold voltage. 

The parameter $\alpha$ ranges from $1$, complete velocity saturation, to $2$, no velocity saturation (for example in $1000$nm technology and older). For $0.25\mu$m technology, $\alpha$ is typically in the range $[1.3,1.5]$~\cite{gonzalez1997supply}. Due to the condition that $V_{\rm th}$ is negligible relative to $V$~\cite{wolf2016physics}, we drop $V_{\rm th}$ and thus \eqref{eq:chip} implies $\w \propto V^{\alpha-1}$. 
Equivalently, the CPU at frequency $\w$ has voltage 
\begin{equation}\label{eq:Vpropto}
  V \propto \w^{1/(\alpha-1)}.   
\end{equation}

We note that it will be convenient to parameterize the CPU speed by the execution time of a cycle batch. Specifically, a batch $x$ executed at frequency $\w_x$ will have execution time $\tau_x=c/\w_x$. Since the energy consumption of this  batch is equal to the power $P$ times the batch duration $\tau_x$, it follows from \eqref{eq:dvs} and \eqref{eq:Vpropto} that the energy consumed  to process batch $x$ is
\begin{align}
P\cdot\tau_x =C_LV^2 \w_x\tau_x= cC_LV^2 \propto \w_x^{\frac{2}{\alpha-1}},\quad x\in[b].
\end{align}
By choosing the energy unit to absorb scaling constants, the energy needed to process a cycle batch in time $\tau_{x}$ is 
\begin{align}\label{eq:energy}
\Ebatch(\tau_x)\triangleq\left(\frac{1}{\tau_x}\right)^{\frac{2}{\alpha-1}},\quad x\in[b].
\end{align}

\subsection{CPU Scheduling Decisions}
Upon completion of the computational task associated with an update,
the scheduler decides on \textit{when} to generate its next update task and \textit{how fast} (i.e., CPU speed scaling) to execute it. 
Specifically, for the $n$th task, the scheduler must decide 
\begin{itemize}
    \item \textit{Waiting (Sleep)} time: 
    Upon completion of task $n-1$, the task generator will sleep for time $Z_n\in\mathbb{R}_+$ prior to generating task $n$. 
    \item \textit{CPU Scaling}: 
    The CPU processes the first batch of $c$ cycles at frequency $\w_{n,1}$, the second batch at frequency $\omega_{n,2}$, and so on until the task is completed. 
    Equivalently, each cycle batch $x$ requires execution time $\tau_{n,x}=c/\w_{n,x}$ and we use these execution times to specify the CPU scaling policy.
    \begin{subequations}
    \label{eqn:tau-defns}
\begin{itemize}
        \item \textit{In the UTS case}, the CPU scaling policy for task $n$ is specified by the vector of execution times 
        \begin{equation}
        \boldsymbol{\tau}_n=
        \begin{bmatrix}
            \tau_{n,1} & \tau_{n,2} &\cdots&\tau_{n,b}.
        \end{bmatrix}  
        \end{equation}
        Due to the unpredictable task size, the CPU processes each task $n$ batch by batch. The system first selects time $\tau_{n,1}$ to execute batch $1$.
        When batch $1$ is finished, the system either learns that task $n$ is completed, or the system assigns $\tau_{n,2}$ for the execution time of batch $2$. 
        This sequential batch-by-batch CPU speed adaptation continues until the entire task is completed. This allows the system to adapt dynamically in that $\tau_{n,j}$ for batch $j$ is chosen with the knowledge that $X_n\ge j$.
        \item \textit{In the PTS case}, we treat the observed value of $X_{n}$ of task $n$ as a sub-state and we introduce the $b\times b$ lower triangular matrix $\boldsymbol{\Tau}_{n}$ to denote the CPU scheduling matrix.
 When the number of batches $X_{n}=x$ in task $n$ is observed, row $x$ of $\boldsymbol{\Tau}_{n}$ determines the CPU strategy. Specifically, with $\ij{\Tn}$ denoting element $i,j$ of the matrix $\Tn$, the observation $X_n=x$ selects the CPU scaling policy 
\begin{equation}
 \hat{\boldsymbol{\tau}}_{n}(x)
    = \begin{bmatrix}
    \ij[x,1]{\Tn} & \ij[x,2]{\Tn} &\cdots & \ij[x,x]{\Tn}.
    \end{bmatrix}
\end{equation}
Notably, the action set $\boldsymbol{\Tau}_{n}$ already incorporates the sub-state $X_{n}$.
    \end{itemize}       
    \end{subequations}
    We emphasize that for both UTS and PTS, the scaling decision for task $n$, namely $\boldsymbol{\tau}_n$ for UTS or $\boldsymbol{\Tau}_{n}$ for PTS, is specified at the completion of the prior updating task.
\end{itemize}
   
Henceforth, we use $\bs{a}^u_n\triangleq (Z_n,\boldsymbol{\tau}_n)$ and $\bs{a}^p_n\triangleq (Z_n,\boldsymbol{\Tau}_{n})$ to denote the scheduling actions for update $n$ in the UTS and PTS cases. Furthermore, we denote the sets of all feasible decisions in the respective cases by
\begin{subequations}
\begin{IEEEeqnarray}{rCl}
    \mathcal{A}^u &\triangleq&\set{(Z_n,\boldsymbol{\tau}_n)\colon Z_n\in\mathbb{R}_+, \tau_{n,x}\in[\tau_{\min},\tau_{\max}],x\in[b]},\IEEEeqnarraynumspace\\
\mathcal{A}^p&\triangleq&\set{(Z_n,\Tn)\colon Z_n\in\mathbb{R}_+, \Tn\in \mathcal{L}_b(\tau_{\min},\tau_{\max})},
\end{IEEEeqnarray}
\end{subequations}
where $\mathcal{L}_b(\tau_{\min},\tau_{\max})$ is the set of all 
 $b\times b$ lower triangular  matrices with lower triangular entries in the interval $[\tau_{\min},\tau_{\max}]$.

\begin{figure}[t]
\centering
\includegraphics[width=0.7\linewidth]{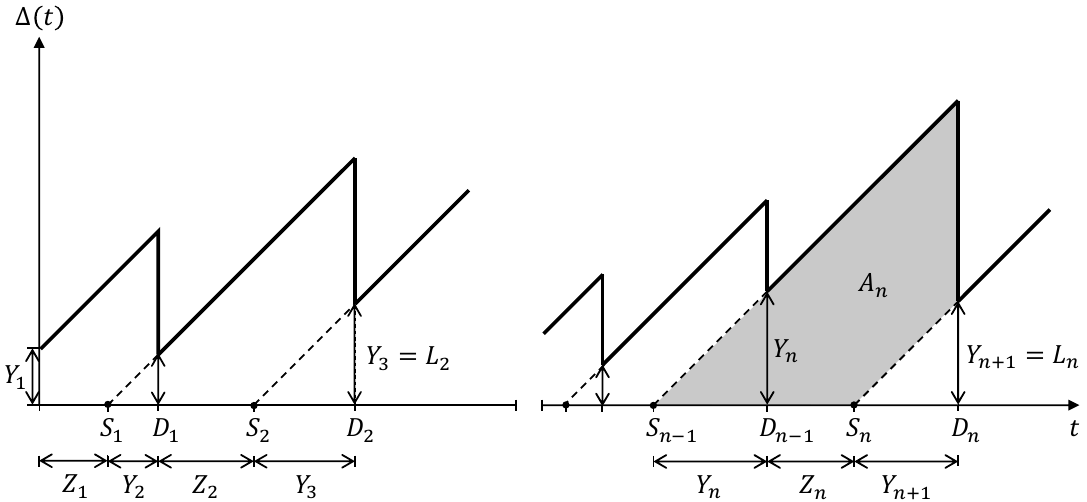}
\caption{Age of information $\Delta(t)$ evolution in time.}
\label{fig:age}
\end{figure}

\subsection{Age of Information}
\label{subsec:aoi}
To measure the information freshness, we consider the \textit{age of information (AoI)} metric from the perspective of the destination monitor~\cite{yates2021age}. 
In our system model, updates are generated  
as a 
stochastic process at times $\set{S_n \colon n=1,2,\ldots}$. The $n$th update is time-stamped $S_n$ since it is derived from the measurement of an external process of interest at that time instant. Each update $n$ requires processing by the CPU before being delivered to the destination monitor. Based on the CPU scheduling policy, the overall computation time for update $n$ is 
\begin{align}\label{eq:Ln}
    L_n=
    \begin{cases}
        L^{\ba_n^u}(X_n)=\sum_{x=1}^{X_n}\tau_{n,x}, & \UTS,\\
        L^{\bap_n}(X_n)=\sum_{x=1}^{X_n} \ij[X_n,x]{\Tn}, & \PTS.
    \end{cases}
\end{align}
so that update $n$ is delivered to the monitor at time $D_n = S_n+L_n$. Let $\mathcal{Y}\triangleq [\tau_{\min},b\tau_{\max}]$ denote the set of all possible values of $Y_n$.
Hence, at the monitor, the freshest delivered update at time $t$ has time-stamp 
\begin{align}
U(t)\triangleq \max_{n} \set{S_n| D_n \leq t},
\end{align}
and the age of information (AoI) is \cite{kaul2012real}
\begin{align}
    \Delta(t)\triangleq t- U(t).
\end{align}
As depicted in Fig.~\ref{fig:age}, the AoI process $\Delta(t)$ has a characteristic sawtooth shape, growing at unit rate in the absence of an update but dropping at each time $D_n$ to $L_n$, the age of the update received at that time instant. 
For update $0$, we set $D_{0}=0$ without loss of generality. 
After the delivery of update $n-1$, the CPU waits for a time $Z_n$ before generating update $n$.

AoI analysis was initiated in \cite{kaul2012real}, which  defined the average AoI as the time-average $\lim_{T\to\infty} \frac{1}{T}\int_0^T\Delta(t)\,dt$. This time-average was evaluated by graphically decomposing the integral into areas $A_n$, as shown in Fig.~\ref{fig:age}. Following this approach, we define epoch $n$ as the time interval $[S_{n-1},S_{n})$ and we observe from Fig.~\ref{fig:age} that associated with this epoch is the area 
\begin{IEEEeqnarray}{rCl} 
A_n=A(Y_n,Z_n,L_n)&\triangleq&
\frac{1}{2}\left(Y_{n}+Z_{n}+L_{n}\right)^{2}
-\frac{1}{2}L_{n}^{2}\nn
&=& (Y_n+Z_n)L_n +\frac{1}{2}(Y_n+Z_n)^2.\IEEEeqnarraynumspace\label{eq:Q-defn} 
\end{IEEEeqnarray}
It follows that the long-term average AoI is thus given by \cite{yates2021age}:
\begin{align}
{\Delta}^{(\text{ave})}\triangleq&\limsup_{N\rightarrow \infty}\frac{ \sum_{n=1}^N \mathbb{E}[A(Y_n,Z_n,L_{n})]}{\sum_{n=1}^N \mathbb{E}[Y_{n}+Z_{n}]}. \label{average-AoI}
\end{align}

\begin{figure}[t]
\centering
\includegraphics[width=0.5\linewidth]{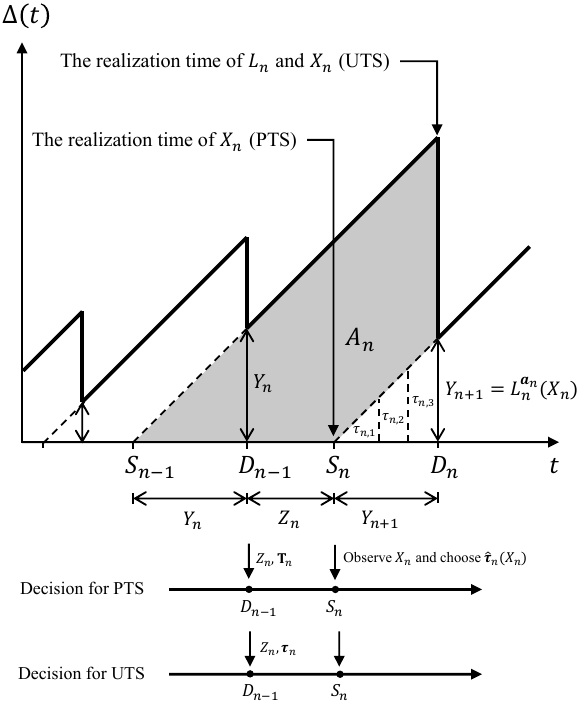}
\caption{Age of information $\Delta(t)$ evolution in time and the time-slot for batch demand $X_{n}$ is realized at the beginning of task $n+1$ and after the completion of task $n+1$.}
\label{fig:two_cases_dynamic}
\end{figure}

Our goal is to minimize the average AoI ${\Delta}^{(ave)}$ by controlling the sequence of actions, including waiting times and computation times. 
We frame the age-minimal CPU scheduling problem in the context of both unpredictable and predictable task sizes, formulating it as a general SMDP problem. 
Specifically, the details are as follows:
\begin{enumerate} 
\item In the UTS case, depicted in Fig.~\ref{fig:two_cases_dynamic}, the system decides the waiting time $Z_n$ and the CPU scaling strategy $\boldsymbol{\tau}_{n}$  based only on the observation of $Y_n$.
    The task size $X_{n}$ is unknown at the initiation of task $n$ in UTS and the system only becomes aware of the values of $L_n$ and $X_n$ at time $D_n$ when task $n$ is completed. 

    \item In the PTS scenario also depicted in Fig.~\ref{fig:two_cases_dynamic}, the system decides the waiting time $Z_n$ and the CPU scaling strategy $\bs{\Tau}_{n}$  based only on the observation of $Y_n$.
    After learning the task size $X_{n}$ when update $n$ is generated at time $S_{n}$, the system chooses the corresponding $X_{n}$-th row of $\bs{\Tau}_{n}$, i.e., $\hat{\bs{\tau}}_{n}(X_{n})$ as the CPU scaling policy for task $n$.
\end{enumerate}

\subsection{Problem Formulation}
\subsubsection*{Scheduling Policies} Let $\pi_u\triangleq (\bau_0,\bau_1, \bs{a}^u_2,\ldots,\bs{a}^u_N)$ and $\pi_p\triangleq (\bs{a}^p_0,\bs{a}^p_1, \bs{a}^p_2,\ldots,\bs{a}^p_N)$ denote the scheduling policies in UTS and PTS, respectively.

Since an update computation requires cycles statistically, 
a cycle batch is executed with a certain probability. Specifically, it follows from \eqref{eq:energy} and \eqref{eqn:tau-defns} that processing update $n$ 
requires energy
\begin{align}\label{eq:Wn}
    \Etask_{n}=
    \begin{cases}
    \sum\limits_{j=1}^{X_n} \Ebatch(\tau_{n,j}), & \UTS,\\
        \sum\limits_{j=1}^{X_n} \Ebatch(\ij[X_n,j]{\Tn}), & \PTS.
    \end{cases}
\end{align}
Recalling the definition~\eqref{eq:Fbar} that $\Fbar(x)={\rm Pr}[X_n\ge x]$, 
the expected energy needed to process update $n$ is~\footnote{Inspection of Figure~\ref{fig:two_cases_dynamic} shows that the energy $\Etask_n$ expended on update $n$ actually occurs in epoch $n+1$, although the commitment to expend $\Etask_n$ is made in epoch $n$. This one epoch delay in energy expenditure makes no difference in the average energy expended over time.}
\begin{align}\label{eq:Etask-defn}
\E[\Etask_{n}]&=\begin{cases}
\sum\limits_{x=1}^b\Fbar(x)\Ebatch(\tau_{n,x}), & \UTS,\\
	\sum\limits_{x=1}^bf(x)
 \sum_{j=1}^x \Ebatch(\ij[x,j]{\Tn}), & \PTS.
	\end{cases}
\end{align}

We consider the class of \textit{causal} policies, in which decisions are based on the history $(Y_1,Y_{2},\ldots,Y_{n})$ for both PTS and UTS cases.
Let $\Pi_P$ and $\Pi_U$ denote the set of all feasible policies $\pi_p$ and $\pi_u$ in the PTS and UTS cases, respectively.
For both cases of $\Pi\in\{\Pi_{P},\Pi_U\}$, we then formulate the \textit{age-minimal CPU scheduling problem} as 
\begin{subequations}
\label{eq:minimize_age}
\begin{align}
\label{eq:gamma^*}
\gamma^*\triangleq\min_{\pi\in\Pi}&\quad\limsup_{N\rightarrow \infty}\frac{ \sum_{n=1}^N \mathbb{E}[A(Y_n,Z_n,L_n)]}{ \sum_{n=1}^N \mathbb{E}[Y_n+Z_{n}]}, \\
    {\rm s.t.}&\quad\limsup_{N\to\infty}\frac{\sum_{n=1}^N\E[\Etask_{n}]}{\sum_{n=1}^N\mathbb{E}\left[Y_n+Z_{n}\right]} \leq \bar{P},\label{eq:avePower}
\end{align}
\end{subequations} 
where $\gamma^*$ is the optimum objective value (i.e., the minimal AoI) of Problem \eqref{eq:minimize_age} and
$\bar{P}$ is the maximum average power.

To enable the following discussion to be applicable to both UTS and PTS, we use $\ba_n$ as a generic notation for the action, corresponding to  $\bs{a}^u_n= (Z_n,\boldsymbol{\tau}_n)$ for UTS or  $\bs{a}^p_n=(Z_n,\boldsymbol{\Tau}_{n})$ for PTS.
At the start of epoch $n$, the system observes the service of task $n-1$, which delivers an update to the  monitor of age $Y_n=L_{n-1}$. Given $Y_n$ and the policy $\pi(\ba_{n}\vert Y_n)$, it follows from \eqref{eq:Ln} that the next state $Y_{n+1}=L_n$ is specified by the action $\ba_{n}$ and the task size $X_{n}$, which is independent of the past history.
This implies that the transition probability to $Y_{n+1}$ depends only on the previous state $Y_n$. Thus $Y_n$ is the state of an embedded Markov chain.
In addition, we observe that epoch $n$ has duration $Y_n+Z_n$ with associated age cost $A_n$ in~\eqref{eq:Q-defn} and  and energy cost $\Etask_n$ in~\eqref{eq:Wn}.
Since the decision $\ba_n$ is made after observing  the state $Y_{n}$ of the embedded Markov chain,
our problem is a constrained SMDP with an uncountable state space~\cite{ross1970average}.

Since SMDP problems are a generalized form of MDP appropriate for modeling continuous-time discrete-event systems~\cite{sutton1999between}, Problem~\eqref{eq:minimize_age} belongs to the class of infinite-state and action SMDPs with constraints. It is known to be difficult to solve such a problem.
Most of the existing studies (e.g., \cite{feinberg1994constrained,feinberg2002constrained})
consider constrained SMDPs with finite state and action sets. To the best of our knowledge, only \cite{sun2017update} has considered a constrained SMDP problem with an uncountable state space, but only in the special case in which the state transition probabilities do not depend on decisions.

We use $\Pi_{\rm SR}$ to denote the set of stationary randomized policies satisfying $\bs{a}_n\in\mathcal{A}$ for all $n\in\mathbb{N}$. 
Different from~\cite{sun2017update}, the state transition probabilities of our SMDP depend on decisions.
Motivated by~\cite{saldi2017asymptotic}, we characterize the optimality of stationary policies in the following theorem.
\begin{theorem}[Stationarity]\label{theorem_stationary}
    There always exists a stationary randomized policy $\pi\in\Pi_{\rm SR}$ that is asymptotically optimal for Problem~\eqref{eq:minimize_age}.
\end{theorem}
\begin{IEEEproof}
    See Appendix~\ref{proof:stationarity}.
\end{IEEEproof}

By employing a stationary policy, $Y_n$, $Z_n$, $L_n$ and $\Etask_n$ are asymptotically stationary 
and we can drop the index $n$.
Therefore, Problem~\eqref{eq:minimize_age} can be simplified to 
\begin{subequations}\label{eq:minimize_age_stat}
\begin{align}
\min_{\pi\in\Pi_{\rm SR}}&\quad\frac{\mathbb{E}[A(Y,Z,L)]}{\mathbb{E}[Y+Z]}, \label{eq:yzl}\\
    {\rm s.t.}&\quad\frac{\E[\Etask]}{\mathbb{E}\left[Y+Z\right]} \leq \bar{P}.
\end{align}
\end{subequations}

\section{PTS and UTS Unified Problem Formulation}\label{sec:pbs}
In this section, we prove the existence of an optimal constant batch speed policy for the PTS system. This will enable a unified problem formulation for the PTS and UTS cases. 

Under this unified formulation, we develop a series of techniques to derive the optimal solutions. 
In particular, we transform the problem with a fractional cost into an equivalent one with an average cost and prove the existence of a stationary deterministic policy that is optimal.
Finally, we summarize the key steps in a value-iteration-based algorithm with a convergence guarantee.

With the observation of the task size $X=x$ in the PTS system, we demonstrate in the following lemma that processing each batch at a constant rate $\tau_x$ performs as effectively as any dynamic rate $\hat{\bs{\tau}}(x)$ policy.
\begin{lemma}[PTS Lemma]\label{Lemma_constant}
With the observation of task size $X=x$, there exists an optimal  CPU schedule for the PTS system with constant speed $\tau_x$
such that 
\begin{align}
\ij[x,1]{\bs{\Tau}}=\ij[x,2]{\bs{\Tau}}=\cdots=\ij[x,x]{\bs{\Tau}}=\tau_x.
\end{align}
\end{lemma}
\begin{IEEEproof}
Given $X=x$ and a CPU schedule that yields a service time $T$, the schedule with constant speed $\tau_x=T/x$ completes at the same time but convexity of the energy function $\Ebatch(\cdot)$ implies the constant speed policy consumes less energy.     See details in Appendix~\ref{proof:constant}.
\end{IEEEproof}
It follows from Lemma~\ref{Lemma_constant} that we can restrict our search to stationary randomized PTS policies $\pi_p\in \Pi_{\rm SR,P}$ of the form  $(Z,\tau_1,\tau_2,\cdots,\tau_b)$ that specify the same speed $\tau_x$ for all batches whenever the task size is $X=x$. Thus, for both PTS and UTS, the CPU scheduling policy can be specified by the vector $\btau=(\tau_1,\tau_2,\cdots,\tau_b)$. Furthermore, the action at each step is specified by the vector $\ba=(Z,\btau)$ belonging to the action space
\begin{IEEEeqnarray}{rCl}
\mathcal{A} &\triangleq& \set{(Z,\btau)\colon Z\in\mathbb{R}_+, \tau_{x}\in[\tau_{\min},\tau_{\max}],x\in[b]},\IEEEeqnarraynumspace
\end{IEEEeqnarray} 

We emphasize that the interpretation of $\btau$ depends on whether the system is PTS or UTS. In the PTS system, the number of batches $X=x$ is specified at the beginning of the job, and speed $\tau_x$ is employed for all $x$ batches. For UTS, $X$ is unknown at the start of the job and the system proceeds sequentially with processing batch $1$ at speed $\tau_1$, batch $2$ at speed $\tau_2$ and so on.

In terms of the task size $X$ and the policy $\btau$, it follows from \eqref{eq:Ln} that the overall computation time for an update  is 
\begin{align}\label{eq:L2}
    L(X;\btau)=
    \begin{cases}
X\tau_X, & \PTS,\\
\sum_{x=1}^{X}\tau_{x}, & \UTS.
    \end{cases}
\end{align}
By averaging over the task size $X$, the expected processing time is
\begin{align}\label{eq:Lbar}
\Lbar(\btau)=\E[L(X;\btau)]&=\begin{cases}
\sum\limits_{x=1}^bf(x) x\tau_x, & \PTS,\\
\sum\limits_{x=1}^b\Fbar(x)\tau_{x}, & \UTS.
	\end{cases}
\end{align}
Similarly, 
it follows from \eqref{eq:Wn} and \eqref{eq:Etask-defn} that processing an update of size $X$ 
requires energy
\begin{align}\label{eq:Wtau}
    \Etask(X;\btau)=
    \begin{cases}
X\Ebatch(\tau_X), & \PTS,\\
\sum_{x=1}^{X} \Ebatch(\tau_{x}), & \UTS.    \end{cases}
\end{align}
which has expected value
\begin{align}
\label{eq:Wbar}
\Wbar(\btau)&=\E[\Etask(X;\btau)]=\begin{cases}
\sum\limits_{x=1}^bf(x) x \Ebatch(\tau_x), & \PTS,\\
\sum\limits_{x=1}^b\Fbar(x)\Ebatch(\tau_{x}), & \UTS.
	\end{cases}
\end{align}

Problem~\eqref{eq:minimize_age_stat} 
can now be simplified as
\begin{subequations}\label{eq:problem_known_constant}
    \begin{align}
\gamma^*=\min_{\pi\in\Pi_{\rm SR}}&\quad\frac{\mathbb{E}[A(Y,Z,L(X;\btau)])}{\mathbb{E}[Y+Z]},\label{eq:problem_known_constant-objective}\\
        {\rm s.t.}&\quad\frac{
\mathbb{E}\left[
        W(X;\btau)
        \right]}{\mathbb{E}\left[Y+Z\right]} \leq \bar{P}.
\label{eq:problem_known_constant-constraint}
    \end{align}
\end{subequations}
In Problem~\eqref{eq:problem_known_constant}, the objective and constraint depend on the action $\ba=(Z,\btau)$, a random vector that may depend on the prior service time $Y$, as defined by the policy $\pi\in\Pi_{\rm SR}$.

\section{Average-Cost Transformation}
Since Problem~\eqref{eq:problem_known_constant} is a constrained SMDP with an uncountable state space and a non-convex fractional objective, it is not directly solvable. We will overcome this challenge by employing the Dinkelbach fractional programming method \cite{ren2005markov,dinkelbach1967nonlinear,serfozo1979equivalence}  to transform Problem~\eqref{eq:problem_known_constant} into an average-cost SMDP, which facilitates the subsequent optimization framework.

For Problem~\eqref{eq:problem_known_constant}, 
with motivation by \eqref{eq:Q-defn} 
and incorporating averaging over the task size $X$ as needed, we define the age cost function
\begin{subequations} \label{gfunc}
\begin{align}
    g_{1}(y,\ba)&=(y+z)\Lbar(\btau)+\frac{1}{2}\left(y+z\right)^2,\\
    \intertext{the epoch duration cost function}
    g_{2}(y,\ba)&=y+z,
\intertext{and the power-based cost function}
g_{3}(y,\ba)&=\Wbar(\btau)-(y+z)\bar{P}.\label{g3p}
\end{align}
\end{subequations}
With these definitions, the fractional objective 
\eqref{eq:problem_known_constant-objective} can be rewritten as $\mathbb{E}[g_{1}(Y,\ba)]/\mathbb{E}[g_{2}(Y,\ba)]$ and the power constraint 
\eqref{eq:problem_known_constant-constraint} 
becomes $\mathbb{E}[g_{3}(Y,\ba)]\leq 0$.

\subsubsection*{Dinkelbach's Transformation}
We use the fractional programming technique to overcome the nonconvexity of the objective $\mathbb{E}[g_{1}(Y,\ba)]/\mathbb{E}[g_{2}(Y,\ba)]$ by optimizing the expected value of the difference 
\begin{equation}\label{eq:Gdefn}
    \g{Y}{\ba} \triangleq g_1(Y,\ba)-\gamma g_2(Y,\ba),
\end{equation}
where $\gamma$ is the Dinkelbach variable.
With this definition, the Dinkelbach reformulation is
\begin{subequations}\label{eq:dinkel_pbs}
\begin{align}
    J(\gamma)=\min_{\pi\in\Pi_{\rm SR}}\quad &\mathbb{E}[\g{Y}{\ba}],\label{eq:J_gamma_pbs}\\ 
    {\rm s.t.}\quad &\mathbb{E}[g_{3}(Y,\ba)]\leq 0, \label{eq:constraint_known}
\end{align}
\end{subequations}

Given the optimal policy $\pi^*$ to Problem~\eqref{eq:problem_known_constant}, we have $\mathbb{E}_{\pi}[\g[\gamma^*]{Y}{\ba}]\geq0$ and $\mathbb{E}_{\pi^*}[\g[\gamma^*]{Y}{\ba}]=0$, indicating that $\pi^*$ is also optimal to Problem~\eqref{eq:dinkel_pbs}. This established the equivalence of Dinkelbach's transformation in the stationary policy space.
The following lemma~\cite[Theorem]{dinkelbach1967nonlinear}
guarantees the equivalence of the transformed problem.
\begin{lemma}\cite[Theorem]{dinkelbach1967nonlinear}\label{Lemma_Dinkel}
When $\gamma^*$ is the optimum objective value of Problem \eqref{eq:problem_known_constant}, Problem \eqref{eq:dinkel_pbs} with $\gamma=\gamma^*$ is equivalent to Problem \eqref{eq:problem_known_constant}
and $J(\gamma^*)=0$. 
\end{lemma}

Lemma \ref{Lemma_Dinkel} shows the possibility of equivalently reformulating the fractional objective into one with an average cost. In addition, the Dinkelbach's method suggests that the optimum objective $\gamma^*$ of Problem \eqref{eq:problem_known_constant}  satisfies  $J(\gamma^*)=0$ and hence conducting a line search over $\gamma$ leads to the optimal policy $\pi^*$.
We note that the average-cost objective is essential for the value iteration that will be presented in Section~\ref{subsec:algorithm_pbs}. 

\subsubsection*{Dual Problem} 
Furthermore, we use the Lagrangian duality theory to handle the constraint condition in Eq.~\eqref{eq:constraint_known}.
We note that our analysis is significantly 
different from traditional finite dimensional optimization problems, as we adopt the
Karush-Kuhn-Tucker (KKT) theorem for infinite space~\cite{butnariu2000totally} in our method. With $\lambda$ denoting the dual variable, we define
\begin{equation}
    \newG{Y}{\ba}\triangleq  \g{Y}{\ba}+\lambda g_{3}(Y,\ba)
\end{equation}
and consider the \textit{dual function}
\begin{align}
d(\gamma,\lambda)=\min\limits_{\pi\in\Pi_{\rm SR}}\mathbb{E}\left[\newG{Y}{\ba}\right]. 
\label{dual-function}
\end{align}
With respect to Problem \eqref{eq:dinkel_pbs}, the \textit{dual problem} is given by
\begin{align}\label{eq:dual_pbs}
    \max_{\lambda\geq 0}~d(\gamma,\lambda).
\end{align}
Given the optimal $\gamma^*$, the optimal dual variable is thus
\begin{align}
\lambda^*\triangleq\arg\max\limits_{\lambda\geq 0}d(\gamma^*,\lambda).
\end{align}

\subsubsection*{Value Function} 
To further characterize the optimality condition of Problem \eqref{eq:dinkel_pbs}, given the initial state $Y_1=y$, we define the value function under the stationary 
policy $\pi$ as 
\begin{align}\label{eq:define_Rp}
R^{\pi}(y)\triangleq\limsup_{N\to\infty}\frac{1}{N}&\mathbb{E}_{\pi}\Bigl[\sum\limits_{n=1}^N \bigl(\newG{Y_n}{\ba_n}
\vert Y_1=y\Bigr].
\end{align}
The optimal value function is then defined as
\begin{align}
    R^{*}(y)\triangleq \min_{\pi} R^{\pi}(y), \label{R-optimalvalue}
\end{align}
with $\pi^*$ denoting an optimal policy. According to~\cite[Definition 2.2]{hernandez1991average}, a policy $\pi^*$ is optimal for Problem \eqref{eq:dinkel_pbs} if and only if
\begin{align}
\label{eq:optima_policy_pbs}
\pi^*=\arg\min\limits_{\pi\in\Pi_{\rm SR}}
    \mathbb{E}&\big[\newG[\gamma^*]{Y}{\ba}
\big].
\end{align}

Therefore, the constrained SMDP problem with the fractional cost is equivalent to an unconstrained SMDP with the average cost~\cite{ren2005markov}. However, Eq.~\eqref{eq:optima_policy_pbs} cannot be directly used to compute the optimal policy because the optimal fractional cost $\gamma^*$ and the optimal Lagrangian dual variable $\lambda^*$ are unknown.

Motivated by \cite[Theorem 8.4.4]{puterman2014markov}, we use $p^{\ba}(l\vert y)$ to denote the probability measure for transitioning to state $l$ when the system takes action $\bs{a}$ in state $y$ in order to characterize the optimal policy in the following theorem.
\begin{theorem}
\label{theorem_optimality_pbs}
There exists an optimal value function $R^*(y)$ such that the optimal policy $\pi^*$ solves
\begin{align}\label{eq:pi(y)_pbs}
R^{*}(y)=~&\min \limits_{\bap\in\mathcal{A}^p}\biggl[\int p^{\ba}(l\vert y)R^{*}(l)dl 
+\newG[\gamma^*]{y}{\ba}
\biggr].
\end{align}
\end{theorem}
\begin{IEEEproof}
    See Appendix~\ref{proof:optimality_pbs}.
\end{IEEEproof}
Theorem~\ref{theorem_optimality_pbs} indicates that the optimal policy is a stationary deterministic policy, i.e., $\pi^*(y)$ is a deterministic function. In addition, Theorem~\ref{theorem_optimality_pbs} also shows the existence of the optimal value function such that the optimal policy can be decomposed across all states $y\in\mathcal{Y}$, i.e., the optimal action corresponding to state $y$ only involves solving a subproblem (as in Eq.~\eqref{eq:pi(y)_pbs}). 
That is, the optimal value function $R^{*}(y)$ is obtained when the right side of the above equation is minimized by choosing an optimal action $\ba^{*}$.
Based on Theorem~\ref{theorem_optimality_pbs}, we obtain the optimal solutions of Eq.~\eqref{eq:dinkel_pbs} via value iteration algorithm in the next section.

\section{Age-minimal CPU Scheduling Algorithm}\label{subsec:algorithm_pbs}
We now describe the key steps of Algorithm~\ref{alg:dual_pbs} which converges to the optimal stationary deterministic age-minimal CPU scheduling policy satisfying Theorem~\ref{theorem_optimality_pbs}.
The global structure of Algorithm~\ref{alg:dual_pbs} consists of four loops, including Generalized Benders Decomposition, Value Iteration, Dual Update, and Dinkelbach Update.

With initialization $R^0(y)=0$ at $m=0$, for any given $\lambda$ and $\gamma$, the value function $R^{m+1}(y)$ is obtained iteratively by solving for each $y\in\mathcal{Y}$,
\begin{align}
R^{m+1}(y)
&=\min\limits_{\ba\in \mathcal{A}}\Big[\int p^{\ba}(l\vert y)R^{m}(l)dl+\newG{y}{\ba}
\Big] \label{value itera1}
\\
&=\min\limits_{\ba\in \mathcal{A}}\Big[
\E[R^m(L(X;\bs{\tau}))]
+\newG{y}{\ba} 
\Big]. \label{value itera2}
\end{align}
In \eqref{value itera2},  the system determines the action $\ba=(z,\bs{\tau})$ for each state $y$.
In this process, the next task size is $X=x$ with probability $f(x)$ and this implies next system state is $l=L(x;\btau_x)$. In particular, it follows from \eqref{eq:L2} that 
\begin{IEEEeqnarray}{rCl}
\E[R^m(L(X;\btau))]&=&\begin{dcases}
    \sum_{x=1}^b f(x)R^m(x\tau_x), &\PTS,\\
    \sum_{x=1}^b f(x)R^m\biggl(\sum_{k=1}^x \tau_k\biggr),& \UTS.
\end{dcases}
\IEEEeqnarraynumspace
\end{IEEEeqnarray}
This value iteration transform is the key step to Algorithm~\ref{alg:dual_pbs}.

We note that Sun \textit{et al.} in~\cite{sun2017update} studied a similar class of problem but did not require a value function mainly because the considered state transition probabilities were independent of decisions. 
Specifically, to obtain the value function, we define the objective in Eq.~\eqref{value itera2} for iteration $m$ as the Q-function
\begin{equation}\label{eq:Q-pbs}
Q^m(y,\ba)= \E[R^m(L(X;\btau))]+\newG{y}{\ba}.
\end{equation}
We minimize \eqref{eq:Q-pbs}
for each state $y$ by finding the optimal policy 
$\bs{a}^{m}(y)=(z^m(y),\bs{\tau}^m(y))$, which can be expressed as
\begin{align}\label{eq:gradient_pbs}
(z^m(y),\bs{\tau}^m(y))\in\arg\min\limits_{z,\boldsymbol{\tau}} Q^m(y,(z,\boldsymbol{\tau})).
\end{align}
In addition, \eqref{eq:gradient_pbs} has the fixed point $Q^*(y,\ba)$, given as
\begin{equation}\label{eq:Qp-star}
    Q^*(y,\ba) =\E[R^*(L(X;\btau))]+\newG[\gamma^*,\lambda]{y}{\ba}.
\end{equation}
Because the Hessian matrix of second-order partial derivatives of $Q^m(y,\ba)$ is positive semidefinite for all $\bs{\tau}$ with fixed $z$ and $\mathcal{A}$ is a nonempty compact set, we employ a generalized Benders decomposition~\cite{geoffrion1972generalized} method to solve Problem~\eqref{eq:gradient_pbs}. Keeping in mind that the action $\ba=(z,\bs{\tau})$ specifies the waiting time $z$ and the CPU schedule $\bs{\tau}$, Problem~\eqref{eq:gradient_pbs} is equivalently transformed as
\begin{align}\label{eq:benders_pbs}
\min\limits_{z\geq 0}~\min\limits_{\bs{\tau}\in[\tau_{\min},\tau_{\max}]^{b}}~{Q}^m(y,(z,\bs{\tau})).
\end{align}

\begin{algorithm}[t]
\caption{Value Iteration for PTS or UTS}
\label{alg:ValueItera}
\KwIn{Dual variable $\lambda$ and Dinkelbach's variable $\gamma$\;}
Initialize the convergence criteria $\epsilon_{\ba}$ and $\epsilon_R$\;
Set $m=0$ and $R^0(y)=0$ for all $y\in\mathcal{Y}$\; 
\Repeat{$\Vert R^{m+1}(\cdot)-R^m(\cdot)\Vert\leq\epsilon_R$}{Set $m \leftarrow m+1$ and $\kappa=0$\label{alg:begin}\; 
\Repeat{$\Vert Q^m(y,(z[\kappa+1],\bs{\tau}[\kappa]))-Q^m(y,(z[\kappa],\bs{\tau}[\kappa]))\Vert_2\leq \epsilon_{\ba}$}{Set $\kappa \leftarrow \kappa+1$\; 
Compute $\frac{\partial R^m(y)}{\partial y}$ according to~\eqref{eq:par_R_par_y}\; 
Fixing $z[\kappa]$, obtain $\bs{\tau}^*$ via projected gradient descent of $\nabla_{\bs{\tau}}Q^m(y,(z[\kappa],\bs{\tau}))$ in~\eqref{eq:gradient_bartau}\;
Set $\bs{\tau}[\kappa]=\bs{\tau}^*$\;
Obtain $z^*$ via projected gradient descent of $\partial Q^m(y,(z,\bs{\tau}[\kappa]))/\partial z$ in~\eqref{eq:gradient_z}  \label{alg:end}\;
Set $z[\kappa+1]=z^*$\;
}
Update the value function $R^{m+1}(y)$ based on~\eqref{value itera2}\;
}
\end{algorithm}

In lines \ref{alg:begin}-\ref{alg:end} of Algorithm~\ref{alg:ValueItera}, we solve Problem~\eqref{eq:benders_pbs} using a gradient-based method  that employs the partial derivatives $\partial Q^m/\partial \tau_x$ and $\partial Q^m/\partial z$ of the Q-function 
${Q}^m(y,(z,\bs{\tau}))$. 
We now present these partial derivatives.

We start by observing that it follows from \eqref{value itera2} and the envelope theorem~\cite{takayama1985mathematical}  that 
\begin{align}
\label{eq:par_R_par_y}
    \frac{\partial R^m(y)}{\partial y}=\Lbar(\btau^m(y))+y+z^m(y)-\gamma-\lambda\bar{P}.
\end{align}

Next, we observe that \eqref{gfunc} and \eqref{eq:Q-pbs} imply   
\begin{IEEEeqnarray}{rCl}
\frac{\partial Q^m(y,(z,\btau))}{\partial \tau_x}
=\begin{dcases}
xf(x)\biggl(\frac{\partial R^{m}(x\tau_x)}{\partial y}+y+z
+\frac{2\lambda\tau_x^{\frac{1+\alpha}{1-\alpha}}}{1-\alpha}\biggr), &\PTS, \\
\sum_{k'=x}^b f(k')\frac{\partial R^{m}(\sum_{x=1}^{k'}\tau_x)}{\partial y}+\Fbar(x)\biggl(y+z+\frac{2\lambda\tau_{x}^{\frac{1+\alpha}{1-\alpha}}}{1-\alpha}\biggr), & \UTS.
\end{dcases}\label{eq:gradient_bartau}
\IEEEeqnarraynumspace
\end{IEEEeqnarray}

Furthermore, \eqref{gfunc} and \eqref{eq:Q-pbs} also imply
\begin{equation}
\frac{\partial Q^m(y,(z,\btau))
}{\partial z}= \Lbar(\btau)+y+z-\gamma-\lambda\bar{P}.\label{eq:gradient_z}
\end{equation}

For both the PTS and UTS systems, the generalized Benders decomposition, which alternately updates $\tau_x$ and $z$ while keeping the other fixed, has a provable convergence guarantee~\cite{geoffrion1972generalized}. Finally, from Theorem~\ref{theorem_optimality_pbs}, if $R_p^{m}(y)$ converges to the optimal value function $R^*_p(y)$, then the corresponding policy $\bs{a}^{m}(y)$ converges to the optimal policy $\pi^*$ as well.

\subsection{Dual Update} As the dual problem~\eqref{eq:dual_pbs} is always convex, we adopt a sub-gradient method 
\begin{IEEEeqnarray}{rCl}\label{eq:dual_update_pbs}
    \lambda[\ell+1]&=&\left[\lambda[\ell]-s_{\ell}\left(\bar{P}\mathbb{E}_{\ell}\left[Y+Z\right]-\E_{\ell}[\Etask(X;\btau)]\right)\right]^{\,\mathclap{+}}\IEEEeqnarraynumspace
\end{IEEEeqnarray}
to update the dual variable at iteration index $\ell$ using positive step-size $s_{\ell}$.
Note that  $\mathbb{E}_{\ell}[\cdot]$ 
represents the expectation with respect to $X$, as generated by $f(x)$,  and $(Y, \btau)$, whose joint stationary distribution  is derived from the policy 
$\pi^*_{\ell}$ that minimizes the right side of 
\eqref{dual-function} when $\lambda=\lambda[\ell]$.

Note that the sub-gradient method is a widely-used iterative method for solving general convex optimization problems, and the dual variable $\lambda[\ell+1]$ will converge to the optimal $\lambda^*$ as $\ell\to\infty$~\cite{palomar2006tutorial}.
Therefore, the convergence of the proposed dual update algorithm can be guaranteed.

\subsection{Dinkelbach Variable Update}
Based on Lemma~\ref{Lemma_Dinkel} and the fact that $J(\gamma)$ decreases in $\gamma$, we adopt the bisection method to search for the optimal Dinkelbach variable $\gamma^*$ such that $J(\gamma^*)=0$.
Following Problem \eqref{eq:problem_known_constant}, when Algorithm~\ref{alg:dual_pbs} converges, the minimum average age subject to the power constraint is $\gamma^*$.

\begin{algorithm}[t]
\caption{Age-minimal CPU Scheduling}
\label{alg:dual_pbs}
\KwIn{Convergence criteria $\epsilon_{\gamma}$ and $\epsilon_{\lambda}$ and the step size $s_{\ell}$\;}
Initialize $l=0$ and sufficiently large $u$\;
\Repeat{
$\Vert u-l\Vert\leq\epsilon_{\gamma}$}{
$\gamma:=(l+u)/2$ and set $\ell=0$\;

\Repeat{$|\lambda[\ell+1]-\lambda[\ell]|\leq\epsilon_{\lambda}$}{
Set $\ell\leftarrow \ell+1$\;
Obtain optimal policy $\pi^*(y, \lambda[\ell])$ given $\lambda[\ell]$ from Algorithm~\ref{alg:ValueItera}\; 
Perform the sub-gradient method to update $\lambda[\ell]$ according to~(\ref{eq:dual_update_pbs})\;
}
Compute $J(\gamma)$ according to~\eqref{eq:J_gamma_pbs}\;
\eIf{$J(\gamma)\leq 0$}{$u:=\gamma$}{$l:=\gamma$}
}
\end{algorithm}

\subsection{Convergence Guarantee}
We provide theoretical guarantees of convergence of Algorithm~\ref{alg:dual_pbs} in this subsection.
Traditional dynamic programming methods such as value iteration~\cite{bellman1966dynamic} and policy iteration~\cite{howard1960dynamic} have been commonly used for solving MDP problems.
In particular, value iteration is one of the fundamental and widely adopted algorithms~\cite{bertsekas2012dynamic,pashenkova1996value}.
The key idea behind value iteration is to break down the complex objective problem into sequence of subproblems and solve them iteratively. 
In~\cite{puterman2014markov}, Puterman introduced variants of value iteration to solve various MDP models, including discounted MDP problems and the average reward criterion. 
However, all these methods considered discrete-state spaces and focused primarily on finite-space models. One characteristic of our algorithm is that we provide the convergence of value iteration in Theorem~\ref{theorem_convergence} for our SMDP problem with continuous state space.
\begin{theorem}\label{theorem_convergence}
For given $\lambda$ and $\gamma$, the sequences of value functions $\{R^m(\cdot)\}_{m}$ generated by Algorithm \ref{alg:ValueItera} converges to the optimal value function $R^*(\cdot)$.
\end{theorem}
\begin{IEEEproof}
    See Appendix~\ref{proof:convergence}.
\end{IEEEproof}

Theorem~\ref{theorem_convergence} is nontrivial due to the uncountable state and action spaces. In particular, the state probability distribution in our model depends on decisions as well as the task size.

\section{Analysis}
\label{sec:analysis}
In this section, we employ Theorem~\ref{theorem_optimality_pbs} to derive  structural results for age-minimal CPU scheduling policies. We examine the PTS and UTS systems in  subsections~\ref{subsect-PTS} and~\ref{subsect-UTS}. To improve the clarity of these results, we  define
\begin{align}
    H_{\alpha}(u)\triangleq\left[u^{\frac{\alpha-1}{\alpha+1}}\right]_{\tau_{\min}}^{\tau_{\max}},
\end{align} 
In addition, we explicitly state the following lemma to facilitate its application in the forthcoming analysis.
\begin{lemma}[Differentiability of Optimal Value Function]\label{lemma:diff_Ry}
The optimal value function $R^*(y)$ satisfies
\begin{align}
    R^*(y)=\min_{z,\bs{\tau}} Q^*(y,(z,\bs{\tau}))=Q^*(y,(z^*,\bs{\tau}^*)).
\end{align}
According to the envelope theorem, $R^*(y)$ is differentiable and it follows from \eqref{eq:Qp-star} that
\begin{align}\label{eq:par_Rpl}
    \frac{\partial R^*(y)}{\partial y}&=\frac{\partial Q^*(y,(z^*,\bs{\tau}^*))}{\partial y} \nn
    &=\Lbar(\btau^*(y))+y+z^*(y)-\gamma^*-\lambda^*\bar{P}. 
\end{align}
\end{lemma}

Finally, subsection~\ref{sec: toy example} presents a closed-form expression of the optimal policy in the special case of a deterministic number of cycle batches when the PTS and UTS systems are one in the same.

\subsection{Structure of the Optimal Policy for PTS}
\label{subsect-PTS} 
For the PTS system, we define
\begin{align}
     \psi_x(y)\triangleq \frac{\partial R^*(l)}{\partial l}\bigg|_{l=x\tau_x^*(y)}. \label{eq:psi_pbs}
\end{align}
The optimal structure is stated in the following corollary.
\begin{corollary}[Optimal Structure for PTS]\label{corollary_solution_pbs}
Given the minimal AoI $\gamma^*$ and the optimal dual variable $\lambda^*$, the optimal PTS policy $\pi^*$ 
for all state $y$ satisfies 
\begin{subequations}\label{eq:structure_pbs}
\begin{align}\label{eq:z+y_pbs}
    z^*(y)&=\left[\gamma^*+\lambda^*\bar{P}-y-\Lbar(\btau^*(y))
    \right]^+,\\ 
    \tau_x^*(y)&=H_{\alpha}\left(\frac{2\lambda^*/(\alpha-1)}{\psi_x(y)+y+z^*(y)}\right),~x\in[b], \label{eq:tau_pbs}\\
    \psi_x(y)
    &=\sum\limits_{x'=1}^bf(x')x'\tau_{x'}^*(x\tau^*_x(y))+x\tau^*_x(y)+z^*(x\tau^*_x(y))-\gamma^*-\lambda^*\bar{P},~x\in[b]. \label{eq:psi_pts1}
\end{align}
\end{subequations}
\end{corollary}
\begin{IEEEproof}
    See Appendix~\ref{proof:solution_pbs}.
\end{IEEEproof}

We observe that Eq.~\eqref{eq:z+y_pbs} exhibits a \textit{water-filling-like} solution, where $\gamma^*+\lambda^*\bar{P}$ represents the water-level.
Specifically, the sum $y+\Lbar(\btau^*(y))$
is bounded below by $y$ and therefore for sufficiently large $y$, the waiting time for the next update will be zero.
In contrast, for small $y$ the CPU typically sleeps for some time.  Intuitively, the optimal waiting attempts to direct the generation times of updates towards being equally spaced.
On the other hand, the optimal CPU scaling $\tau_x^*(y)$ is determined by the fixed point relation in \eqref{eq:structure_pbs}.

\begin{figure}[t]
    \centering
    \includegraphics[width=0.6\linewidth]{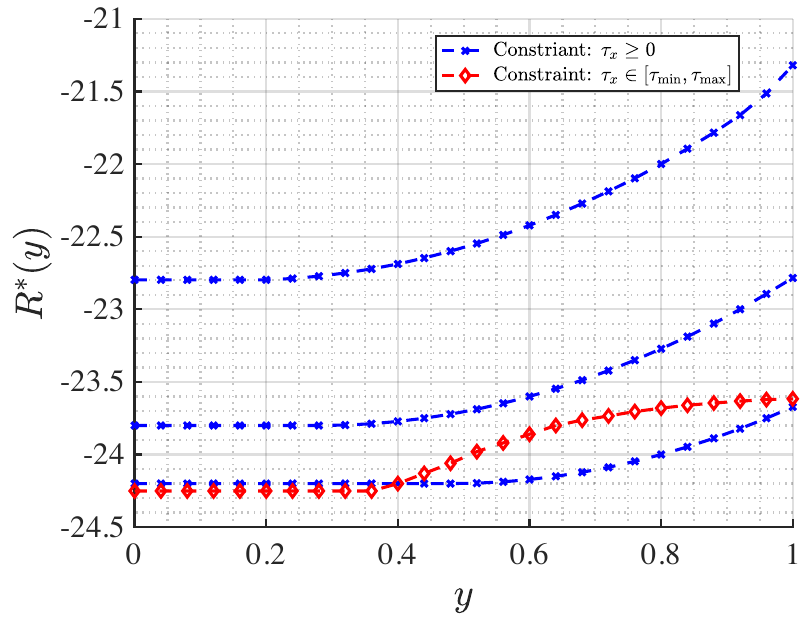}
    \caption{Illustrative examples of convex $R^*(y)$ and non-convex $R^*(y)$ in the PTS system.}
    \label{fig:R_convex}
\end{figure}

To further analyze the optimal CPU strategy in the PTS case, we examine the convexity of the optimal value function $R^*(y)$.
Fig.~\ref{fig:R_convex} provides illustrations of $R^*(y)$ in the PTS system under different conditions. We observe that in the absence of $\tau_{\min}$ and $\tau_{\max}$ constraints, $R^*(y)$ is convex, as depicted by the blue line.
In contrast, when $\tau_{\min}$ and $\tau_{\max}$ constraints are imposed, $R^*(y)$ exhibits non-convex behavior, as shown by the red line.

Following from Corollary \ref{corollary_solution_pbs},
we exploit Jensen's inequality 
to obtain the following properties of the optimal CPU scaling in the PTS case.
\begin{proposition}\label{proposition_pbs}
If $R^*(y)$ is convex, then the optimal  CPU scaling $\{\tau_x^*(y)\}_{x\in[b]}$ for the PTS system satisfies
\begin{align*}
  \frac{x}{x'}\tau_{x}^*(y) \overset{(a)}{\le}  \tau^*_{x'}(y)&\overset{(b)}{\le} \tau^*_{x}(y),\quad x\leq x', x,x'\in[b].
\end{align*}
\end{proposition}
\begin{IEEEproof}
    See Appendix~\ref{proof:proposition_pbs}.
\end{IEEEproof}
When the state at the time of generating a new update is $y$ and the observed task size is $x$, the system chooses the optimal $\tau^*_x(y)$ to process the task, and the total duration is $x\tau_x^*(y)$.
Proposition~\ref{proposition_pbs}(a) indicates that total processing time of an updating task is increasing with the task size $x$, even though   Proposition~\ref{proposition_pbs}(b) indicates that the processor speed, which is the reciprocal of the batch processing time $\tau^*_x(y)$, is increasing with the task size $x$. 
Nevertheless, Proposition~\ref{proposition_pbs}(a) aligns with our understanding that the larger the task, the faster the CPU runs to ensure the efficiency and performance of the networking system.

\begin{proposition}\label{propo_yp}
For the PTS system, there exists an optimal policy $\pi^*$ and a threshold $y_p$ such that 
   \begin{IEEEeqnarray}{rCl}\label{y_p}
y_p&=&\gamma^*+\lambda^*\bar{P}-\sum\limits_{x=1}^bf(x)xH_{\alpha}\left(\frac{2\lambda^*/(\alpha-1)}{\psi_x(y_p)+y_p}\right), \IEEEeqnarraynumspace
\end{IEEEeqnarray}
$z^*(y_p)=0$, and for all $y<y_p$,
\begin{align}
    \frac{\partial \tau^*_x(y)}{\partial y}=0,~\frac{\partial z^*(y)}{\partial y}=-1,~\frac{\partial \psi_x(y)}{\partial y}=0;
\end{align}
and for all $y\geq y_p$, $z^*(y)=0$.
\end{proposition}
\begin{IEEEproof}
    See Appendix~\ref{proof:proposition_yp_exist}.
\end{IEEEproof}
Proposition~\ref{propo_yp} indicates that prior to the threshold $y_p$, the optimal CPU policy $\tau_x^*(y)$ remains constant with respect to $y$, while $z^*(y)$ decreases linearly in $y$ with a slope of $-1$.
Once the optimal waiting strategy $z^*(y)$ reaches $0$ at threshold $y_p$, the system adopts the zero-waiting policy for all $y>y_p$.
The specific optimal solutions are further clarified in the following corollary.
\begin{corollary}\label{Corollary_yp}
For the PTS system, the optimal solution that satisfies Corollary~\ref{corollary_solution_pbs} is given by
\begin{subequations}\label{z23}
\begin{IEEEeqnarray}{rCl}
    z^*(y)&=&\begin{cases}        
    \gamma^*+\lambda^*\bar{P}-y-\Lbar(\btau^*(y_p))
    &y<y_p,\\
         0, & y\geq y_p,\\
         \end{cases}\IEEEeqnarraynumspace\label{eq:zstar:PTS}\\
    \tau_x^*(y)&=&\begin{cases}        H_{\alpha}\left(\frac{2\lambda^*/(\alpha-1)}{\psi_x(y_p)+y_p}\right), &y<y_p,  \\
        H_{\alpha}\left(\frac{2\lambda^*/(\alpha-1)}{\psi_x(y)+y}\right), & y\geq y_p.
    \end{cases}
\end{IEEEeqnarray}
\end{subequations}
\end{corollary}

\subsection{Structure of the Optimal Policy for UTS}
\label{subsect-UTS} 
We now investigate the structure of the optimal policy for the unpredictable task size case.
We first define
\begin{align}
    \zeta_x(y)\triangleq\frac{1}{\Fbar(x)}\sum\limits_{k'=x}^b f(k')\frac{\partial R^*(l)}{\partial l}\bigg|_{l=\sum\limits_{x=1}^{k'}\tau_x^*(y)}. \label{eq:zeta_Ru}
\end{align}
The optimal structure can be derived as the following corollary. 

\begin{corollary}[Optimal Structure for UTS]\label{corollary_solution_ubs}
Given  the minimal AoI $\gamma^*$ and the optimal dual variable $\lambda^*$, the optimal policy $\pi^*$ satisfies that, for all $y\in\mathcal{Y}$,
\begin{subequations}\label{eq:structure}
\begin{align}
    \label{eq:z+y}
    z^*(y)&=\left[\gamma^*+\lambda^* \bar{P}-y-
    \Lbar(\btau^*(y))
    \right]^+,\\ 
    \tau_x^*(y)&=H_{\alpha}\left(\frac{2\lambda^*/(\alpha-1)}{\zeta_x(y)+y+z^*(y)}\right),~x\in[b], \label{eq:tau}\\
    \zeta_x(y)
    &=\frac{1}{\Fbar(x)}\sum\limits_{k'=x}^b f(k')\bigg(\sum\limits_{x'=1}^b \Fbar(x')\tau^*_{x'}\Bigl(\sum_{x=1}^{k'}\tau_x^*(y)\Bigr) \nn
&\quad+\sum_{x=1}^{k'}\tau^*_x(y)+z^*\Bigl(\sum_{x=1}^{k'}\tau^*_x(y)\Bigr)-\gamma^*-\lambda^*\bar{P}\bigg),~x\in[b].\label{eq:zeta}
\end{align}
\end{subequations}
\end{corollary}
\begin{IEEEproof}
    See Appendix~\ref{proof:solution_pbs}.
\end{IEEEproof}

Similar to Eq.~\eqref{eq:z+y_pbs} for the PTS case, Eq.~\eqref{eq:z+y} has a \textit{water-filling-like} solution where $\gamma^*+\lambda^*\bar{P}$ is the water-level.
That is, when 
$y+\Lbar(\btau^*(y))$
is sufficiently large, the next update will be generated and processed immediately; otherwise, the CPU should sleep for a certain amount of time. 
Intuitively, the optimal waiting makes the
generation times of updates towards being equally spaced.
On the other hand, the optimal CPU scaling $\bs\tau^*(y)$ are based on the fixed point relation determined by
\eqref{eq:structure}.

Similarly, the optimal value function $R^*(y)$ in the UTS case exhibits the same pattern as $R^*(y)$ in the PTS case shown in Fig.~\ref{fig:R_convex}; i.e., $R^*(y)$ for UTS  remains convex only when the $\tau_{\min}$ and $\tau_{\max}$ constraints are absent.

From Corollary \ref{corollary_solution_ubs}, 
we can exploit Jensen's inequality to obtain the following characterization of the optimal CPU scaling.
\begin{proposition}\label{proposition_ubs}
If $R^*(y)$ in the UTS system is convex in $y$, then the optimal  CPU scaling $\{\tau_x^*(y)\}_{x\in[b]}$ satisfies
    \begin{align}
          \tau_{x}^*(y)\ge  \tau_{x'}^*(y),\quad x\leq x', x, x'\in[b],
    \end{align}
\end{proposition}
\begin{IEEEproof}
    See Appendix~\ref{proof:proposition_ubs}.
\end{IEEEproof}
For the UTS case, Proposition \ref{proposition_ubs} implies that the CPU starts the task at a low speed and accelerates as the task progresses. Such an acceleration in CPU frequencies is partially consistent with the zero-wait DVS scheme in~\cite{yuan2006energy}.

\begin{proposition}\label{propo_yu}
In the UTS system, there exists an optimal policy $\pi^*$ and a threshold $y_u$ such that
   \begin{IEEEeqnarray}{rCl}\label{y_u}
    y_u&=&\gamma^*+\lambda^*\bar{P}-\sum\limits_{x=1}^b\Fbar(x)H_{\alpha}\left(\frac{2\lambda^*/(\alpha-1)}{\zeta_x(y_u)+y_u}\right),
    \IEEEeqnarraynumspace
\end{IEEEeqnarray}
$z^*(y_u)=0$, and for all $y<y_u$,
\begin{align}
    \frac{\partial \tau_x^*(y)}{\partial y}=0,~\frac{\partial z^*(y)}{\partial y}=-1,~\frac{\partial \zeta_x(y)}{\partial y}=0;
\end{align}
for all $y\geq y_u$, $z^*(y)=0$.
\end{proposition}
\begin{IEEEproof}
    See Appendix~\ref{proof:proposition_yp_exist}.
\end{IEEEproof}

Proposition~\ref{propo_yu} states that the optimal CPU policy $\tau_x^*(y)$ remains constant with respect to $y$ up to the threshold $y_u$, while $z^*(y)$ decreases linearly with a slope of $-1$.
When $z^*(y)$ reaches $0$ at threshold $y_u$, the system employs the zero-waiting policy for all $y>y_u$.
The following corollary further clarifies the specific optimal solutions.

\begin{corollary}\label{corollary_yu}
The optimal solution that satisfies Corollary~\ref{corollary_solution_ubs} is given as
\begin{subequations}\label{eq:ubs_zero}
\begin{IEEEeqnarray}{rCl}
    z^*(y)&=&\begin{cases}
        \gamma^*+\lambda^* \bar{P}-y-\sum\limits_{x=1}^b\Fbar(x)\tau^*_x(y_u), & y<y_u\\
         0, & y\geq y_u,\\
    \end{cases}\IEEEeqnarraynumspace\label{eq:zstar:UTS}\\
    \tau_x^*(y)&=&
    \begin{cases}
    H_{\alpha}\left(\frac{2\lambda^*/(\alpha-1)}{\zeta_x(y_u)+y_u}\right), & y<y_u,  \\
    H_{\alpha}\left(\frac{2\lambda^*/(\alpha-1)}{\zeta_x(y)+y}\right), &y\geq y_u.
    \end{cases}
\end{IEEEeqnarray}
\end{subequations}
\end{corollary}

Equation \eqref{eq:ubs_zero} indicates that when $y<y_u$, $\tau^*_x(y)$ remains constant respect to $y$, while $z^*(y)$ decreases with a slope of $-1$ in $y$. In contrast, when $y\geq y_u$, the optimal waiting policy is zero-wait. We will confirm these observations in Section~\ref{subsection:VIII-C}.

\subsection{Discussion: PTS and UTS Optimal Policies}
\label{subsect:discussion}
Despite the differences in CPU speed policy decision-making, we see from Corollaries~\ref{Corollary_yp} and~\ref{corollary_yu} that the optimal PTS and UTS policies share certain properties. Specifically, from 
\eqref{eq:zstar:PTS} and \eqref{eq:zstar:UTS}, there exists a threshold $\hat{y}\in\set{y_p,y_u}$ satisfying

\begin{equation}
  \hat{y} = \gamma^*+\lambda^* \bar{P}-\Lbar(\btau^*(\hat{y})).
\end{equation}
such that the optimal waiting function is
\begin{equation}
z^*(y)=
[\hat{y}-y]^+. 
\end{equation}
That is, the water-filling-like waiting functions in \eqref{eq:z+y_pbs} for PTS and \eqref{eq:z+y} for UTS are in fact water-filling. 
In addition, we note that this form of optimal waiting strategy is consistent with Bellman's  principle.
Specifically, consider the case when
the prior update is delivered,
the monitor's age is $y$ and the system waits for an optimal waiting time $z^*(y)>0$.  
Suppose the optimal waiting time function $z^*(y)$ permits a partial wait of duration $z'$; that is, after waiting for a time $z'<z^*(y)$ (i.e. in the middle of the waiting period), the system is free to choose a new waiting period. 
At this point, the system will remain idle but the age at the monitor becomes $y+z'$ and the optimal (residual) waiting time is now $z^*(y+z')$. With this second-step optimization, the total waiting time becomes $z'+z^*(y+z')$. 
From Bellman's principle, if $z^*(y)>0$ is the original optimal waiting period, then for all $z'<z^*(y)$,
\begin{equation}
\label{eq:zopt'}
    z^*(y) = z'+z^*(y+z').
\end{equation}
Thus, for all $y$ such that $z^*(y)>0$, it follows from \eqref{eq:zopt'} that
\begin{equation}
    \frac{z^*(y+z')-z^*(y)}{z'}=-1.
\end{equation}
In the limit as $z'\to 0$,  it follows that $dz^*(y)/dy=-1$ if $z^*(y)>0$. This  leads to the conclusion that $z^*(y)=[\hat{y}-y]^{+}$ for some constant $\hat{y}>0$. These observations will be demonstrated in Section \ref{subsection:VIII-B}.

We further observe that this form of optimal waiting function implies that the optimal CPU speed policy $\btau^*(y)$ is unchanging for all $y\le \hat{y}$. In particular, for all $y\le \hat{y}$, the age at the monitor at the completion of the waiting period of length $z^*(y)$ is $y+z^*(y)=\hat{y}$. Since a fresh update goes into service at that time, the subsequent age evolution at the monitor depends only on the age $\hat{y}$ at the start of service and the processing time of the fresh update. Hence the CPU speed policy should be the same for all such updates.

\subsection{Special Case: Fixed Task Size}\label{sec: toy example}
When there is a deterministic number $X_n=b>1$ batches in each update processing task, the probability distribution of the batches becomes $f(k)=0$ for all $k<b$ and $f(b)=1$ in both PTS and UTS cases. Under such a scenario,  
Problem~\eqref{eq:problem_known_constant} for PTS reduces to
\begin{subequations}
\begin{align}
\gamma^*=\min_{z,\tau_b}\quad&b\tau_b+\frac{1}{2}(y+z),  \\
    {\rm s.t.}\quad&b\Ebatch(\tau_b)-\bar{P}(y+z)\leq0.
\end{align}  
\end{subequations}
Similarly, Problem~\eqref{eq:minimize_age_stat} for UTS reduces to 
\begin{subequations}\label{eq:special_uts}
\begin{align}
\gamma^*=\min_{z,\tau_x}\quad&\sum\limits_{x=1}^b\tau_x+\frac{1}{2}(y+z),  \\
    {\rm s.t.}\quad&\sum\limits_{x=1}^b\Ebatch(\tau_x)-\bar{P}(y+z)\leq0.
\end{align}  
\end{subequations}

Notably, when the task size is deterministic, 
the PTS and UTS scenarios become one in the same and 
 the optimal PTS and UTS solutions 
 are identical.
Moreover, by relaxing the $\tau_{\min}$ and $\tau_{\max}$ constraints, the optimal solutions are stated in the following proposition.
\begin{proposition}\label{propo_special}
When $\alpha \in (1,2]$ and there is a deterministic number $X_n=b$ batches in each update processing task, the PTS and UTS systems become identical with optimal batch processing times $\tau^*_b=\tau_x^*$, processor sleep time $z^*$, and the average age of information $\gamma^*$ given by
\begin{subequations}
    \begin{align}
    \tau_x^*&=\tau^*_b=\left((\alpha-1)\bar{P}\right)^{\frac{1-\alpha}{1+\alpha}},\quad x\in[b],\label{eq:tau-soln}  \\
    z^*&=0,  \\
    \gamma^*&=\frac{3}{2}b\left((\alpha-1)\bar{P}\right)^{\frac{1-\alpha}{1+\alpha}}.
\end{align}
\end{subequations}
\end{proposition}
\begin{IEEEproof}
    See Appendix~\ref{sec:proof_proposition_special}.
\end{IEEEproof}

To summarize, when the number of processing batches is deterministic and $\alpha \in (1,2]$, the timely computing system always employs a zero-wait strategy to achieve the minimum average AoI for the average power budget. Furthermore, the optimal execution speed of the CPU is only determined by the power budget $\bar{P}$. In particular, having the CPU work faster and then sit idle before processing another update is suboptimal.  These conclusions are roughly consistent with~\cite{sun2017update}, where it was found that the optimal waiting policy was zero-wait when the service times were deterministic.

We also note that if the $\tau_{\min}$ and $\tau_{\max}$ constraints are further enforced, the zero-wait strategy in Proposition \ref{propo_special} becomes infeasible as $\tau_{\max}$ becomes sufficiently small, as operating at $\tau_{\max}$ with zero-wait would exceed the average power constraint. Instead, the optimal policy involves a waiting period to meet the power constraint.

\section{Numerical Results}
\label{sec:numerical result}
In this section, we present numerical results to i)  validate our analytical characterization  of the structure of the optimal age-minimal CPU scheduling policy, and ii) compare our proposed scheme against existing benchmarks.

\begin{figure*}[t]
\centering
\subfigure[$b=2$]{
\centering
\includegraphics[width=0.31\linewidth]{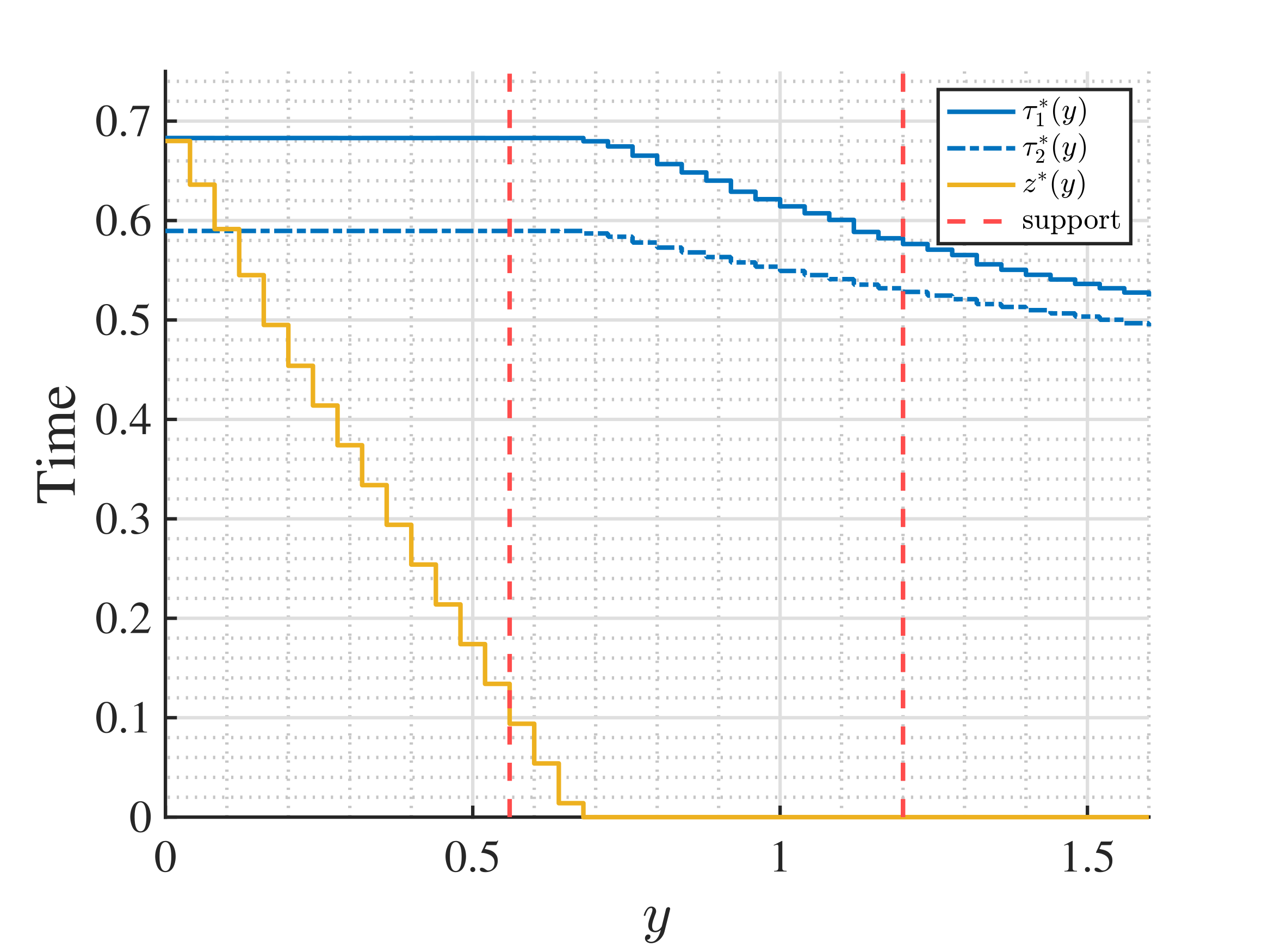}
\label{fig:known_cycle_2}
}
\subfigure[$b=3$]{
\centering
\includegraphics[width=0.31\linewidth]{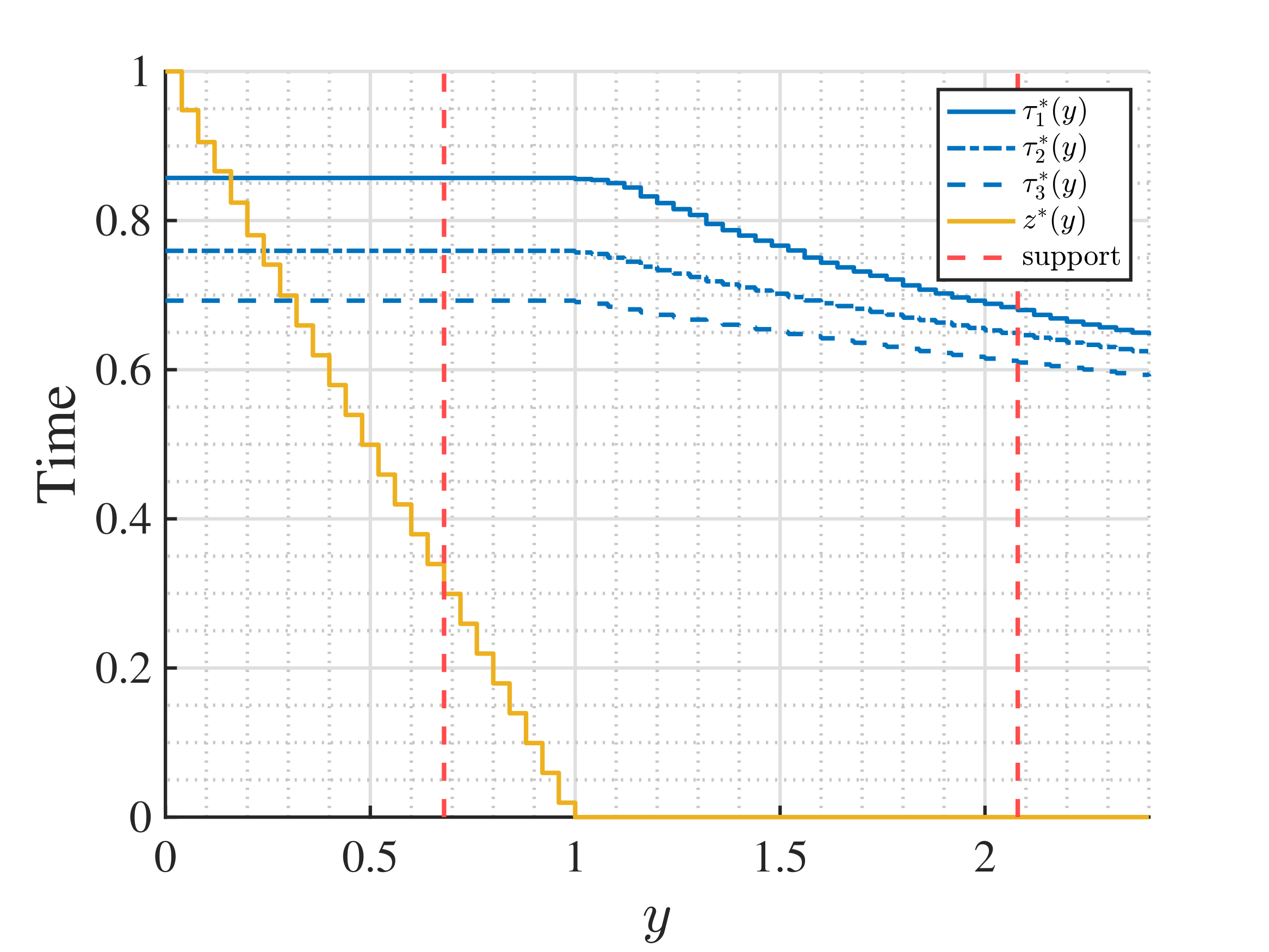}
\label{fig:known_cycle_3}
}
\subfigure[$b=5$]{
\centering
\includegraphics[width=0.31\linewidth]{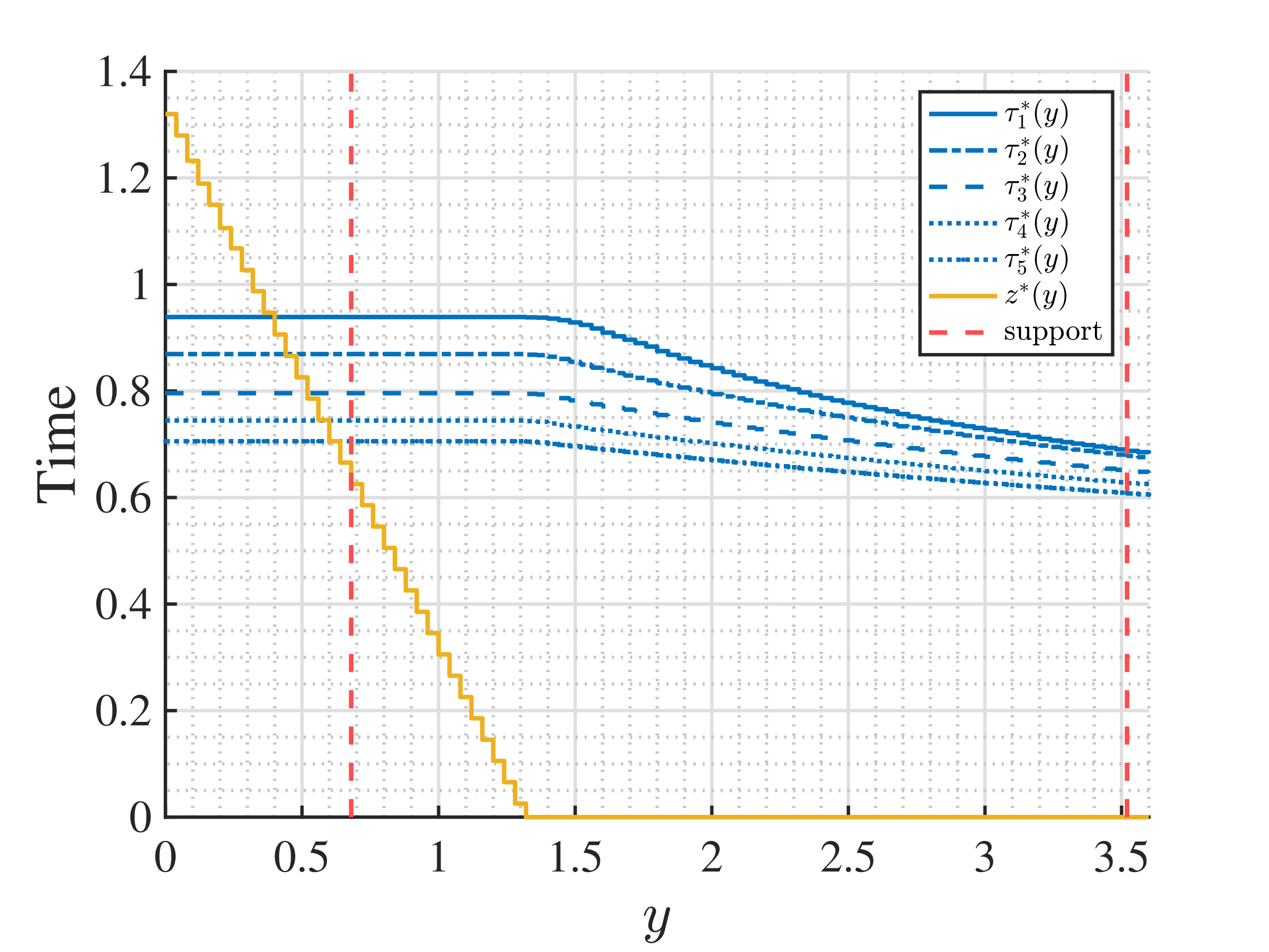}
\label{fig:known_cycle_5}
}
\centering
\caption{Numerical evaluations of the optimal policy with (a) $b=2$, (b) $b=3$ and (c) $b=5$ batches under $\bar{P}=3$ in PTS case.
The region between the two red dashed lines represents the support of the stationary distribution of state $y$.
}
\label{fig:cycle_pbs}
\end{figure*}

\begin{figure*}[t]
\centering
\subfigure[$\bar{P}=15$]{
\centering
\includegraphics[width=0.31\linewidth]{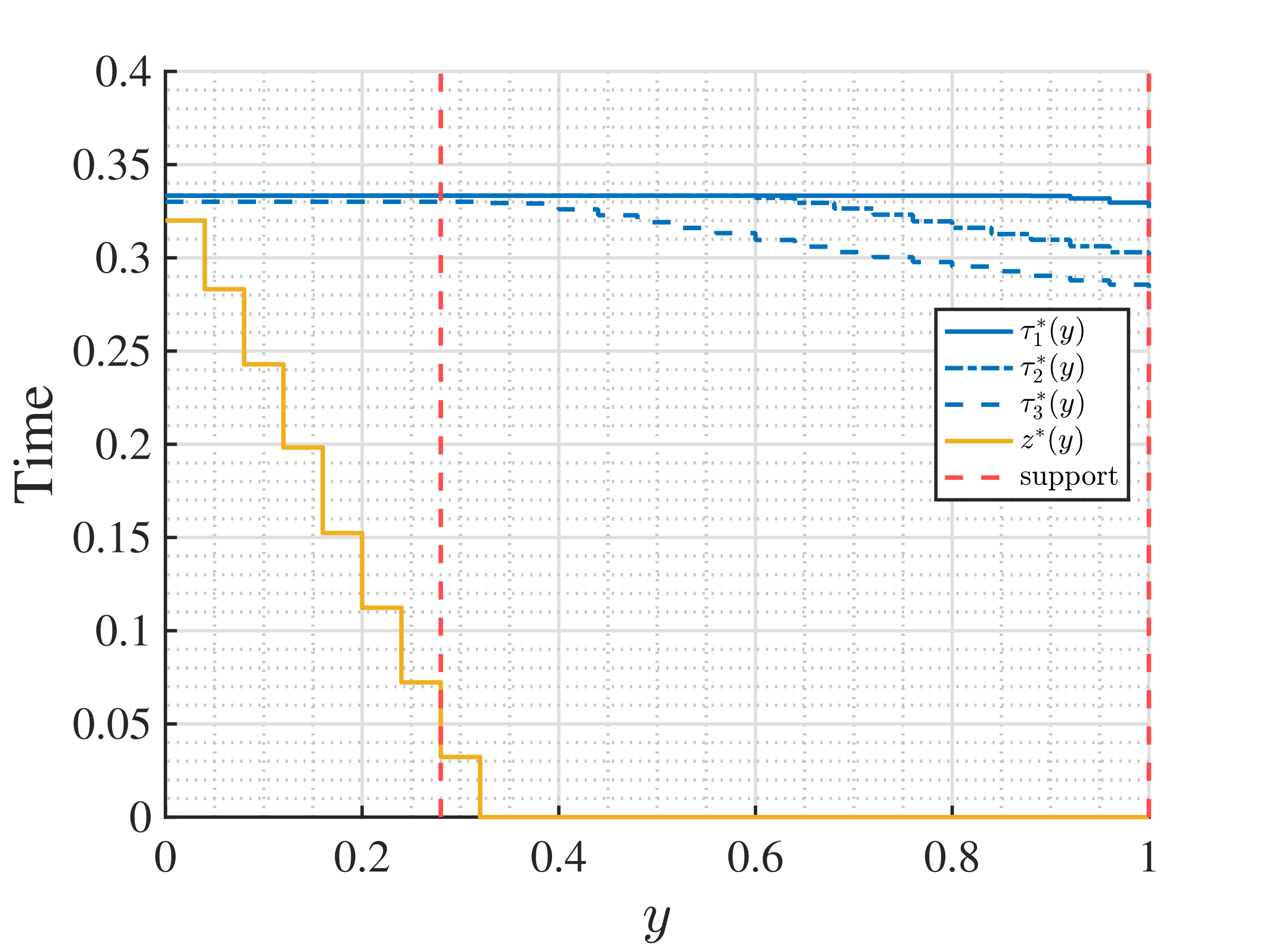}
\label{fig:pbs_b3P6}
}
\subfigure[$\bar{P}=20$]{
\centering
\includegraphics[width=0.31\linewidth]{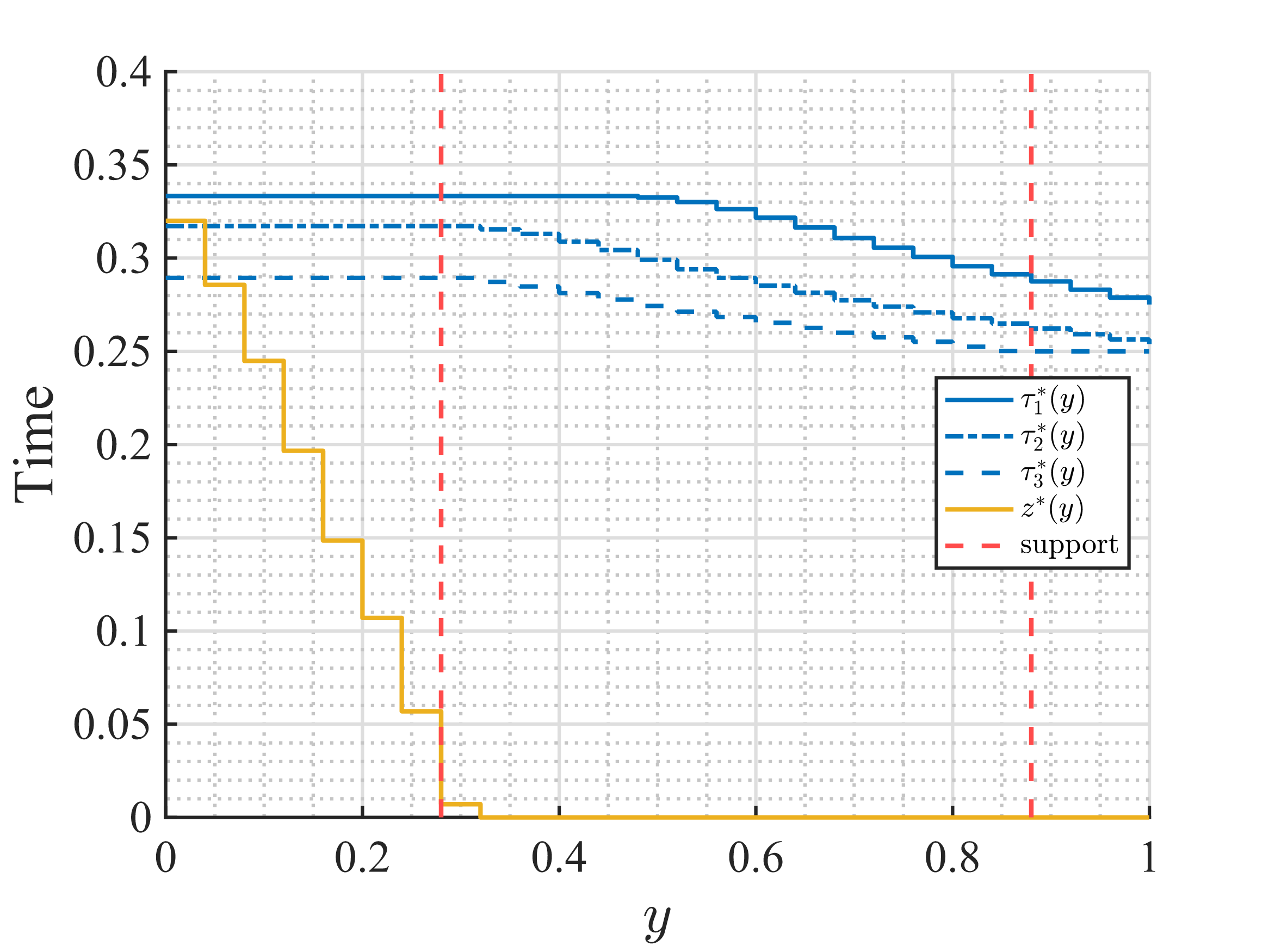}
\label{fig:pbs_b3P8}
}
\subfigure[$\bar{P}=25$]{
\centering
\includegraphics[width=0.31\linewidth]{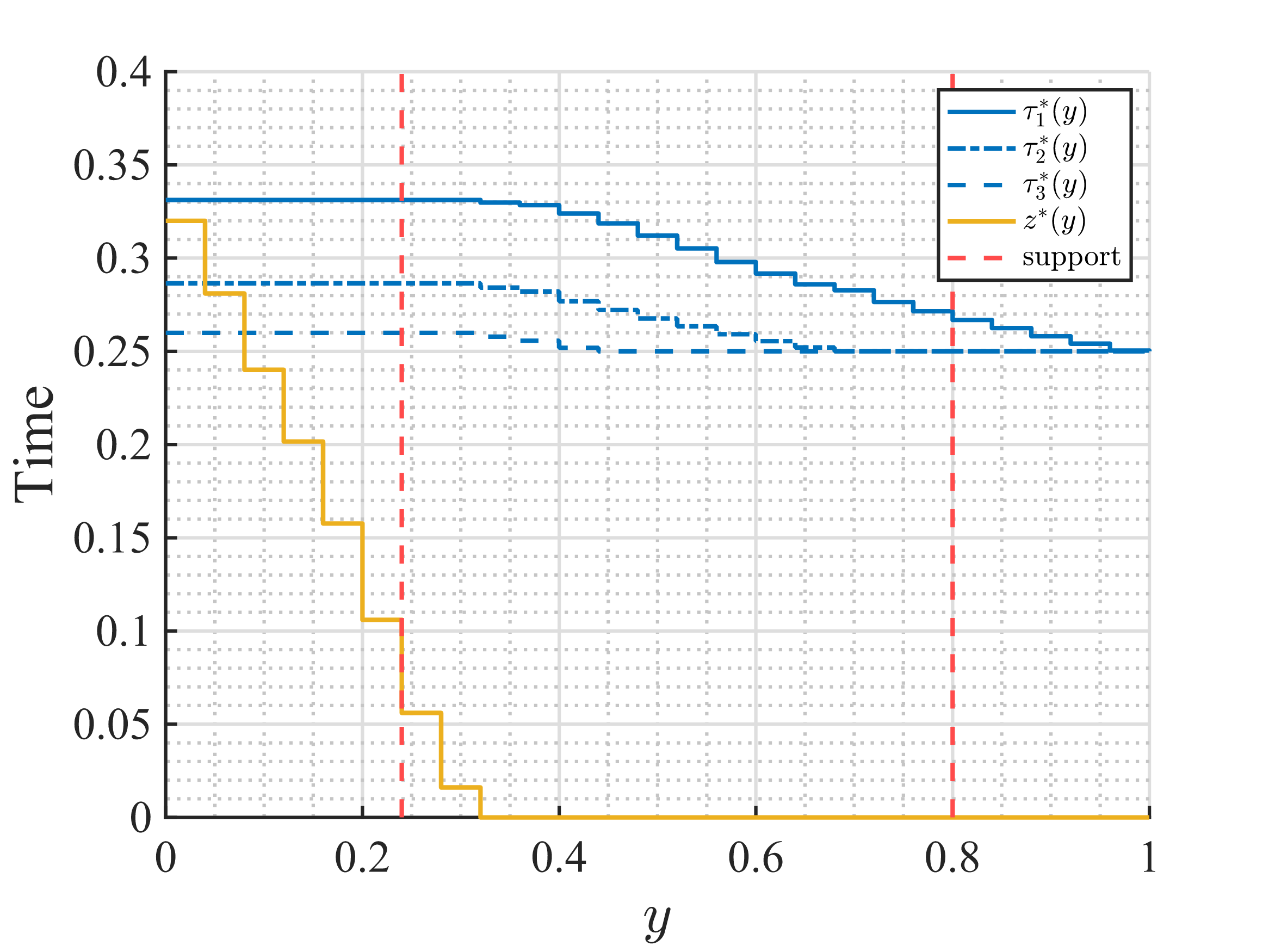}
\label{fig:pbs_b3P10}
}
\caption{Numerical evaluations of the optimal policy under (a) $\bar{P}=15$, (b) $\bar{P}=20$ and (c) $\bar{P}=25$ with $b=3$ batches in PTS case.
The region between the two red dashed lines represents the support of the stationary distribution of state $y$.
}
\label{fig:cycle_pbs_btmax}
\end{figure*}

\subsection{Setup}
We remind the reader that our model of an SMDP in an infinite continuous state space may lead to an unacceptable computational burden in numerical evaluation.
To overcome such a challenge, we consider a \textit{piece-wise constant approximation} that constrains the state $Y$ to a finite interval $[0,\ymax]$ and then quantizes this domain into $\qmax$ continuous intervals each with a fixed length $\Delta_Y=\ymax/\qmax$.
For $q\in[\qmax]$, the value function $R^m(y)$ as well as the actions $\bs{a}^{m}(y)$ remain constant over any given interval 
$\Yint_q=[(q-1)\Delta_Y,q\Delta_Y)$.
Specifically, with $y'_q =(q-1/2)\Delta_Y$ denoting the midpoint of the $q$th interval,
\begin{align}
    &\bs{a}^{m}(y)=\bs{a}^{m}(y'_q),\quad y\in\Yint_q.
    \label{eq:quantizedpolicy}
\end{align}
By this piece-wise constant approximation, we only need to update the value function $R^m(y)$ and therefore the action for $\qmax$ intervals.
Furthermore, such a quantized policy can achieve asymptotic optimality (as stated in~\cite{saldi2017asymptotic}), enabling us to make tradeoffs between computational complexity and optimality.

\subsection{Policy Demonstration for Multi-Batch Demands in PTS}\label{subsection:VIII-B}
To investigate the optimal policy for multi-batch demands in various scenarios for the PTS case, we now present numerical results for different settings of the task size $b$ and power budget $\bar{P}$.

We consider the power consumption model with the chip parameter $\alpha=2$.
In addition, we set
$\tau_{\min}=0$
and arbitrarily large $\tau_{\max}$ to explore the optimal policy for cycle batch demand with $b=2$, $b=3$ and $b=5$, all under the same power budget of $\bar{P}=3$.
Furthermore, for the $b=2$ case, we set $f(1)=0.7=1-f(2)$.
For the $b=3$ and $b=5$ cases, we use an average probability of $f(x)=1/b$ for each $x\in[b]$.

To maintain the same $\Delta_Y=0.04$, we divide $y_{\max}$ into several intervals for each case, with differing values of $y_{\max}$.
For instance, in the $b=2$ case, we set $40$ intervals to quantize $y_{\max}=1.6$.
Finally, we take the mid-point value of each time interval to determine the corresponding optimal batch processing times $\tau^*_x(y)$, and waiting time $z^*(y)$ to draw the step diagram shown in Fig.~\ref{fig:cycle_pbs}.
It is worth mentioning here that we use the blue curve to denote the CPU scaling policy, while the yellow curve indicates the waiting strategy. 

\begin{figure*}[t]
\centering
\subfigure[$b=2$]{
\centering
\includegraphics[width=0.31\linewidth]{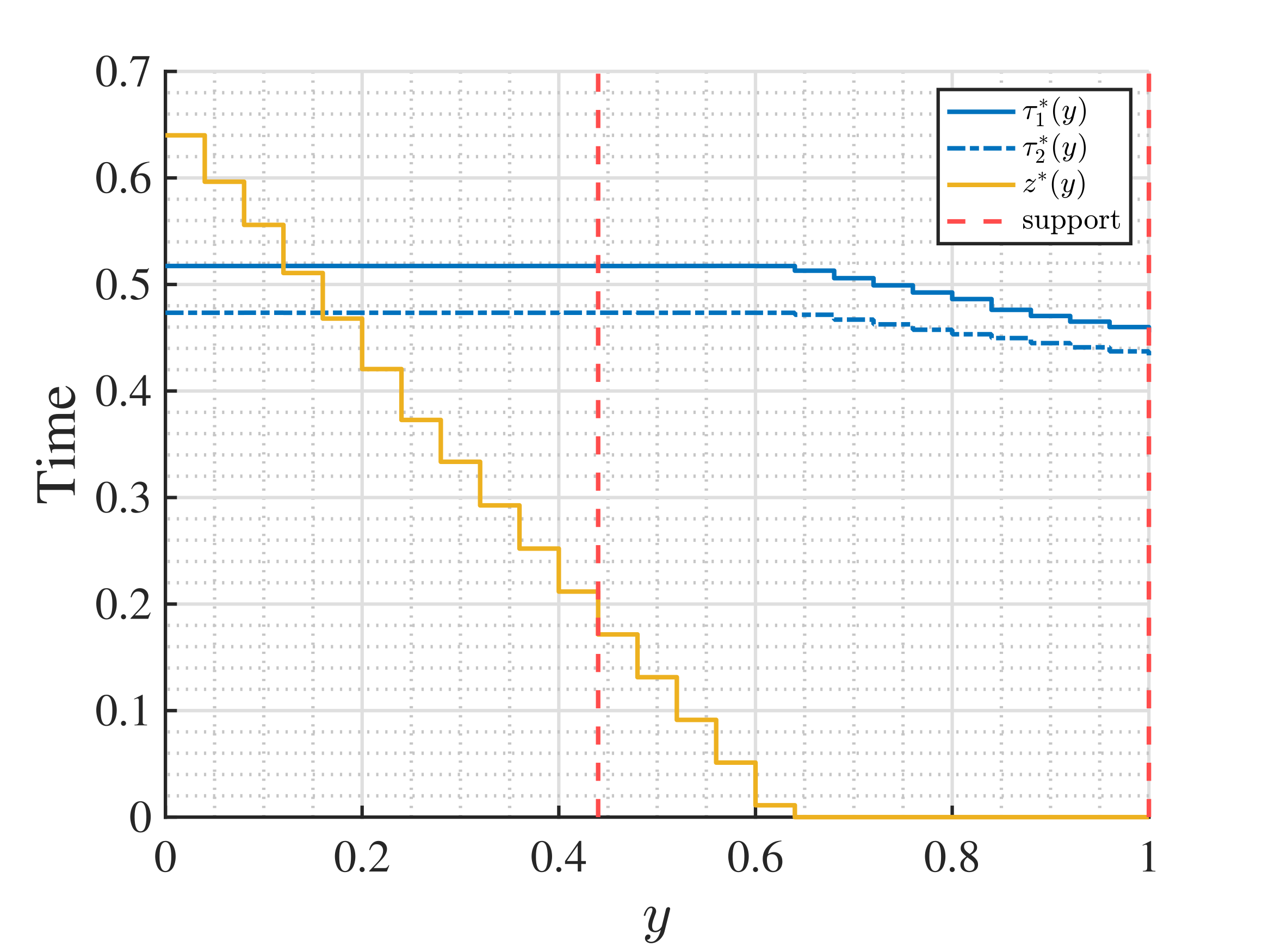}
\label{fig:cycle_2}
}
\subfigure[$b=3$]{
\centering
\includegraphics[width=0.31\linewidth]{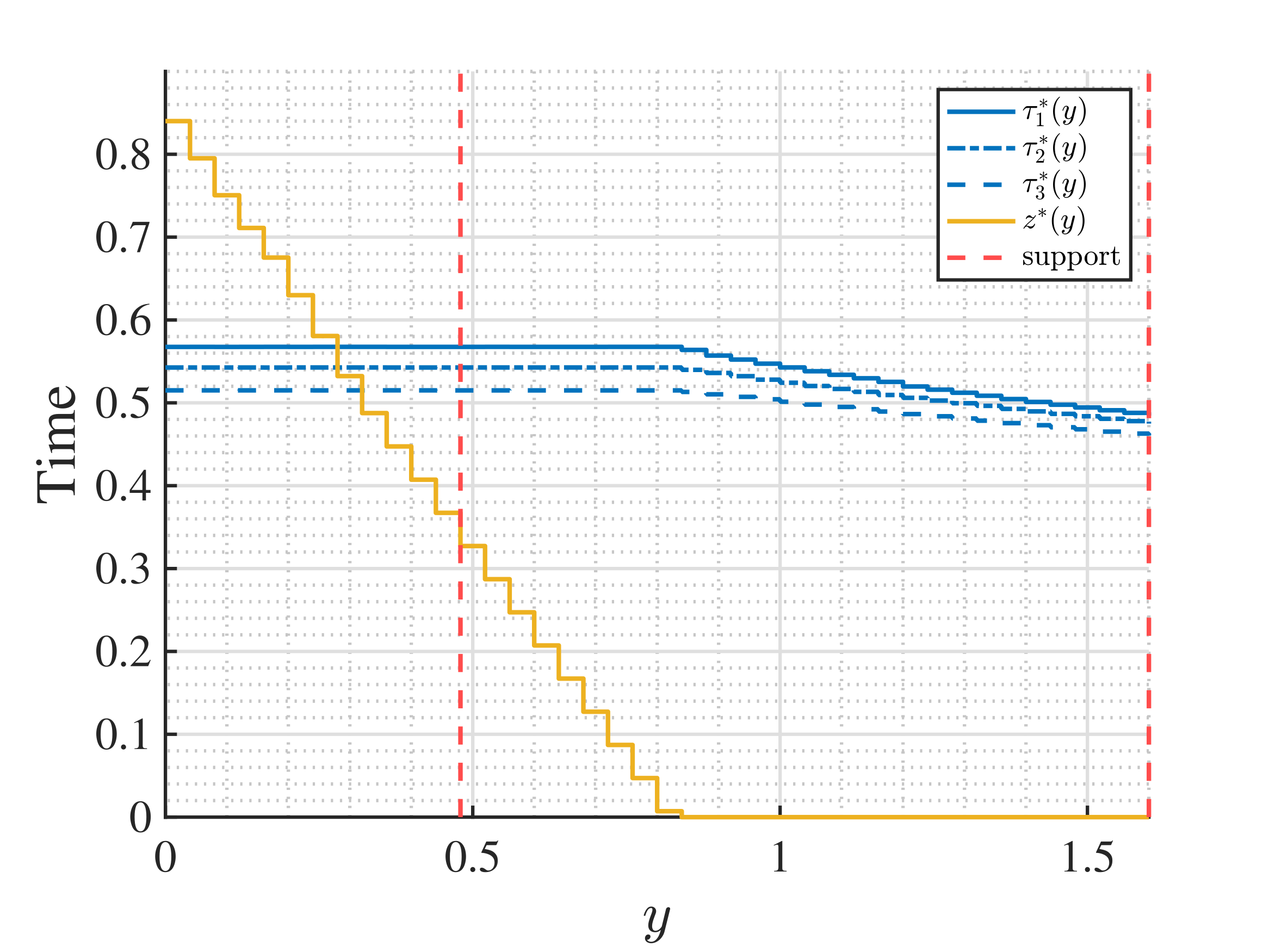}
\label{fig:cycle_3}
}
\subfigure[$b=5$]{
\centering
\includegraphics[width=0.31\linewidth]{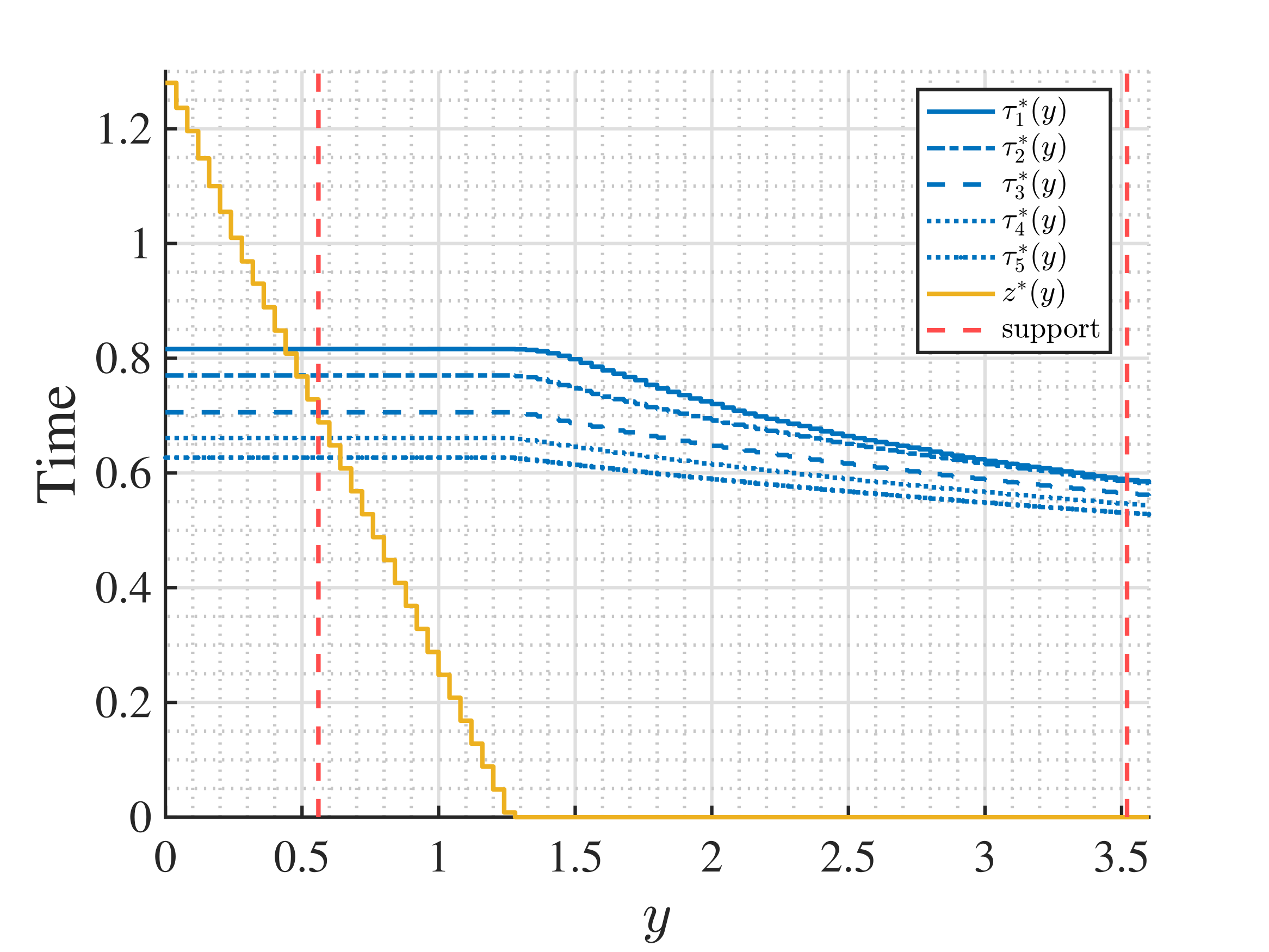}
\label{fig:cycle_5}
}
\centering
\caption{Numerical evaluations of the optimal policy with (a) $b=2$, (b) $b=3$ and (c) $b=5$ batches under $\bar{P}=8$ in UTS case.
The region between the two red dashed lines represents the support of the stationary distribution of state $y$.
}
\label{fig:cycle_ubs}
\end{figure*}

\begin{figure*}[t]
\centering
\subfigure[$\bar{P}=30$]{
\centering
\includegraphics[width=0.31\linewidth]{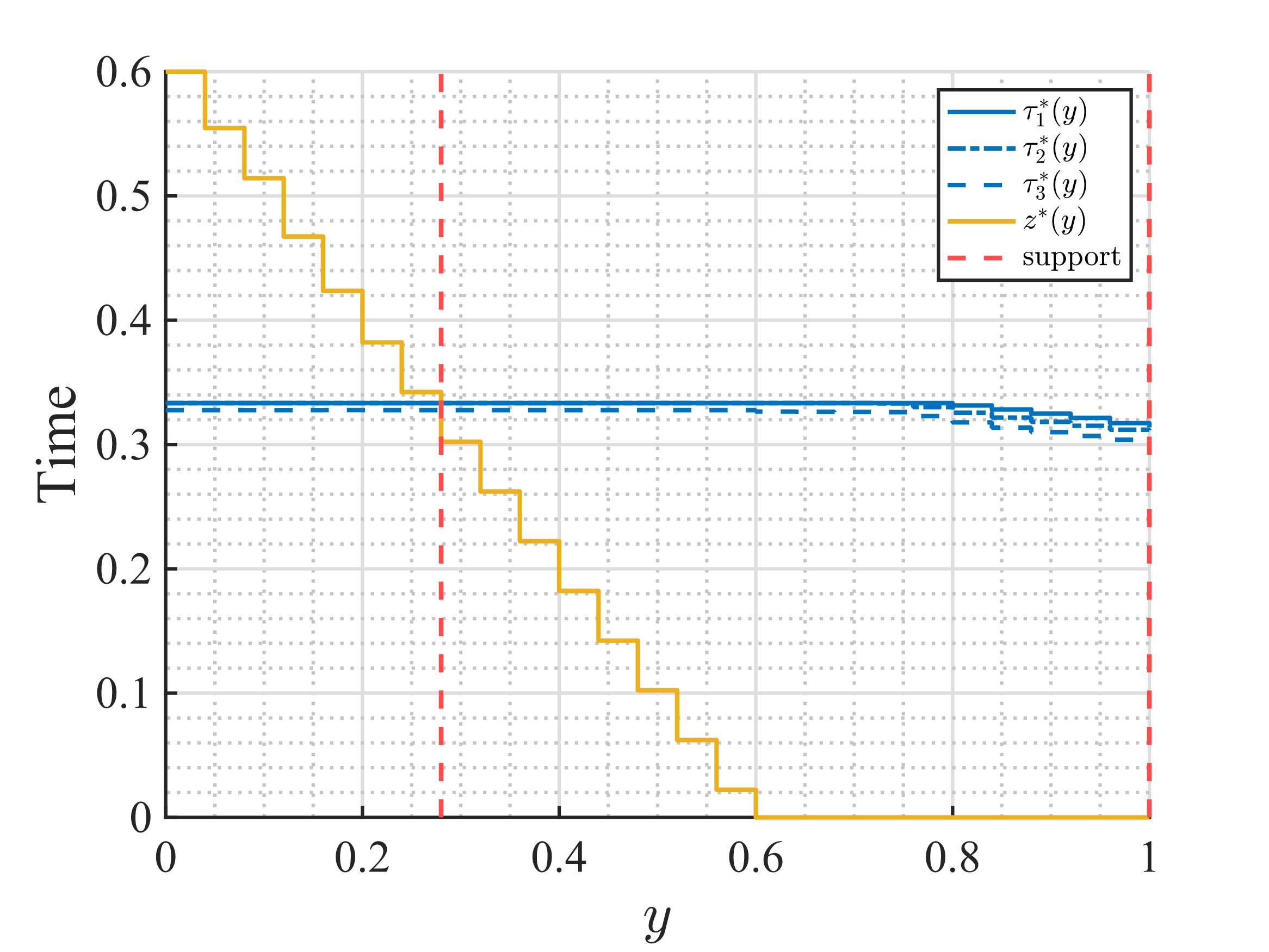}
\label{fig:ubs_b3P3}
}
\subfigure[$\bar{P}=50$]{
\centering
\includegraphics[width=0.31\linewidth]{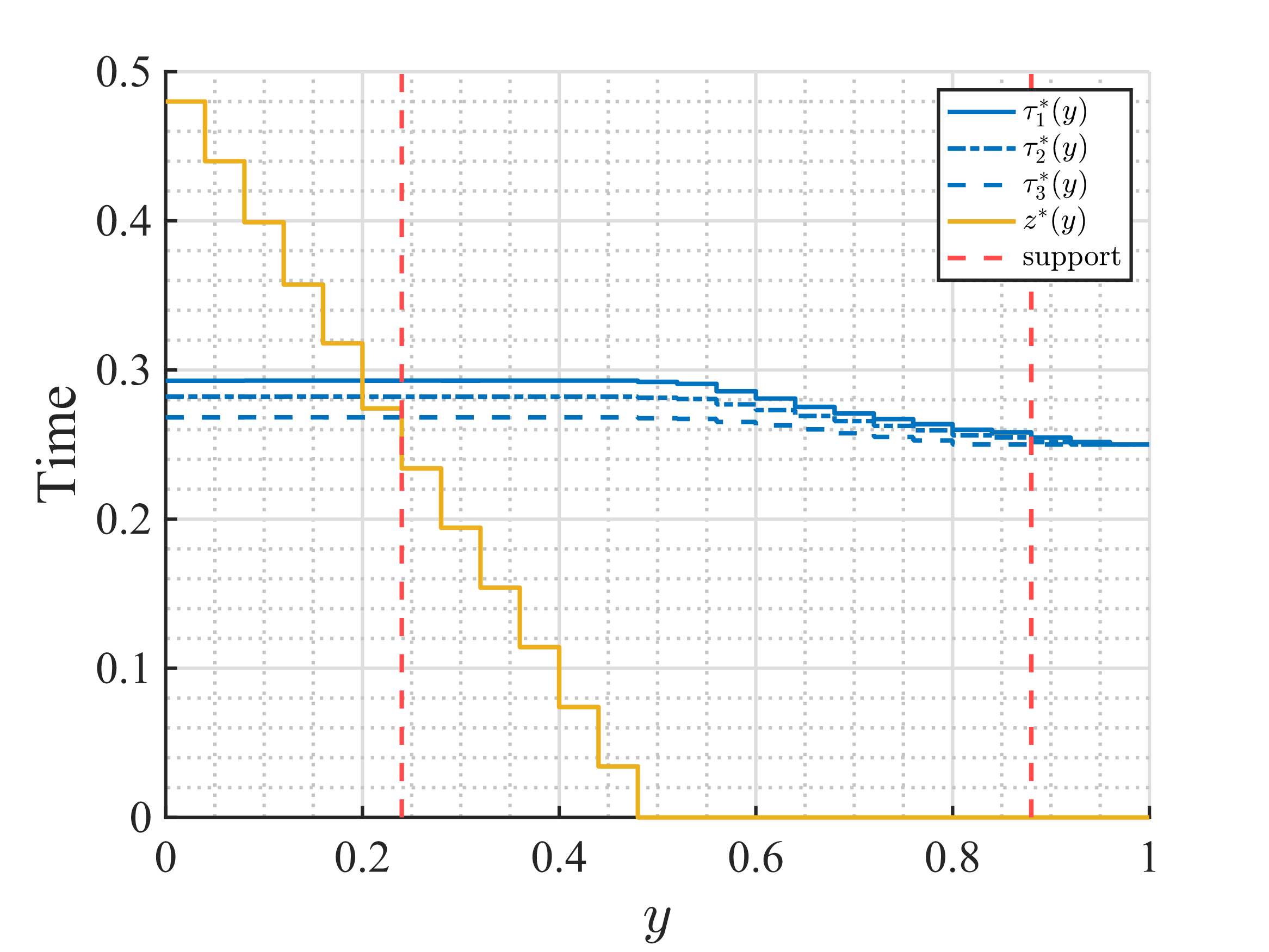}
\label{fig:ubs_b3P5}
}
\subfigure[$\bar{P}=60$]{
\centering
\includegraphics[width=0.31\linewidth]{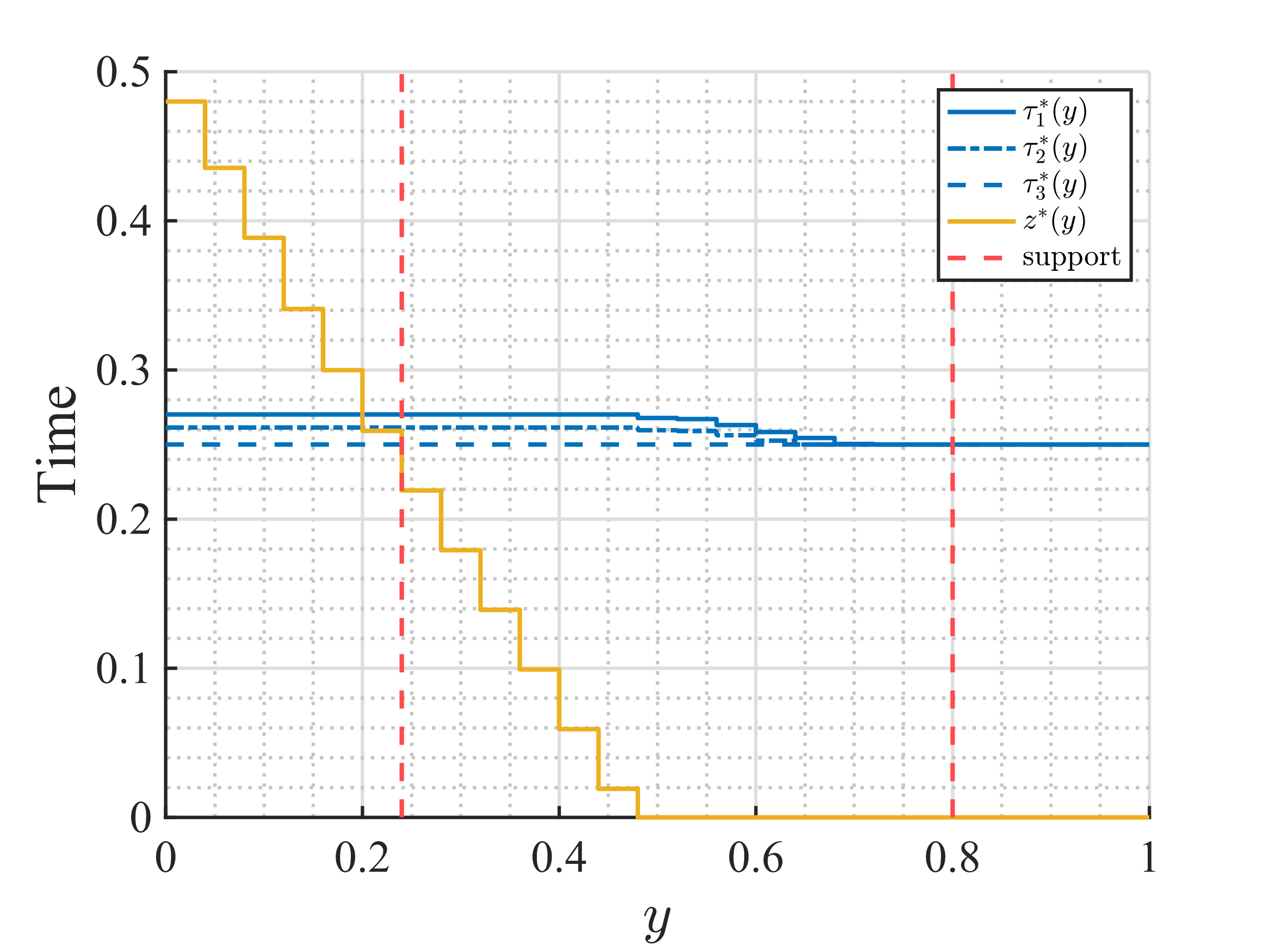}
\label{fig:ubs_b3P6}
}
\centering
\caption{Numerical evaluations of the optimal policy under (a) $\bar{P}=30$, (b) $\bar{P}=50$ and (c) $\bar{P}=60$ with $b=3$ batches in UTS case.
The region between the two red dashed lines represents the support of the stationary distribution of state $y$.
}
\label{fig:cycle_ubs_btamx}
\end{figure*}

In Fig~\ref{fig:cycle_pbs}, we note that $z^*(y)$ decreases in $y$ and appears indistinguishable from a quantized version of the straight line $z^*(y)=(\hat{y}-y)^{+}$ for some constant $\hat{y}>0$.
When $y\le \hat{y}$,
the change in each $\tau_x^*(y)$ is negligible. 
As discussed Section~\ref{subsect:discussion}, this  is as expected because for all $y\le \hat{y}$, the age at the monitor at the end of the waiting period (i.e., when the update begins processing) is always $\hat{y}$. Only when $y\ge \hat{y}$ and $z^*(y)=0$ does $\tau^*_x(y)$ begin to decrease in $y$. 
That is, the CPU only adjusts its speed when the age at the start of processing exceeds the threshold $\hat{y}$.
Moreover, $\tau^*_1(y)\ge\tau^*_2(y)\ge \cdots \ge \tau_5^*(y)$ regardless of the time length $y$ required for the prior update task, which is consistent with Corollary~\ref{corollary_solution_pbs} as well as Proposition~\ref{proposition_pbs}.

We also investigate the optimal policy under the constraints of $\tau_{\min}$ and $\tau_{\max}$ for $b=3$ case.
We keep the same numerical evaluation settings as in the examples of multi-batch demands case, with the exception that we set 
$b\tau_{\max}=y_{\max}=1$. Thus 
$b=3$ implies $\tau_{\max}=\frac{1}{3}$. In addition, we also set $\tau_{\min}=\frac{3}{4}\tau_{\max}=\frac{1}{4}$, a value consistent with $\omega_{\max}=\SI{4}{\giga\hertz}$ and each batch requiring $10^6$ cycles.
With these settings, we explore scenarios with power budgets $\bar{P}=15,20,25$ and draw the step diagrams shown in Fig.~\ref{fig:cycle_pbs_btmax}.

From Fig.~\ref{fig:pbs_b3P6} to Fig.~\ref{fig:pbs_b3P10}, we observe that the optimal waiting strategy $z^*(y)$ maintains the same decreasing trend in $y$ as in the multi-batch demands case without the $\tau_{\min}$ and $\tau_{\max}$ constraints, as shown in Fig.~\ref{fig:cycle_pbs}. 
In Fig.~\ref{fig:pbs_b3P10}, with a power budget $\bar{P}=25$, $z^*(y)$ drops to $0$, and all CPU strategies $\tau^*_1(y)$, $\tau^*_2(y)$ and $\tau^*_3(y)$ begin to decrease simultaneously when $y$ exceeds $0.32$. 
However, the optimal $\tau^*_2(y)$ and $\tau^*_3(y)$ will stop decreasing when $y$ is greater than $0.68$ and $0.44$, respectively, due to the $\tau_{\min}$ limitation.
On the other hand, when the power budget $\bar{P}$ is too small, the optimal CPU solutions will be constrained by $\tau_{\max}$. 
For example, when $\bar{P}=15$, even as $z^*(y)$ drops to $0$, $\tau^*_1(y)$ and $\tau^*_2(y)$ remain unchanged for a period due to the $\tau_{\max}$ constraint.
Similarly, when $\bar{P}=20$, $\tau^*_1(y)$ is nondecreasing 
for some time, also constrained by $\tau_{\max}$.

\subsection{Policy Demonstration for Multi-Batch Demands in UTS}
\label{subsection:VIII-C}
In this subsection, we start with a setting that satisfies $f(1)=0.7, f(2)=0.3$ and limit the power budget to $\bar{P}=8$ as an example to explore the optimal solutions in UTS case.
Considering the power consumption model with $\alpha=2$, we set the maximum value of state $y$ to $y_{\max}=1$ and quantize $y$ into $25$ intervals each of length $\Delta_Y=0.04$. Finally, we take the mid-point value of each time interval and find its corresponding optimal solutions $\tau^*_1(y)$, $\tau^*_2(y)$, and $z^*(y)$
to draw the step diagram shown in Fig.~\ref{fig:cycle_2}.

We observe from Fig.~\ref{fig:cycle_2} that $z^*(y)$ decreases in $y$, and drops to $0$ when $y$ is greater than $0.64$. The support of $y$ is in the interval of $[0.44,1]$,
as marked by red dashed line boundaries. 
Specifically, when $y<0.64$,  $\tau^*_1(y)$ and $\tau^*_2(y)$ remain constant while $z^*(y)$ decreases linearly with a slope of $-1$. 
However, when $y\geq0.64$, the optimal waiting strategy is zero-wait and $\tau^*_1(y),\tau^*_2(y)$ begin to decrease with respect to $y$.
Moreover, $\tau^*_1(y)\ge \tau^*_2(y)$ regardless of the value of the time length $y$ required for the prior task, which is consistent with Corollary~\ref{corollary_solution_ubs} as well as Proposition~\ref{proposition_ubs}.

We also extend the experiment to the case of cycle batch demand $b=3$ and $b=5$. 
We keep the same numerical evaluation settings as in the example with cycle batch numbers of $2$, except that we set the maximum time value $y$ required for the task to $1.6$, $3.6$, and set the average probability $f(x)=1/b$.
In Fig.~\ref{fig:cycle_3} and Fig.~\ref{fig:cycle_5}, we notice that $z^*(y)$ maintains the same decreasing case as in Fig.~\ref{fig:cycle_2}, and drops to $0$ when $y$ is greater than $0.84$ and $1.28$. At the same time, the values of cycle batch execution time start to decrease. That is, the CPU only changes its speed for processing updates that were generated with zero wait.
In addition, the result of decreasing cycle batch execution time remains valid from Fig.~\ref{fig:cycle_2} to Fig.~\ref{fig:cycle_3} and Fig.~\ref{fig:cycle_5}.

To explore the effect of $\tau_{\min}$ and $\tau_{\max}$ on the optimal solutions in the UTS case, we keep the same setting as in the PTS case with the $\tau_{\min}$ and $\tau_{\max}$ constraints except for different power budgets $\bar{P}=30,50,60$. Similar to the PTS case, the optimal CPU strategy is constrained by $\tau_{\max}$ when the power budget is as low as $\bar{P}=30$ in Fig.~\ref{fig:ubs_b3P3} and is constrained by $\tau_{\min}$ when the power budget is as large as $\bar{P}=60$ in Fig.~\ref{fig:ubs_b3P6}.

\subsection{AoI Comparison over Different Policies}
\label{sec:aoi_comparison}
To evaluate the AoI performance improvements achieved by our proposed PTS and UTS schemes, we consider three benchmarks: the zero-wait constant speed policy, the zero-wait DVS strategy in~\cite{yuan2006energy}, and the optimal-wait constant speed scheme in~\cite{sun2017update}, described in the following:
\begin{itemize}
    \item \textit{Zero-wait constant speed}: Under this policy, the CPU is never idle (i.e., it follows a \textit{zero-waiting} strategy) and works at constant speed. Thus, for each task $n$, $Z_n=0$ and the processor runs at the same speed  $\omega$ for each cycle batch $x$. The AoI minimizing $\omega$ is the maximum speed $\omega^*$ within the power budget $\bar{P}$. That is, each cycle batch $x$ has the same optimal execution time $\tau^*$.    
     With the same power consumption model as in Section~\ref{subsec:CPU power}, the power constraint \eqref{eq:avePower} implies $\tau_x^*$ satisfies
\begin{align}\label{eq:constant_speed}
\frac{N\mathbb{E}\left[X\right]\Ebatch(\tau^*)}{N\mathbb{E}\left[X\right]\tau^*}\le \bar{P}.
\end{align}
This implies 
$\tau^*=\bar{P}^{-\frac{\alpha-1}{\alpha+1}}$. Furthermore, since $Y_n=\tau^*X_n$, it follows from~\eqref{eq:Q-defn} and~\eqref{average-AoI} that the constant speed policy achieves average AoI
\begin{align}\label{eq:AoI-constantspeed}
    \Delta^{(\text{ave})}
    =\tau^*\left[\frac{\text{var}[X]}{2\E[X]}
    +\frac{3\E[X]}{2}\right].
\end{align}
We observe in \eqref{eq:AoI-constantspeed} that for a given expected CPU demand $\E[X]$, the constant speed policy suffers an AoI penalty associated with the variance $\text{var}[X]$ in CPU demand. 
As was previously observed in \cite{sun2017update}, which employed a system model consistent with constant CPU speed, the waiting time $Z_n$ mitigates this age penalty by not wasting computational resources on updates for which the age reduction will be small. In our numerical results, we will see that the policy combination of both waiting and CPU speed offers a two-fold improvement. As in \cite{sun2017update}, waiting avoids wasting computation resources but now CPU scheduling can also reduce the variance in the CPU demand $X$.

\begin{figure*}[t]
\centering
\label{fig:compare_aoi}
\subfigure[]{
\centering
\includegraphics[width=0.31\linewidth]{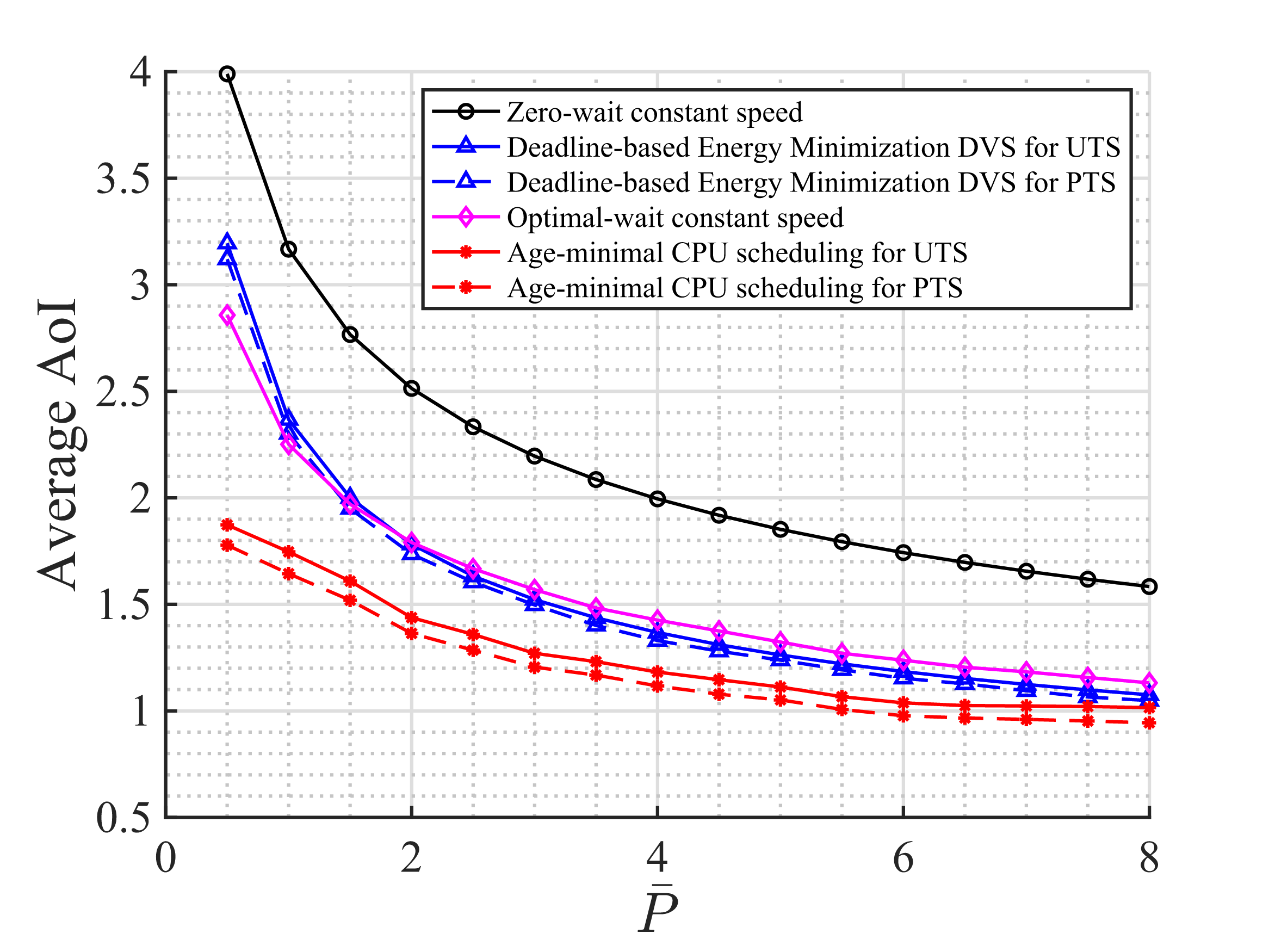}
\label{fig:compare_aoi_power}
}
\subfigure[]{
\centering
\includegraphics[width=0.31\linewidth]{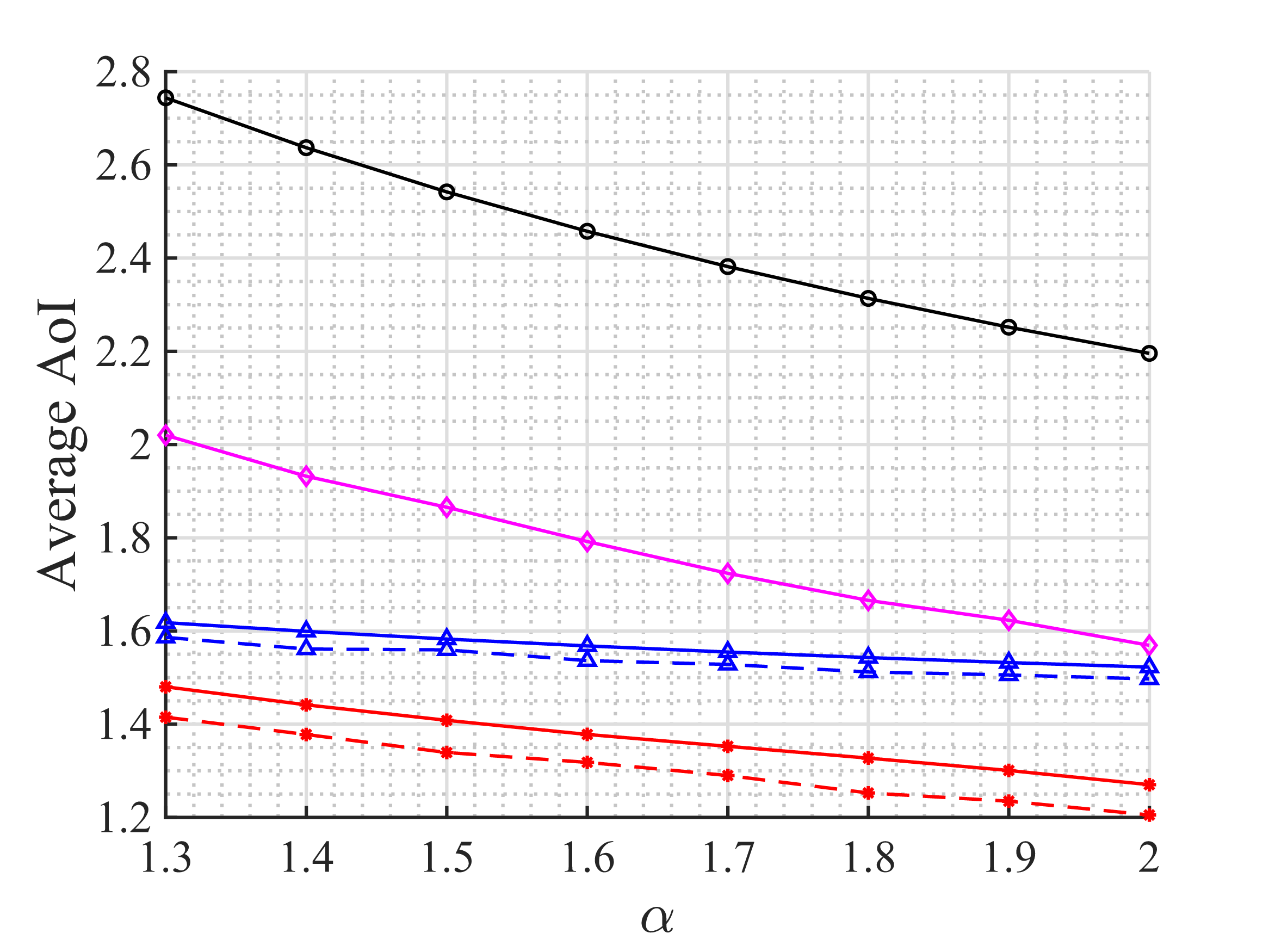}
\label{fig:compare_aoi_alpha}
}
\subfigure[]{
\centering
\includegraphics[width=0.31\linewidth]{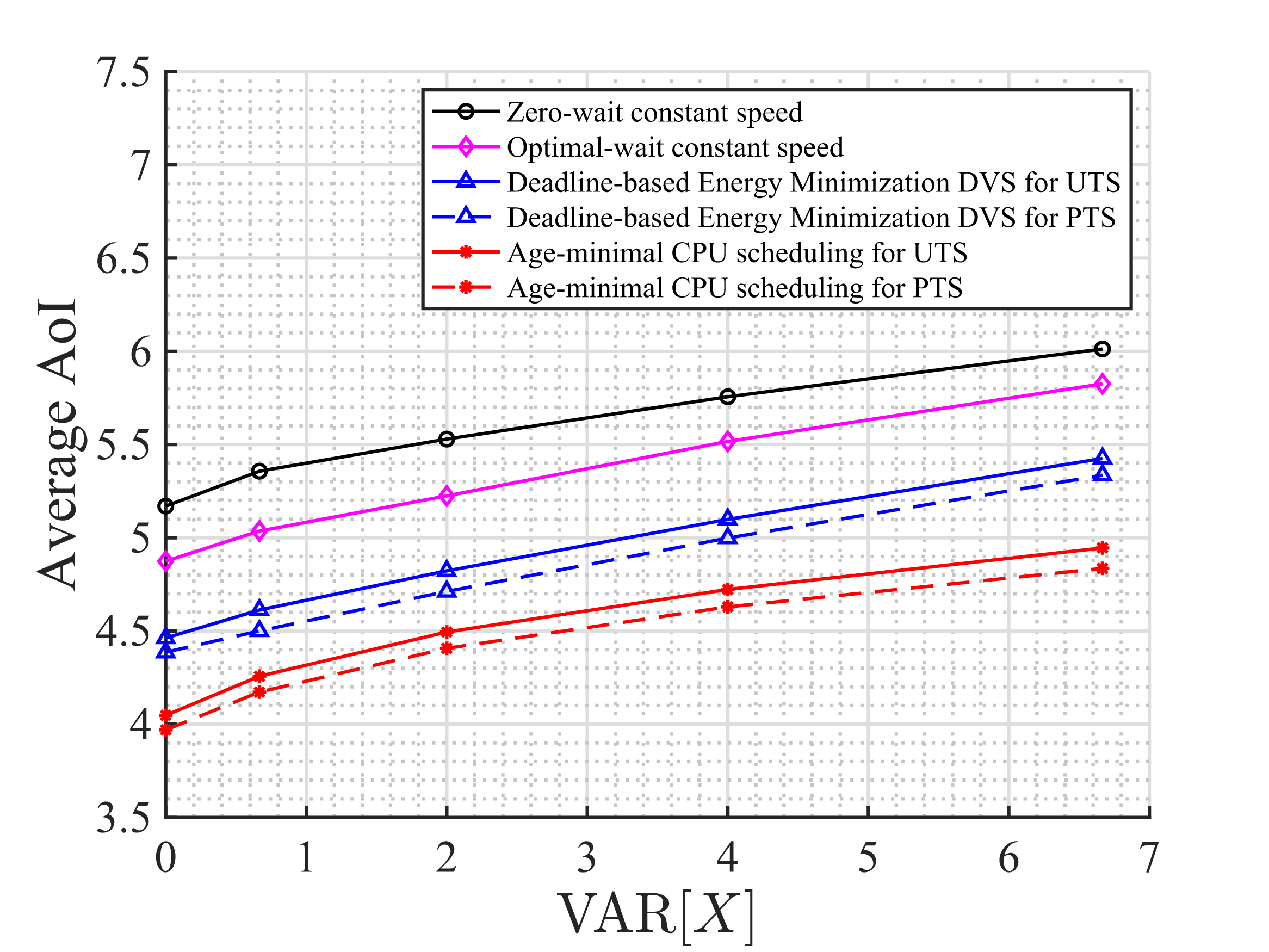}
\label{fig:totalAoI_var}
}
\centering
\caption{Average AoI comparison of six policies as functions of (a) the average $\bar{P}$, (b) the velocity saturation constant $\alpha$ and (c) the variance of task size $X$ which follows the uniform distribution.}
\end{figure*}
    
     \item \textit{Deadline-based Energy Minimization DVS~\cite{yuan2006energy} for UTS}: Under this benchmark policy, a batch cycle time $\tau_x$ is to be found for each batch $x$ in an update processing task such that the total expected energy consumption of these allocated cycles is minimized subject to a guarantee that the task execution time does not exceed a deadline $T$. 
     In addition, we consider zero-wait in this case.
Therefore, the speed schedule for a task can be presented as
     \begin{subequations}
     \begin{align}  \min_{\bs{\tau}}&\quad\sum\limits_{x=1}^b\Fbar(x)\Ebatch(\tau_x),\\
         {\rm s.t.}&\quad\sum\limits_{x=1}^b\tau_x\leq T,
     \end{align}
     \end{subequations}
     For a fair comparison, we tune the deadline $T$ so that the average power consumption meets the power budget $\bar{P}$.  

      \item \textit{Deadline-based Energy Minimization DVS for PTS}:  We now examine a DVS policy for the PTS case, a system that was not considered in \cite{yuan2006energy}. Given the deadline constraint $T$ and the observation of the task size $X$, the energy consumption of the total task is minimized by executing each batch of the task with the same speed: $\tau=\tau^*(X)=T/X$. 
      The achieved minimal energy consumption is  thus $X\Ebatch(\tau)$. To facilitate a fair comparison, we adjust the deadline $T$ to satisfy the average power constraint 
    \begin{align}    
    \frac{\mathbb{E}\left[X\Ebatch(\tau^*(X))\right]}{T}=\bar{P}.
    \end{align}

     \item \textit{Optimal-wait constant speed~\cite{sun2017update}}: Under this policy, the schedule chooses an optimal waiting strategy $z(\cdot)$ given an allowed update frequency $f_{\max}$ towards minimizing the average AoI. Due to the existence of the stationary randomized policy~\cite[Theorem 1]{sun2017update}, such an update frequency constraint condition can be expressed as
     \begin{align}
     \label{eq:update_constraint}
         {\rm s.t.} \quad\mathbb{E}[Y+z(Y)]\geq \frac{1}{f_{\max}}.
     \end{align}
     
     In the comparison, we apply the same power consumption model as in Section~\ref{subsec:CPU power}. Therefore, for each task, the processor works at a constant speed $\omega$ which corresponds to the same batch execution time $\tau$. Followed with Eq.~\eqref{eq:avePower}, constraint~\eqref{eq:update_constraint} becomes
     \begin{align}  
       {\rm s.t.}\quad\mathbb{E}[Y+z(Y)]\geq \frac{\mathbb{E}[X]\Ebatch(\tau)}{\bar{P}}.
     \end{align}  
    We explain the details to obtain the optimal solutions in Appendix~\ref{sec:alg_optimal-wait}. Since $Y_n=\tau X_n$, the optimal update policy achieves average AoI
     \begin{align}
     \label{eq:update_aoi}
         \Delta^{\rm (ave)}=\frac{\mathbb{E}[(\tau X+z(Y))^2]}{2\left(\mathbb{E}[\tau X]+\mathbb{E}[z(Y)]\right)}+\mathbb{E}[\tau X].
     \end{align}
\end{itemize}

To understand the performance gains of our proposed scheme, we compare the performance of the other four benchmarks under a $b=2$ batches scenario.

In Fig.~\ref{fig:compare_aoi_power}, we fix the power consumption model as $\alpha=2$ and then plot the AoI with the power budget $\bar{P}$ varying from $\bar{P}=0.5$ to $\bar{P}=8$. As expected, the average AoI is decreasing in $\bar{P}$ for all six schemes. 
We next observe that the optimal age-minimal CPU scheduling policy significantly reduces the AoI compared to the other policies. For example, when $\bar{P}=5$, 
age-minimal CPU scheduling for UTS achieves an AoI that is $40\%$ lower than zero-wait constant speed, $16\%$ lower than optimal-wait constant speed, and $12\%$ lower than deadline-based energy minimization DVS for UTS. 
This AoI reduction expands with decreasing $\bar{P}$ to $53\%$, $34\%$ and $41\%$ respectively at $\bar{P}=0.5$.
In the meanwhile, age-minimal CPU scheduling for PTS achieves an AoI that is $43\%$ lower than zero-wait constant speed, $21\%$ lower than optimal-wait constant speed, and $15\%$ lower than deadline-based energy minimization DVS for PTS at $\bar{P}=5$. 
Furthermore, this AoI reduction expands with decreasing $\bar{P}$ to $55\%$, $38\%$ and $43\%$ respectively at $\bar{P}=0.5$.
\textit{That is, compared to the other four benchmarks, our proposed strategy obtains greater benefits when the power budget is tighter.}

Comparing the deadline-based energy minimization DVS policy with the optimal-wait constant speed policy, we find that the impact of CPU scheduling on AoI increases with the power budget. 
Moreover, the DVS policy for the PTS case yields greater AoI benefits than for the UTS case within the same power budget $\bar{P}$ due to the additional information of task size $X$. This advantage is also evident in our proposed age-minimal CPU scheduling policy for PTS and UTS cases.
Furthermore, the realization time of the task size $X$ plays a more vital role when facing a stricter power budget.
Alternatively, when all six policies are required to meet the same AoI target $\gamma^*$, age-minimal CPU scheduling reduces power consumption. For example, at $\gamma^*=1.79$, our proposed age-minimal CPU scheduling policy reduces energy consumption by over $50\%$ and $75\%$ for the UTS and PTS cases, respectively.

To characterize the effect of the power consumption model on AoI, 
Fig.~\ref{fig:compare_aoi_alpha} fixes the power budget at $\bar{P}=3$ and varies $\alpha$ from $1.3$ to $2$ with a step size of $0.1$.  We notice that all the optimal Dinkelbach's coefficient $\gamma^*$, i.e., the minimum average AoI, decrease in $\alpha$. Due to the optimization problem formulation in Eq.~\eqref{eq:minimize_age}, it is obvious that the value of $\frac{2}{\alpha-1}$ decreases in $\alpha$, which leads to a more lenient constraint. 
That is, the CPU becomes more sensitive to frequency as $\alpha$ decreases.

We further observe in Fig.~\ref{fig:compare_aoi_alpha} that the zero-wait constant speed policy, optimal-wait constant speed scheme, and deadline-based energy minimization DVS strategy for PTS and UTS incur a larger average AoI than our proposed scheme.
Moreover, when $\alpha=2$, the age-minimal CPU scheduling policy for UTS produces a $17\%$ reduction in average AoI compared to the deadline-based energy minimization DVS policy for UTS, a $19\%$ reduction compared to the optimal-wait constant speed policy, and a $42\%$ reduction to the zero-wait constant speed scheduling, respectively.
Meanwhile, the age-minimal CPU scheduling policy for PTS results in a $19\%$ reduction in average AoI compared to the deadline-based energy minimization DVS policy for PTS, a $23\%$ reduction compared to the optimal-wait constant speed policy, and a $45\%$ reduction to the zero-wait constant speed scheduling.

To characterize the effect of the task size distribution on the age of information, we consider the case where task size $X$ follows a uniform distribution in a discrete scenario. Specifically, we fix $\mathbb{E}[X]=5$ and set a maximum cycle task size $b=9$. For instance, when $f(4)=f(5)=f(6)=\frac{1}{3}$, the variance ${\rm Var}[X]=\frac{2}{3}$.
With power budget $\bar{P}=5$, we plot the average age of information against ${\rm Var}[X]$ in Fig.~\ref{fig:totalAoI_var}.
We notice that all the minimum average AoI $\gamma^*$ increases with the variance. Furthermore, our proposed age-minimal CPU scheduling strategy yields greater AoI benefits when the variance of the task size distribution is higher.

\section{Conclusions}
\label{sec:conclusion}
In this paper, we formulate the age-minimal CPU scheduling problem as a constrained SMDP problem with uncountable space to achieve the trade-off between the age of information and energy-saving for both predictable task size (PTS) and unpredictable task size (UTS) cases.

We design an action matrix that enumerates CPU strategies for all task sizes so that the CPU schedule is specified by the observation of the task size for the PTS case.
With this approach, the task size state for PTS is mapped into the action dimension, which unifies the PTS and UTS problem formulations and transforms the unified problem formulation into a standard SMDP.

To simplify the problem and overcome the nonconvexity in this formulation, 
we first prove the existence of a stationary randomized policy for a quantized state space that achieves asymptotic optimality in the uncountable state space as the quantization step size goes to zero, and then transform the problem into an average cost one.
We develop the age-minimal CPU scheduling algorithm to obtain the optimal scheduling policies with provable convergence guarantees.
Our numerical results demonstrate that our proposed schemes provide greater AoI benefits under tighter power constraints.

As the first study on designing and understanding the impact of CPU scheduling on data freshness, there are many future research directions. Since our work assumes a generate-at-will update model, one potential direction is to consider queueing models in which updates are generated by an exogenous arrival process. Furthermore,  it is worth considering a system in which CPU scheduling is implemented by multiple computational servers serving multiple updating sources.

\bibliographystyle{IEEEtran}
\bibliography{Reference_Journal}

\appendices
\section{Proof of Theorem~\ref{theorem_stationary}}\label{proof:stationarity}
Since our model is formulated as a Semi-Markov Decision Process problem with an average cost in an infinite continuous state space, it is challenging to prove the existence of an optimal stationary randomized policy in such an infinite space SMDP.

Notably, the state transition probabilities of our SMDP problem depend on decisions.
To this end, we employ a quantization approach to discretize the state space.

Specifically, we constrain the state $Y$ to a finite interval $[0,\ymax]$ and then quantize this domain into $\qmax$ continuous intervals, each with a fixed length $\Delta_Y=\ymax/\qmax$.
For $q\in[\qmax]$, the actions $\bs{a}(y)$remain constant over any given interval 
$\Yint_q=[(q-1)\Delta_Y,q\Delta_Y)$.
Specifically, with $y'_q =(q-1/2)\Delta_Y$ denoting the midpoint of the $q$th interval,
\begin{align}
    \bs{a}(y)=\bs{a}(y_q'),\quad y\in\Yint_q. 
\end{align}

According to \cite{saldi2017asymptotic}, for average cost MDP problems, adopting a finite approximation method guarantees the existence of a stationary randomized policy that approximates the optimal policy with arbitrary precision.

Therefore, for our formulated problem, there always exists a stationary randomized policy that achieves asymptotic optimality. This enables us to approximate the global optimum by appropriately adjusting the quantization interval. This completes the proof.

\section{Proof of Lemma~\ref{Lemma_constant}}\label{proof:constant}

In this proof, we will focus on the stationary policies $\pi\in\Pi_{\rm SR}$ due to Theorem \ref{theorem_stationary}. 
Let $\pi_p\triangleq\{\pi_{p,Z},\pi_{p,\T}\}\in\Pi_{\rm SR}$ represent the stationary randomized policy within the PTS scenario. 
Here, $\pi_{p,Z}$ denotes the policy for determining the waiting action $Z$ and $\pi_{p,\T}$ denotes the policy for deciding on the CPU scaling action $\T$, both based on a given state $Y$.

Suppose that $\pi_p^*=\{\pi^*_{p,Z},\pi^*_{p,\T}\}$ is the optimal policy to Problem~\eqref{eq:minimize_age_stat}. In the following, we construct a new policy $\bar{\pi}_p \triangleq 
\{\bar{\pi}_{p,Z},\bar{\pi}_{p,\T}\}$ and show that it is optimal as well.  For this new policy, the waiting policy is unchanged; i.e. 
$\bar{\pi}_{p,Z}={\pi}^*_{p,Z}$.
However, for the new CPU scaling policy $\bar{\pi}_{p,\T}$, whenever $\pi^*_{p,\T}$ employs the scaling matrix $\T$, we instead use the lower triangular scaling matrix $\bar{\T}(\T)$ such that the nonzero entries in each row,
\begin{align}
    \ij[x,i]{\bar{\T}(\T)}=\taubar_x(\T)\triangleq\frac{1}{x}\sum_{j=1}^x \T_{x,j},\quad i\leq x,
\end{align}
are all identical. 
For any given current state $Y$ and realized task size $X$, the next state $Y'$ is given by
\begin{align}
    Y'= \sum_{j=1}^{X}\ij[X,j]{\bar{\T}(\T)}=\sum_{j=1}^X \T_{X,j}, \label{Y'}
\end{align}
whether  $\T$ or $\bar{\T}(\T)$ is used for CPU scaling. Hence, \eqref{Y'} implies that the new policy leads to the same transition probability of the embedded Markov chain $Y_n$, ${\rm Pr}(Y_{n+1}|Y_n)$.
Therefore, we conclude that the new policy maintains the same objective value as the optimal one, i.e., 
\begin{align}
    \frac{\mathbb{E}_{(Z,\T)\sim\pi^*}[A(Y,Z,L)]}{\mathbb{E}_{(Z,\T)\sim\pi^*}[Y+Z]}=\frac{\mathbb{E}_{(Z,\T)\sim\bar{\pi}}[A(Y,Z,L)]}{\mathbb{E}_{(Z,\T)\sim\bar{\pi}}[Y+Z]}.
\end{align}
We next check the energy feasibility of $\bar{\pi}_p$. 
By Jensen's inequality, we further conclude that given the current state $Y=y$ and task size $X=x$, the required processing energy is reduced since
\begin{align}
    x\Ebatch(\taubar_x)=x\Ebatch\Bigl(\frac{1}{x}\sum_{j=1}^x\ij[x,j]{\T}\Bigr)&\leq x \sum_{j=1}^x\frac{1}{x}\Ebatch(\ij[x,j]{\T}).
\end{align}
Therefore,  $\bar{\pi}_{p}$ is a feasible solution to Problem~\eqref{eq:minimize_age_stat}.
\begin{align}
\mathbb{E}[\Ebatch(\taubar_x)]&=\mathbb{E}_{X}\mathbb{E}_{\T\sim\bar{\pi}_{p,\T}}\bigg[\sum_{j=1}^X \Ebatch(\ij[X,j]{\T})\bigg]\nn
&\leq \mathbb{E}_{X}\mathbb{E}_{\T\sim\pi^*_{p,\T}}\bigg[\sum_{j=1}^X \Ebatch(\ij[X,j]{\T})\bigg],
\end{align}
which indicates that $\bar{\pi}_{p}$ achieves an expected energy consumption no larger than that of the optimal solution $\pi_{p}^*$. Therefore, $\bar{\pi}_{p}$ is also a feasible solution to Problem~\eqref{eq:minimize_age_stat}.

\section{Proof of Theorem~\ref{theorem_optimality_pbs}}\label{proof:optimality_pbs}
We use ${\rm Pr}\{l\vert y,\bs{a}\}$ to denote the transition probability which has a corresponding probability distribution density $p^{\bs{a}}(l\vert y)$. To this end, ${\rm Pr}[l\vert y,\bs{a}]\triangleq\int p^{\bs{a}}(l\vert y)\,dl$ for all $y\in\mathcal{Y}$ and $\bs{a}\in\mathcal{A}^\xi$.
By Scheffe's Theorem~\cite{hernandez2012adaptive} and using the identity $\vert s-t \vert=s+t-2\min[s,t]$, we can write 
\begin{align}
    \Vert {\rm Pr}\{l\vert y,\bs{a}\}-{\rm Pr}\{l\vert y',\bs{a'}\} \Vert=\int \vert p^{\bs{a}}(l\vert y)-p^{\bs{a'}}(l\vert y') \vert\, dl& \nn
    =2-2\int\min\left(p^{\bs{a}}(l\vert y), p^{\bs{a'}}(l\vert y')\right)\,dl&, 
\end{align}
which leads to the existence of a scalar $\eta\in(0,1)$ that satisfies $\Vert {\rm Pr}\{l\vert y,\bs{a}\}-{\rm Pr}\{l\vert y',\bs{a'}\} \Vert\leq 2\eta$.

Combining \cite[Condition 1]{hernandez1991average}  and \cite[Theorem 4.1]{hernandez1991average}, we can obtain that there exists a constant $\rho^*$ and a function $R(y)$ such that $(\rho^*, R^*(y))$ is a solution to the optimality equation
\begin{align}\label{eq:theorem4.1}
\rho^*+R^*(y)=\min\limits_{\bs{a}\in\mathcal{A}}&\biggl[\int p^{\bs{a}}(l\vert y)R^*(l)dl +\newG[\gamma^*]{y}{\bs{a}} 
\biggl],
\end{align}
where $\rho^*$ is just the optimal cost of an SMDP under the policy. 
With Dinkelbach's method, the optimal average cost $\rho^*$ for our problem is given by 
\begin{IEEEeqnarray}{rCl}
\rho^*&=&\min_{\bs{a}\in\mathcal{A}}\lim\limits_{N\to\infty}\mathbb{E}\biggl[\frac{1}{N}\sum\limits_{n=1}^N\g[\gamma^*]{Y_n}{\bs{a}}\biggr].
\IEEEeqnarraynumspace
\end{IEEEeqnarray}
Following from Equation~\eqref{eq:optima_policy_pbs},
we substitute the optimal average cost $\rho^*=0$ into Eq.~\eqref{eq:theorem4.1}, 
$R^*(y)$ is the optimal solution to the equation as follows
\begin{align}
    R^*(y)&=\min_{\bs{a}\in\mathcal{A}}\Big[\int p^{\bs{a}}(l\vert y)R^*(l)dl
+\newG[\gamma^*,\lambda]{y}{\ba}
\Big].
\end{align}
Thus, given a stationary randomized policy $\pi\in\Pi_{{\rm SR}}$ in the PTS and UTS cases, Theorem~\ref{theorem_optimality_pbs} is proved.

\section{Proof of Theorem~\ref{theorem_convergence}}
\label{proof:convergence}
Defining the operator
\begin{align}
\mathbb{T}R(y)=\min\limits_{\bs{a}\in\mathcal{A}}
\int &p^{\bs{a}}(l\vert y)R(l)\,dl 
+\newG{y}{\bs{a}},
\end{align}
the sequence of values generated by the value iteration \eqref{value itera2} 
can be written as 
\begin{align}
    R^{n+1}(y)=\mathbb{T}R^n(y)=\mathbb{T}^{n+1}R^0(y).
\end{align}
We will now that $\mathbb{T}$ is a contraction mapping.
For value functions $R(y)$ and $R'(y)$, define $y^*=\arg\max_y\{\mathbb{T}R(y)-\mathbb{T}R'(y)\}$ and $y_*=\arg\min_y\{\mathbb{T}R(y)-\mathbb{T}R'(y)\}$. Using $\ba_1^*$ to denote the optimal action for $R(y^*)$ and $\ba_2^*$ for $R'(y^*)$, we can derive
\begin{align}
    \mathbb{T}R(y^*)-\mathbb{T}R'(y^*)
    &=\bigg[\int p^{\ba_1^*}(l\vert y)R(l)dl+\newG{y^*}{\ba_1^*}
    \bigg]\nn
    &\qquad-\bigg[\int p^{\ba_2^*}(l\vert y)R'(l)dl+\newG{y^*}{\ba_2^*}
    \bigg]\nn
    &\leq\int p^{\ba_2^*}(l\vert y)R(l)dl-\int p^{\ba_2^*}(l\vert y)R'(l)dl.
\end{align}
Similarly, using $\ba_3^*$ to denote the optimal action for $R(y_*)$ and $\ba_4^*$ for $R'(y_*)$, we have
\begin{align}
    \mathbb{T}R(y_*)-\mathbb{T}R'(y_*)
    &=\bigg[\int p^{\ba_3^*}(l\vert y)R(l)dl+\newG{y_*}{\ba_3^*}
    \bigg]\nn
    &\qquad-\bigg[\int p^{\ba_4^*}(l\vert y)R(l)dl+\newG{y_*}{\ba_4^*}
    \bigg]\nn
    &\geq\int p^{\ba_3^*}(l\vert y)R(l)dl-\int p^{\ba_3^*}(l\vert y)R'(l)dl.
\end{align}

Therefore, the infinity-norm of $\mathbb{T}R(y)-\mathbb{T}R'(y)$ is given as
\begin{align}
    \Vert \mathbb{T}R(y)-\mathbb{T}R'(y) \Vert_\infty
    \leq \int p^{\ba_2^*}(l\vert y)\left[R(l)-R'(l)\right]dl-\int p^{\ba_3^*}(l\vert y)\left[R(l)-R'(l)\right]dl.
\end{align}
Since $p^{\ba}(l|y)$ denotes the transition probability distribution from state $y$ to $l$ which is determined by the action $\ba$ as well as the CPU cycle batch demand, we define
\begin{align}
\label{eq:factor}
    B(\ba_2^*,\ba_3^*;l)=\min\left\{p^{\ba_2^*}(l\vert y),p^{\ba_3^*}(l\vert y)\right\}.
\end{align}
For any $R$ and $R'$, we further define
\begin{align}
    \Lambda(R-R')&=\min\limits_y \left[R(y)-R'(y)\right], \\
    \Upsilon(R-R')&=\max\limits_y \left[R(y)-R'(y)\right].
\end{align}
By using the span seminorm, we define the span of $R(y)-R'(y)$, denoted $\spansp(R(y)-R'(y))$, which is given by
\begin{align}
    \spansp(R(y)-R'(y))=\Upsilon(R-R')-\Lambda(R-R').
\end{align}
Thus, for any $R(y)$ and $R'(y)$, we can obtain that
\begin{align}
    \spansp&(\mathbb{T} R(y)-\mathbb{T} R'(y))=\int p^{\ba_2^*}(l\vert y)\left[R(l)-R'(l)\right]dl-\int p^{\ba_3^*}(l\vert y)\left[R(l)-R'(l)\right]dl \nn
    &=\int \left[p^{\ba_2^*}(l\vert y)-B(\ba_2^*,\ba_3^*;l)\right]\left[R(l)-R'(l)\right]dl-\int \left[p^{\ba_3^*}(l\vert y)-B(\ba_2^*,\ba_3^*;l)\right]\left[R(l)-R'(l)\right]dl \nn  
   &\leq\int \left[p^{\ba_2^*}(l\vert y)-B(\ba_2^*,\ba_3^*;l)\right]dl\Upsilon(R-R')-\int \left[p^{\ba_3^*}(l\vert y)-B(\ba_2^*,\ba_3^*;l)\right]dl\Lambda(R-R') \nn 
    &=\left[1-\int B(\ba_2^*,\ba_3^*;l)dl\right]\spansp(R(y)-R'(y)).
\end{align}
Following~\cite[Theorem 6.6.2]{puterman2014markov}, the operator $\mathbb{T}:R(y)\rightarrow R(y)$ is a span contraction mapping as there exists a $\phi$, $0\leq\phi<1$ satisfies
\begin{align}
    \spansp(\mathbb{T}R(y)-\mathbb{T}R'(y))\leq\phi \spansp(R(y)-R'(y))
\end{align}
for all $R(y)$ and $R'(y)$ and the convergence of value iteration is guaranteed whenever $\mathbb{T}$ is a span contraction.
To this end, 
\begin{align}
    \spansp&(R^{n+m}(y)-R^n(y)) \nn
    &\leq \sum\limits_{k=0}^{m-1}\spansp(R^{n+k+1}(y)-R^{n+k}(y))\nonumber \\
    &=\sum\limits_{k=0}^{m-1}\spansp(\mathbb{T}^{n+k}R^1(y)-\mathbb{T}^{n+k}R^0(y)) \nonumber \\
    &\leq\sum\limits_{k=0}^{m-1}\phi^{n+k}\spansp(R^1(y)-R^0(y)) \nonumber \\
    &=\frac{\phi^n(1-\phi^m)}{(1-\phi)}\spansp(R^1(y)-R^0(y)).
\end{align}
Since $0\leq\phi<1$, $\spansp(R^{n+m}(y)-R^n(y))$ can be made arbitrarily small. 
This implies that there exists a fixed point $R^*(y)$ such that $\mathbb{T}R^*(y)=R^*(y)$. Consequently, for any initial $R^0(y)$, the sequence $R^n(y)$ will converge to $R^*(y)$.

\section{Proof of Corollary~\ref{corollary_solution_pbs} and Corollary~\ref{corollary_solution_ubs}}
\label{proof:solution_pbs}
From Equations~\eqref{eq:pi(y)_pbs} and~\eqref{value itera2}, the optimal value function, given the optimal Dinkelbach variable $\gamma^*$ and the optimal dual variable $\lambda^*$, is 
\begin{align}
R^*(y)=\min\limits_{\ba\in \mathcal{A}}\Big[&\E[R^m(L(X;\bs{\tau}))]+\newG[\gamma^*,\lambda^*]{y}{\ba}.
\label{Eq-Rp*}  
\end{align} 

Following the definition of $Q^*(y,(z,\bs{\tau}))$ in Equation~\eqref{eq:Qp-star}, the Lagrangian of \eqref{eq:benders_pbs} is given by
\begin{IEEEeqnarray}{rCl}
\mathcal{L}(y,(z,\bs{\tau}), \mu, \vartheta, \varphi)
    &=& Q^*(y,(z,\bs{\tau}))-\mu z+\vartheta(\tau_x-\tau_{\max})-\varphi(\tau_x-\tau_{\min}).\IEEEeqnarraynumspace
\end{IEEEeqnarray}
We now present the Karush-Kuhn-Tucker (KKT) conditions.
For the PTS system, 
\begin{subequations}
\begin{IEEEeqnarray}{rCl}
\frac{d\mathcal{L}}{d\tau_x}
&=&xf(x)\bigg(\frac{\partial R^*(x\tau_x)}{\partial y}+y+z+\frac{2\lambda^*}{1-\alpha}\tau_x^{\frac{1+\alpha}{1-\alpha}}\bigg)+\vartheta-\varphi=0, \label{eq:dLdtau-PTS}
\end{IEEEeqnarray}
and for the UTS system,
\begin{IEEEeqnarray}{rCl}
\frac{d\mathcal{L}}{d\tau_x}&=&\sum\limits_{k'=x}^b f(k')\frac{\partial R^*(\sum_{x=1}^{k'}\tau_x)}{\partial y}
 +\Fbar(x)(y+z)\nn
 &&\quad-\frac{2\lambda^*\Fbar(x)}{\alpha-1}\left(\frac{1}{\tau_x}\right)^{\frac{\alpha+1}{\alpha-1}}+\vartheta-\varphi=0.\IEEEeqnarraynumspace \label{eq:dLdtau-UTS}
 \end{IEEEeqnarray}
 \end{subequations}
In addition, for both PTS and UTS systems,
 \begin{subequations}
\begin{IEEEeqnarray}{rCl} 
    \frac{d\mathcal{L}}{dz}&=&\Lbar(\btau)+y+z-\gamma^*-\lambda^*\bar{P}-\mu=0, \label{eq:dLdz}\\
      \mu&\geq& 0,~z\geq0,~\mu\cdot z=0,\label{eq:mu}\\
    \vartheta&\geq& 0,~\tau_x\leq\tau_{\max},~\vartheta\cdot (\tau_x-\tau_{\max})=0,\label{eq:vartheta}\\
    \varphi&\geq&0,~\tau_x\geq\tau_{\min},~\varphi\cdot (\tau_x-\tau_{\min})=0. \label{eq:varphi}
\end{IEEEeqnarray}
\end{subequations}
Based on the differentiability of the optimal value function $R^*(y)$ as stated in Lemma~\ref{lemma:diff_Ry},
the optimal waiting strategy $z^*(y)$ for each given $y$ is obtained by considering the following two cases:
\begin{itemize}
    \item 
    If $\mu=0$, then by \eqref{eq:dLdz}, we obtain
    \begin{align}\label{eq:mu_z^*(y)}
        z^*(y)=\gamma^*+\lambda^*\bar{P}-y-\Lbar(\btau^*(y)).
    \end{align}
    \item  
    If $\mu>0$, then by \eqref{eq:mu}, $z^*(y)=0$.
\end{itemize}
The optimal CPU scaling policy $\bs{\tau}^*$ for each given $y$ is obtained by considering the following three cases:

\smallskip\par\noindent{\textit{Case 1}}: $\vartheta=\varphi=0$:
For the PTS system, it follows from \eqref{eq:dLdtau-PTS} and the definition of $\psi_x(y)$ in \eqref{eq:psi_pbs} that
    \begin{align}\label{eq:tau_convex}
        \tau_x^*(y)&=\left(\frac{2\lambda^*/(\alpha-1)}{\frac{\partial R^*(x\tau_x^*(y))}{\partial y}+y+z^*(y)}\right)^{\frac{\alpha-1}{\alpha+1}}\nn
        &=\left(\frac{2\lambda^*/(\alpha-1)}{\psi_x(y)+y+z^*(y)}\right)^{\frac{\alpha-1}{\alpha+1}},~x\in[b].
    \end{align}
      This verifies \eqref{eq:tau_pbs} for the PTS system. Again from the definition of $\psi_x(y)$ in \eqref{eq:psi_pbs}, the substitution of $x\tau_x^*(y)$ for $y$ in \eqref{eq:par_Rpl} yields \eqref{eq:psi_pts1} for the PTS system.
    
    For the UTS system, \eqref{eq:dLdtau-UTS} and the definition of $\zeta_x(y)$ in \eqref{eq:zeta_Ru}  yield
    \begin{IEEEeqnarray}{rCl}
      \tau_x^*(y)&=&\Bigg(\frac{2\lambda^*/(\alpha-1)}{\frac{1}{\Fbar(x)}\sum\limits_{k'=x}^b f(k')\frac{\partial R^*(\sum_{x=1}^{k'}\tau_x(y))}{\partial y}+y+z^*(y)}\Bigg)^{\frac{\alpha-1}{\alpha+1}} \nn
        &=&\left(\frac{2\lambda^*/(\alpha-1)}{\zeta_x(y)+y+z^*(y)}\right)^{\frac{\alpha-1}{\alpha+1}},~x\in[b].  
    \end{IEEEeqnarray}
    This verifies \eqref{eq:tau} for the UTS system.
    From the definition of $\zeta_x(y)$ in \eqref{eq:zeta_Ru}, the substitution of $\sum\limits_{x=1}^{k'}\tau_x^*(y)$ into \eqref{eq:par_Rpl} yields \eqref{eq:zeta} for the UTS system.
    
\par\smallskip\noindent\textit{Case $2$}: If $\vartheta>0$, then by \eqref{eq:vartheta}, $\tau^*_x(y)=\tau_{\max}$.

\smallskip\par\noindent\textit{Case $3$}: If $\varphi>0$, then by \eqref{eq:varphi}, $\tau^*_x(y)=\tau_{\min}$.

\smallskip 

\noindent This completes the derivation of  the optimal solution for both the PTS and UTS cases as stated in Corollaries~\ref{corollary_solution_pbs} and~\ref{corollary_solution_ubs}, respectively.

\section{Proof of Proposition~\ref{proposition_pbs}}
\label{proof:proposition_pbs}
We start by proving part (a) by contradiction. Suppose that $x\tau^*_x(y)> x'\tau^*_{x'}(y)$ for some $x\leq x'$. Note that since $x\le x'$, this directly implies  $\tau^*_x(y)> \tau^*_{x'}(y)$. Due to 
the convexity of $R^*(y)$ in $y$, 
the derivative $\frac{\partial R^*(l)}{\partial l}$ increases in $l$.
Combined with $x\tau^*_x(y)> x'\tau^*_{x'}(y)$, \eqref{eq:psi_pbs} yields $\psi_x(y)\geq\psi_{x'}(y)$.
However, it then follows from \eqref{eq:tau_pbs} that $\tau^*_x(y)\leq \tau^*_{x'}(y)$, which is a contradiction.
Therefore, we must have $x\tau^*_x(y)\leq x'\tau^*_{x'}(y)$ for all $x\leq x'$. 

For part (b), since $\frac{\partial R^*(l)}{\partial l}$ is increasing in $l$, it follows from \eqref{eq:psi_pbs} that $ \psi_x(y)\leq\psi_{x'}(y)$ for all $x\leq x'$. Thus we conclude from \eqref{eq:tau_pbs} that $\tau^*_x(y)\geq \tau^*_{x'}(y)$. This completes the proof.

\section{Proof of Proposition~\ref{propo_yp} and Proposition~\ref{propo_yu}}
\label{proof:proposition_yp_exist}
To avoid repetitive proofs, we introduce the variable $\xi\in\{p,u\}$, where $\xi=p$ corresponds to the PTS case and $\xi=u$ corresponds to the UTS case. In the following proof, $\xi$ will be used to represent the general case and streamline the presentation.

To establish the existence of the threshold $y_\xi$ in both the PTS and UTS cases, we first define $\tilde{z}^*(y)$ according to \eqref{eq:z+y_pbs} and \eqref{eq:z+y} as 
\begin{align}\label{eq:def_tildez}
    \tilde{z}^*(y)\triangleq\gamma^*+\lambda^*\bar{P}-y-\Lbar(\btau^*(y)).
\end{align}

In the following Section~\ref{subsect:threshold}, we will prove there exists a positive threshold $y_{\xi}$ 
such that $\tilde{z}^*(y_{\xi})=0$. 
Furthermore, Section~\ref{subsect:existence} will establish the existence of the optimal solutions as specified in Propositions~\ref{propo_yp} and~\ref{propo_yu}.
\subsection{The Existence of Threshold $y_\xi$}
\label{subsect:threshold}
Applying the envelope theorem, the first-order derivative of $\newG{y}{\ba^{*}(y)}$ with respect to $y$ 
is given by
\begin{align}\label{eq:R>=0}
    \frac{\partial \newG{y}{\ba^{*}(y)}}{\partial y}
    \overset{(a)}{=}z^*(y)-\tilde{z}^*(y)\geq0,
\end{align}
where (a) follows from the equation for $z^*(y)$ in both PTS and UTS cases, given in \eqref{eq:z+y_pbs} and \eqref{eq:z+y}, respectively, as well as the definition of $\tilde{z}^*(y)$ in \eqref{eq:def_tildez}.
Due to the non-negative dual variable $\lambda^*$ and 
\begin{align}
    \frac{\partial(g_3(y,\bs{a}^{*}(y))}{\partial y}=-\bar{P},
\end{align}
we must have
\begin{align}
    \frac{\partial \g[\gamma^*]{y}{\bs{a}^{*}(y)}}{\partial y}&\geq \lambda^* \bar{P}\geq0, \label{Eq.123}
\end{align}
implying that $\g[\gamma^*]{y}{\bs{a}^{*}(y)}$ is non-decreasing in $y$. 
Recalling that the possible value of state $Y$ is constrained to the finite interval $[y_{\min},\ymax]\subseteq [\tau_{\min},b\tau_{\max}]$,
Lemma \ref{Lemma_Dinkel} implies
\begin{align}
      J(\gamma^*)=&\int_{y_{\min}}^{y_{\max}}\g[\gamma^*]{y}{\bs{a}^{*}(y)}\sigma(y)dy=0, \label{Eq.124}
\end{align}
where $\sigma(y)$ is the probability density function of the stationary distribution of $y$ under the optimal policy $\pi^*$.

From  \eqref{Eq.123} and \eqref{Eq.124},  we have for $y\in(0,y_{\min}]$,
\begin{align}
    \g[\gamma^*]{y}{\bs{a}^{*}(y)}\leq0.
\end{align}
It  then follows from \eqref{gfunc} and \eqref{eq:Gdefn} that
\begin{align}
    (z^*(y)+y)\Lbar(\btau^*(y))+\frac{1}{2}(z^*(y)+y)^2-\gamma^*(z^*(y)+y)\leq0, \forall y\in(0,y_{\min}].
\end{align}
Since $z^*(y)+y>0$ for all $y\in(0,y_{\min}]$, dividing both sides by $z^*(y)+y$ yields
\begin{align}\label{eq:112}
    \Lbar(\btau^*(y))+\frac{1}{2}(z^*(y)+y)&\leq\gamma^*, \forall y\in(0,y_{\min}],
\end{align}
and this implies
\begin{align}\label{eq:Lbar-bound}
    \Lbar(\btau^*(y))<\gamma^*, \forall y\in(0,y_{\min}].
\end{align}
From \eqref{eq:def_tildez} we have
\begin{align}
    \tilde{z}^*(y)+y=\gamma^*+\lambda^*\bar{P}-\Lbar(\btau^*(y)), \forall y\in(0,y_{\min}].
\end{align}
Choosing $\epsilon=\min(\lambda^* \bar{P}, y_{\min})$ and combining with \eqref{eq:Lbar-bound} leads to $\tilde{z}^*(\epsilon)>0$.

Furthermore, \eqref{eq:def_tildez} implies $\lim_{y\to\infty}\tilde{z}^*(y)<0$.
Since $\tilde{z}^*(y)$ is continuous in $y$, there must exist at least one threshold $y_\xi$ such that $\tilde{z}^*(y_\xi)=0$.

\subsection{The Existence of the Solution}
\label{subsect:existence}
We begin by noting that while multiple optimal policies may exist, the optimal value function $R^*(y)$ is unique. This uniqueness arises because the optimal value function is derived from the optimization in \eqref{R-optimalvalue}.
We denote by $\mathcal{Z}^*(y)$ and $\mathcal{T}^*(y)$ the sets of all optimal waiting actions and the CPU scheduling actions, respectively, 
specified by such policies at state $y$. In other words, $\mathcal{Z}^*(y)$ and $\mathcal{T}^*(y)$ comprise all values that an optimal policy $z^*(\cdot)$ and $\boldsymbol{\tau}^*(y)$ may assign to $y$.
We select $(z\in \mathcal{Z}^*(y),y)$ to maximize $z+y$ under the condition that $z>0$ and define the threshold $\hat{y}$ as 
\begin{align}
    \hat{y} &=\sup (z+y)  \nn
    &{\rm s.t.}\  (z,y)\in\{(z,y): z\in \mathcal{Z}^*(y), z>0\}.
    \label{Eq-yp}
\end{align}
Next, we categorize and analyze the two scenarios: $y<\hat{y}$ and $y\geq \hat{y}$, for both the PTS and UTS cases.

\textit{(i)} $y<\hat{y}$: 
We first establish that $z^*(y)>0$ for all $y<\hat{y}$.
We observe from \eqref{eq:z+y_pbs} (for PTS) and \eqref{eq:z+y} (for UTS) and the definition of $\tilde{z}^*(y)$ in \eqref{eq:def_tildez} that $z^*(y)=[\tilde{z}^*(y)]^+$.
Combining Equation \eqref{eq:par_Rpl} from Lemma~\ref{lemma:diff_Ry} with \eqref{eq:def_tildez}, we obtain
\begin{align}\label{eq:par_R_positive0}
    \frac{\partial R^*(y)}{\partial y}&=\Lbar(\btau^*(y))+y+z^*(y)-\gamma^*-\lambda^*\bar{P}\nn
    &=z^*(y)-\tilde{z}^*(y)\geq0,
\end{align}
which implies that $\psi_x(y)\ge0$ and $\zeta_x(y)\ge0$, based on their definitions in \eqref{eq:psi_pbs} and \eqref{eq:zeta_Ru}, respectively.
Next, the second order derivative of $R^*(y)$ with $y$ when $z^*(y)=0$ is given by
\begin{align}\label{eq:partial2_R}
    \frac{\partial^2 R^*(y)}{\partial y^2}&=\frac{\partial \Lbar(\btau^*(y))}{\partial y}+1\nn
    &=\begin{cases}
\sum\limits_{x=1}^bf(x) x\frac{\partial\tau_x}{\partial y}+1, & \PTS,\\
\sum\limits_{x=1}^b\Fbar(x)\frac{\partial\tau_x}{\partial y}+1, & \UTS.
	\end{cases}
\end{align}
Here we need to consider the PTS and UTS cases separately.

Recall that the chip parameter $\alpha\in(1,2]$,
 it follows for the PTS system  from \eqref{eq:tau_pbs} that the first-order derivative of $\tau^*_x(y)$ with respect to $y$ is given as
\begin{align}\label{eq:par_tau_pts}
    \frac{\partial \tau^*_x(y)}{\partial y}=\underbrace{\Bigl(\frac{2\lambda^*}{\alpha-1}\Bigr)^{\frac{\alpha-1}{\alpha+1}}}_{\geq0}\underbrace{\frac{1-\alpha}{1+\alpha}}_{<0}\underbrace{\Bigl(\psi_x(y)+y\Bigr)^{\frac{-2\alpha}{1+\alpha}}}_{\geq0}\Bigl(\frac{\partial^2 R^*(y)}{\partial y^2}x\frac{\partial \tau_x^*(y)}{\partial y}+1\Bigr).
\end{align}
Define
\begin{align}
    A_x=\Bigl(\frac{2\lambda^*}{\alpha-1}\Bigr)^{\frac{\alpha-1}{\alpha+1}}\frac{1-\alpha}{1+\alpha}\Bigl(\psi_x(y)+y\Bigr)^{\frac{-2\alpha}{1+\alpha}}\le 0,
\end{align}
and Equation \eqref{eq:par_tau_pts} becomes
\begin{align}
    \frac{\partial \tau^*_x(y)}{\partial y}=\frac{A_x}{1-A_x\frac{\partial^2 R^*(y)}{\partial y^2}x}.
\end{align}
If $R^*(y)$ is convex in $y$, we have $\frac{\partial^2 R^*(y)}{\partial y^2}\ge0$, implying
\begin{align}\label{eq:sum_ave_partau_pts}
    \sum\limits_{x=1}^bf(x)x\frac{\partial \tau^*_x(y)}{\partial y}=\sum\limits_{x=1}^bf(x)x\frac{A_x}{1-A_x\frac{\partial^2 R^*(y)}{\partial y^2}x}\le0.
\end{align}
Substituting \eqref{eq:sum_ave_partau_pts} into \eqref{eq:partial2_R}, we obtain $0\le\frac{\partial^2 R^*(y)}{\partial y^2}\le1$ in the PTS case.

For the UTS case, it follows from \eqref{eq:tau} that the first-order derivative of $\tau^*_x(y)$ with respect to $y$ is
\begin{align}\label{eq:par_tau_uts}
    \frac{\partial \tau^*_x(y)}{\partial y}=&\underbrace{\Bigl(\frac{2\lambda^*}{\alpha-1}\Bigr)^{\frac{\alpha-1}{\alpha+1}}}_{\geq0}\underbrace{\frac{1-\alpha}{1+\alpha}}_{<0}\underbrace{\Bigl(\zeta_x(y)+y\Bigr)^{\frac{-2\alpha}{1+\alpha}}}_{\geq0}\Bigl(\frac{1}{\Fbar(x)}\sum\limits_{k'=x}^b f(k')\frac{\partial^2 R^*(y)}{\partial y^2}\frac{\partial \tau_x^*(y)}{\partial y}+1\Bigr).
\end{align}
Similar to the PTS case, by defining
\begin{align}
    B_x=\Bigl(\frac{2\lambda^*}{\alpha-1}\Bigr)^{\frac{\alpha-1}{\alpha+1}}\frac{1-\alpha}{1+\alpha}\Bigl(\zeta_x(y)+y\Bigr)^{\frac{-2\alpha}{1+\alpha}}\le0,
\end{align}
Equation \eqref{eq:par_tau_uts} becomes
\begin{align}
    \frac{\partial \tau^*_x(y)}{\partial y}=\frac{B_x}{1-\frac{B_x}{\Fbar(x)}\sum\limits_{k'=x}^b f(k')\frac{\partial^2 R^*(y)}{\partial y^2}},
\end{align}
which yields
\begin{align}\label{eq:sum_ave_partau_uts}
    \sum\limits_{x=1}^b\Fbar(x)\frac{\partial \tau^*_x(y)}{\partial y}= \sum\limits_{x=1}^b\Fbar(x)\frac{B_x}{1-\frac{B_x}{\Fbar(x)}\sum\limits_{k'=x}^b f(k')\frac{\partial^2 R^*(y)}{\partial y^2}}\le0.
\end{align}
Substituting \eqref{eq:sum_ave_partau_uts} into \eqref{eq:partial2_R}, we obtain $0\le\frac{\partial^2 R^*(y)}{\partial y^2}\le1$ in the UTS case.

Therefore, for both PTS and UTS cases, it follows from \eqref{eq:partial2_R} that  $-1\le\frac{\partial \Lbar(\btau^*(y))}{\partial y}\le0$. 
Given the definition of $\tilde{z}^*(y)$ in \eqref{eq:def_tildez}, the first-order derivative of $\tilde{z}^*(y)$ with $y$ is given as
\begin{align}
    \frac{\partial \tilde{z}^*(y)}{\partial y}=-1-\frac{\partial \Lbar(\btau^*(y))}{\partial y}\le0.
\end{align}
This indicates that $\tilde{z}^*(y)$ continues to decrease once it drops to $0$ and cannot become positive afterward.
Therefore, we have $z^*(y)>0$ and thus $\tilde{z}^*(y)=z^*(y)$ for all $y<\hat{y}$.
It follows from Equation \eqref{eq:par_R_positive0} that $\frac{\partial R^*(y)}{\partial y}=0$,
which implies that $R^*(y)$ remains constant for all $y<\hat{y}$.

We now adopt the shorthand $(z', y')$ to denote an optimal solution to \eqref{Eq-yp} at a specific $y'<\hat{y}$. That is,  $(z'=z^*(y'),\boldsymbol{\tau}^*(y'))\in \mathcal{Z}^*(y')\times \mathcal{T}^*(y')$, satisfying $z^*(y')>0$. 
We construct a new policy $(z(y),\boldsymbol\tau(y)=\{\tau_x(y)\}_{x=1}^b)$ such that
\begin{align}
    z(y)=z^*(y')+y'-y>0,~\tau_x(y)=\tau^*_x(y'), \forall y<\hat{y}.\label{eq:z+y-y'}
\end{align}
Next, we demonstrate by contradiction that $(z(y),\boldsymbol{\tau}(y))\in \mathcal{Z}^*(y)\times \mathcal{T}^*(y')$ for all $y<\hat{y}$.  
Suppose that $(z(y),\bs{\tau}(y))\notin \mathcal{Z}^*(y)\times \mathcal{T}^*(y')$ for some state $y<\hat{y}$. This would imply
\begin{align}\label{eq:R<Q_y'}
    R^*(y)<Q^*(y,(z(y),\bs{\tau}(y))).
\end{align}
Since $z'+y'=z(y)+y$ and $\tau^*_x(y')=\tau_x(y)$ as stated in \eqref{eq:z+y-y'}, it follows from the definition of $Q^*(y,\ba)$ in \eqref{eq:Qp-star} that
\begin{align}\label{eq:Q_y'}
  R^*(y')= Q^*(y',(z',\bs{\tau}^*(y')))
  =Q^*(y,(z(y),\bs{\tau}(y))).
\end{align}
Combining \eqref{eq:R<Q_y'} and \eqref{eq:Q_y'}, we obtain
$R^*(y)<R^*(y')$,
which contradicts the established equality $R^*(y')=R^*(y)$ for all $y,y'<\hat{y}$ based on $\frac{\partial R^*(y)}{\partial y}=0$.
Therefore, $(z(y),\bs{\tau}(y))$ must be optimal for all states $y<\hat{y}$. Moreover, it follows from \eqref{eq:z+y-y'} that there exists an optimal policy for all $y<\hat{y}$ such that
\begin{align}
    \frac{\partial z(y)}{\partial y}=-1,~\frac{\partial \tau_x(y)}{\partial y}=0.
\end{align}

\textit{(ii)} $y\geq \hat{y}$:
Suppose that there exists $y\geq\hat{y}$ such that $z^*(y)>0$. This would imply $y+z^*(y)>\hat{y}$, which would contradict the definition of $\hat{y}$ in \eqref{Eq-yp}. 
Thus, for all $y\geq \hat{y}$, the only optimal waiting action in $\mathcal{Z}^*(y)$ is $z^*(y)=0$.

Therefore, we have constructed an optimal policy in which the derivative of $z^*(y)$ is $-1$ for all $y<\hat{y}$ and $z^*(y)=0$ for all $y\geq \hat{y}$.

\section{Proof of Proposition~\ref{proposition_ubs}}
\label{proof:proposition_ubs}
From Corollary~\ref{corollary_solution_ubs}, we  observe that it is sufficient to prove that 
$\zeta_x(y)$ is nondecreasing in $x$. 
We start by adopting the shorthand notations 
\begin{align}
l(k)\triangleq\sum\limits_{x=1}^{k}\tau_x^*(y),~~
\Rhat_k\triangleq\frac{\partial R^*(l)}{\partial l}\bigg|_{l=l(k)}.\label{eq:shorthand}
\end{align}
Due to the convexity of $R^*(y)$, $\partial R^*(y)/\partial y$ increases in $y$. Moreover, since $l(k)$ is increasing in $k$,  it follows that  $\Rhat_k$ is increasing in $k$. 
With the additional shorthand $w_{j,k}\triangleq f(k)/\Fbar(j)$, 
it follows from~\eqref{eq:zeta} and~\eqref{eq:shorthand} that 
\begin{align}
    \zeta_j(y)&=
\sum\limits_{k=j}^b w_{j,k}
    \Rhat_{k}.\label{eq:zeta_j}
   \end{align}
Moreover, \eqref{eq:zeta_j} implies
\begin{align}
\zeta_{j+1}(y)-\zeta_j(y)
&=\sum_{k=j+1}^b [w_{j+1,k}-w_{j,k}]\Rhat_k-w_{j,j}\Rhat_j\nn
&= w_{j,j}\Bigl[\sum_{k=j+1}^{b}
w_{j+1,k}\Rhat_k -\Rhat_j\Bigr]\label{zetadiff2}\\
&= w_{j,j}\sum_{k=j+1}^{b}
w_{j+1,k}(\Rhat_k -\Rhat_j)\ge0.\label{zetadiff3}
\end{align}
Note that \eqref{zetadiff3} holds since $\sum\limits_{k=j+1}^b w_{j+1,k}=1$ and  $\Rhat_k$ is nondecreasing.

\section{Proof of Proposition~\ref{propo_special}}
\label{sec:proof_proposition_special}
When each update processing task has deterministic size $X_n=b$, the probability distribution of the task size is given by $f(1)=f(2)=\cdots=f(b-1)=0$ and $f(b)=1$. 
In such a scenario, the UTS and PTS cases are equivalent. 
Hence, we only analyze this special scenario in the PTS case, in which the problem formulation in~\eqref{eq:dinkel_pbs} simplifies to
\begin{subequations}\label{eq:pts_b}
\begin{align}
\gamma^*=\min_{z,\tau_b}\quad&b\tau_b+\frac{1}{2}(y+z),  \\
    {\rm s.t.}\quad&b\Ebatch(\tau_b)-\bar{P}(y+z)\leq0,\\
    &z\geq 0,~\tau_{\max}\geq\tau_b\geq\tau_{\min}.
\end{align}  
\end{subequations}
Since each update task consists of $b$ batches which require a total processing time of $b\tau_b$, the value of state $y$ remains $y=b\tau_b$. Therefore, Problem~\eqref{eq:pts_b} becomes
\begin{subequations}\label{eq:pts_bb}
\begin{align}
\gamma_p^*=\min_{z,\tau_b}\quad&\frac{3}{2}b\tau_b+\frac{1}{2}z,  \\
    {\rm s.t.}\quad&b\Ebatch(\tau_b)-\bar{P}(b\tau_b+z)\leq0,\label{P_b}\\
    &z\geq 0,~\tau_{\max}\geq\tau_b\geq\tau_{\min}.
\end{align}  
\end{subequations}
Based on~\eqref{P_b}, we derive that
\begin{align}
    z=\frac{b\Ebatch(\tau_b)}{\bar{P}}-b\tau_b,
\end{align}
which transforms Problem~\eqref{eq:pts_bb} as
\begin{subequations}
\begin{align}
\gamma^*=\min_{z,\tau_b}\quad&\frac{1}{2}\frac{b\Ebatch(\tau_b)}{\bar{P}}+b\tau_b,  \label{Delta_pts_b}\\
&z\geq0,~\tau_{\max}\geq\tau_b\geq\tau_{\min}.
\end{align}  
\end{subequations}
Since \eqref{Delta_pts_b} is convex in $\taubar_b$, it follows from Equation~\eqref{eq:energy} that the optimal $\tau^*_b$ can be obtained from
\begin{align}
    \frac{b}{2\bar{P}}\frac{2}{1-\alpha}(\tau^*_b)^{\frac{1+\alpha}{1-\alpha}}+b=0,
\end{align}
which implies the optimal solution of Proposition~\ref{propo_special}.

\section{Algorithm of Optimal-Wait Constant Speed.}
\label{sec:alg_optimal-wait}
On the basis of~\cite[Theorem 4]{sun2017update}, the optimal solution is given by
\begin{align}\label{eq:z_Y}
    z(y)=[\beta-y]_0^{Z_{\max}},
\end{align}
where 
$Z_{\max}$ is the upper bound of $z$, and $\beta>0$ satisfies
\begin{align}
    \mathbb{E}[Y+z(Y)]=\max\left(\frac{\mathbb{E}[X]\Ebatch(\tau)}{\bar{P}},\frac{\mathbb{E}[(Y+z(Y))^2]}{2\beta}\right).
\end{align}
\begin{algorithm}[t]
\caption{The Optimal-Wait Constant Speed Algorithm}
\label{alg:update_wait}
\KwIn{$\ell=0$, sufficiently large $u$. tolerance $\epsilon$;}
Given $\bar{P}$ and its corresponding $[\tau_{\min},\tau_{\max}]$\;
\For{$\tau\in[\tau_{\min},\tau_{\max}]$}{
\Repeat{$u-\ell\leq\epsilon$\label{alg-bisections}}{
$\beta:=(\ell+u)/2$\label{alg-bisection}\;
$o:=\mathbb{E}[Y+z(Y)]-\max\left(\frac{\mathbb{E}[X](1/\tau)^{\frac{2}{\alpha-1}}}{\bar{P}},\frac{\mathbb{E}[(Y+z(Y))^2]}{2\beta}\right)$\;
\eIf{$o\geq 0$}{$u:=\beta$}{$l:=\beta$}\
}
Update $z(Y)$ according to~\eqref{eq:z_Y}\;
Compute the average AoI $\gamma$ according to~\eqref{eq:update_aoi}\;
}
search minimal AoI as $\gamma^*$\;
\end{algorithm}
In lines~\ref{alg-bisection}-\ref{alg-bisections} of Algorithm~\ref{alg:update_wait}, we adopt the Bisection method in~\cite{sun2017update} to obtain the optimal $\beta$.
In addition, we set the value of $\bar{P}$ from $0.5$ to $8$, each $\bar{P}$ has a corresponding $[\tau_{\min},\tau_{\max}]$.
For each $\tau\in[\tau_{\min},\tau_{\max}]$, we find its corresponding optimal $\beta^*$ thus the optimal $z^*(Y)$ and calculate the average AoI. 
Finally, we choose the pair that achieves the minimum average AoI as the optimal policy at that power.

\end{document}